\documentclass[letterpaper,12pt]{report}
	\usepackage[utf8x]{inputenc}
	\usepackage{amsmath}
	\usepackage{amssymb,bm}
	\usepackage{pst-all}
	\usepackage{pstricks-add}
	\usepackage{pst-3dplot}
	\usepackage[dvips]{graphicx}
	
	\renewcommand{\theequation}{\thesection.\arabic{equation}}
	\renewcommand{\vec}{\bm}

\title{
Practical Guide to the \\
Symbolic Computation of Symmetries of Differential Equations
\footnote{
These notes were origionally created by Professor Steinberg and discussed
in the MACSYMA Newsletter in 1990. This version of the notes was created
by Professor Marinho in 2014. Professor Marinho has also updated the
code for doing the symbolic computation of symmetries which is available
for maxima/wxmaxima version of the original MACSYMA.}
}
\author{
Stanly Steinberg, Professor \\
Department of Mathematics and Statistics \\
University of New Mexico \\
Albuquerque, New Mexico, 87131, USA \\
E-mail: stanly@math.unm.edu
\and
Rubens de Melo Marinho Junior, Professor \\
Physics Department  \\
Instituto Tecnológico de Aeronáutica \\
Praça Mal. Eduardo Gomes \\
50 - Vila das Acácias \\
São José dos Campos - SP, 12228-900, Brazil \\
E-mail: marinho@ita.br or marinho.rubens@gmail.com}

\begin{document}
\maketitle
\tableofcontents

\setcounter{equation}{0}
\chapter{TRANSFORMATION GROUPS}

\section{Introduction}

The concept of symmetry is basic to mathematics and its applications
and has many interpretations for many different objects.
In this monograph, the objects under discussion will be systems of 
linear and nonlinear
ordinary or partial differential equations.  A symmetry will be a mapping
of the solution space of the system into the solution space of the system,
that is, a mapping that leaves the solution space invariant.
In the first few chapters the mapping will be restricted to transformations
that involve both the independent and dependent variables used in the system
of differential equations.  This idea of a symmetry includes, as a special
case, a change of variables that maps a differential equation into itself,
that is, leaves the differential equation invariant. 
The observation that rotations leave Laplace's equation invariant then
provides an example of the type of symmetries that will interest us.
Later, the notion of
a symmetry will be generalized to mappings that depend on the derivatives of
the solutions of the system of differential equations.

Many systems of differential equations have obvious symmetries that involve
translations, scalings of the variables, rotations and other geometric
transformation.  Most such symmetries can be found by inspection or by doing
some elementary calculation.  We will be interested in problems that have
{\it hidden symmetries}, that is, symmetries that cannot be found using
elementary techniques.  Of course, our method will produce the elementary
as well as the hidden symmetries.

The purpose of this paper is the description of a set of computer programs
written in the MACSYMA/VAXIMA symbol manipulation language.  For elementary
examples, one of these programs uses, as input,
a description of the differential
equation and produces, as output, the infinitesimal symmetries.  For more
complicated examples, the programs need some human help and thus the
programs are designed to make such human intervention easy.  Because of the
precision demanded by the computer coding, we have devoted the first
chapter of this monograph to a careful description of the methods that we use
to compute the symmetries.  This discussion was based on a minimum of
mathematical concepts in an attempt to make this material accessible to the
widest possible audience.

Perhaps one of the greatest impediments to the use of symmetries is the volume
of trivial algebra required to investigate even a modest problem.  We
believe we have provided software that reduces this algebra to an absolute
minimum!  Once the symmetries have been computed, then it is important to use
the symmetries to obtain some useful information about differential
equations.  No applications are included in this monograph.  However, a
perusal of our references should convince the reader that there are extensive
applications of these ideas.  We have included some programs to help with
applications.  Included are programs to transform differential equations
to new coordinate frames and programs to help calculate similarity
solutions of differential equations.

In principle, the symmetries of a system can be calculated by assuming a general
form for the symmetry and then solving the invariance condition for the
symmetry.  In practice this problem is intractable.  It has been discovered
that these computational difficulties can be over come by restricting the
notion of symmetry still further.
The crucial concept is that of a one parameter group and its infinitesimal.
Recall that two transformations can be applied sequentially to form a new
transformation which is called the composition of the two transformations.
Because only invertible transformations are considered, the composition of
transformations forms an algebraic group.  This explains one word in the 
terminology.  A one parameter group of transformations is a set of
transformations that depend on one real parameter and, moreover, addition
in the parameter is equivalent to composition of the transformations.  The
problem of directly computing the one parameter groups is as intractable
as computing a single transformation.

The idea that is responsible for the power of the group method is to
differentiate with respects to the parameter.  The derivative, at the origin,
with respects to the parameter produces the infinitesimal
group.  The derivative, at the origin, of the invariance condition with respects to the
parameter yields a system of {\it linear} partial differential equations 
for the infinitesimal symmetry.
These equations are called the {\it determining} equations.
If these
equations can be solved for the infinitesimal symmetries, then the one
parameter groups can be found from the infinitesimal groups by solving a
system of ordinary differential equations.  The process of deriving the one
parameter group from the infinitesimal group is called {\it exponentiation}.
The reasons for this terminology will become clear later.  The process of
finding the infinitesimal symmetries is called a mess.

Before the difficulties of finding the infinitesimal symmetries are described,
we note that for many problems that are intuitively symmetric, the infinitesimal
and one parameter groups of symmetries can, in fact, be computed.  On the other
hand, it is worth noting that the, discrete symmetries such as the
transformations that are reflections in one axis are {\it not}
computable by these methods.  Even though not all important symmetries are
computable by these methods, the symmetries that can be computed have many
important applications.

As was noted above, the problem of finding the symmetries of a system of
{\it nonlinear} (or linear) ordinary or partial differential equations 
becomes a problem of solving a system of {\it linear}
partial differential equations.  The system of partial differential equations
that determine the infinitesimal symmetries tends to have the following
properties.  The system of equations is large 
and overdetermined, that is,
there are more equations than unknowns.  Because the equations are linear their
solutions form a linear space.  Thus the infinitesimal symmetries form a
linear space.  Usually these linear spaces are essentially finite dimensional.
Their are theorems to this effect for some problems where the original
equation is linear, see Chandler [56] and Ovsiannikov [18].

We believe that one reason that these methods have not seen more applications
is that the problems of
deriving the determining equations and then solving equations for
for the infinitesimal symmetries is lengthy and tedious.
Other authors have provided programs to compute the determining equations
[27, 28, 30, 31, 32, 33], however, we believe that 
we were the first to add programs for solving the determining equations 
[33].  Because of the use of computers it is essential that
all definitions be clear and that all computational procedures be described precisely.
On the other hand, questions of differentiability of functions, convergence
of series, global definition of objects and many other favorite questions of
analysts play no significant role in the problems of computing symmetries.

If the reader is interested in some of the technical assumptions, then it is
worth noting that the material that is discussed here is local, that is, all
computations are carried out in some sufficiently small neighborhood of a
given point.  The assumption that all functions are analytic in a neighborhood
of the given point will cover most of the applications that are of interest.
If we are considering only neighborhoods of points, then the precise
definitions of the transformation group is quite technical because the
transformation group may move the given point or may move the neighborhood
that is being considered.  On the other hand, it is not possible to assume
that the transformation groups are defined on all of Euclidean space because
some of the more interesting groups have singularities.

These difficulties are not of great concern for us because we are interested in
computing the infinitesimal symmetries and the infinitesimal symmetries are
well defined in the neighborhood of some given point.  Thus we may take the point
of view that the one parameter groups are used to {\it motivate}
a careful definition of the infinitesimal groups and thus a technically
precise definition of one parameter group is not needed.  When we need to
compute a one parameter group from an infinitesimal group we will do this by
solving a system of ordinary differential equations and consequently the theory
of ordinary differential equations provides a firm foundation for these
calculations.

One place where the standard mathematical terminology is not sufficiently
precise is in ordinary differential equations.  Thus, in the simplest
situation of one first order ordinary differential equation, the equation is
frequently written
\begin{equation}
\frac{dy}{dx} = a(x,y)
\end{equation}
and the solution is written
\begin{equation}
y = y(x).
\end{equation}
Here the letter  $ y $  stands both for a real variable and the function that
is to be the solution of the differential equation.  Such minor abuses of
notation will blow away our computer programs.  Consequently we will change
the notation as follows.  Let  $ a(x,y) $  be a function, the variables
$ (x,y) $  and  $ y = f(x) $  be a function of  $ x $.  The differential
equation is then written
\begin{equation}
\frac{df(x)}{dx} = a(x,f(x)).
\end{equation}
We hope the reader will bear with such fine tuning of the notation.

This monograph assumes that the reader is familiar with multivariate calculus,
linear algebra, ordinary differential equations and partial differential
equations.  Some of the material needed from these subjects are reviewed in
the text or in the appendices.  It is not assumed that the reader has a
background in rigorous analysis or Lie group theory.  Not requiring Lie group theory
as background distinguishes this development from many other developments.
The use of the computer codes requires some familiarity with the MACSYMA/VAXIMA
symbol manipulation programs.

The general organization of this monograph is as follows.  The remainder
of Chapter 1 is devoted to introducing notation, developing background
material and setting up the procedure for calculating symmetries.  Reviews of
some well known material that is particularly important for our discussion
is presented in the appendices.  Chapter 2 covers the calculation of symmetries
of ordinary differential equations.  Chapter 3 is independent of Chapter 2 and
describes the calculation of the symmetries of partial differential equations.
In Chapter 4 the symmetries are allowed to depend on the derivatives of the
of the solutions of the differential equations being studied.
We call such transformations {\it jet} transformations although they are frequently
called Lie-Backlund transformations.  These transformations include the classical
contact transformations.

The remainder of Chapter 1 is organized as follows.  Section 2 sets up the
notation and introduces one parameter groups of point transformations and their
infinitesimals.  At this point there is no need to distinguish between dependent
and independent variables so they are thought of as a single set of variables.
Point transformations induce a natural action on functions of the variables.
In Section 3 {\it Lie series} are used to describe this action.  The solutions of
systems of differential equations are curves, surfaces or hyper surfaces in
higher dimensional spaces.  Section 4 describes the action of one parameter
groups and their infinitesimal on curves and surfaces.  Once the infinitesimal
symmetries for a system of differential equations are found then many
applications hinge on finding invariant functions and canonical coordinates
for the group so this is discussed in Section 5.  Finally, in Section 6,
we describe the method for calculating the infinitesimal symmetries of a system
of differential equations.

\section{One Parameter Groups and Infinitesimal Groups}

In this section a one parameter group of transformations
and the infinitesimal group of a one parameter group of transformations
are defined.
Also, it is shown that the infinitesimal group determines the one parameter
group.  The transformations will map points in  $ n $  dimensional
Euclidean space  $ R^n $  into itself.  In  $ R^n $ we will denote points by
\begin{equation}
\vec{v} = (x_1, x_2, \cdots ,x_n), \vec{\nu} = ( \xi_1, \xi_2,
\cdots , \xi_n).
\end{equation}
Of course,  $ x_i $  and  $ \xi_i $  are real variables.
A transformation will be denoted by a capital letter, say  $ \vec{G} $,
and then we can write
\begin{equation}
\vec{\nu} = \vec{G} ( \vec{v} ), \nu_i = G_i (x_i, \cdots , x_n ).
\end{equation}

Many of our examples will be given in low dimensional spaces so we introduce some
special notation.  If  $ n=1 $, then we will frequently use the notation
\begin{equation}
\xi = G(x)
\end{equation}
while if  $ n=2 $, we will frequently use
$$ \vec{v} = (x,y), \vec{\nu} = ( \xi , \eta ) $$
\begin{equation}
\vec{\nu} = \vec{G} ( \vec{v} ),
\end{equation}
$$ \xi = G_1 (x,y), \eta = G_2 (x,y) $$
and if  $ n=3 $, we will frequently use
$$ \vec{v} = (x,y,z), \nu = ( \xi , \eta , \zeta ), $$
\begin{equation}
\vec{\nu} = \vec{G} ( \vec{v} ), \\
\end{equation}
$$ \xi = G_1 (x,y,z), \eta = G_2 (x,y,z), \zeta = G_3 (x,y,z). $$
We do not always use these letters for variables, but we always use the
same style for labeling points.

A one parameter group of transformation is a set of transformations that
depends on one real parameter and the dependence on this parameter has
certain {\it exponential}
properties.  We usually denote the parameter by  $ \epsilon $, the group of
transformations by  $ \vec{G} ( \epsilon ) $ and the action of the group on 
points by
\begin{equation}
\vec{\nu} ( \epsilon ) = \vec{G}( \epsilon )( \vec{x} ) = \vec{G} 
( \epsilon , \vec{x} ).
\end{equation}
The intuitive idea behind the group properties is given by thinking of
$ \vec{G} ( \epsilon ) $  as being the exponential of ``something'',
\begin{equation}
\vec{G} ( \epsilon ) = e^{\epsilon L}.
\end{equation}
Then
$$ \vec{G} (0) = e^{0L} = 1, $$
\begin{equation}
\vec{G} ( \epsilon_1 ) \vec{G} ( \epsilon_2 ) = e^{ \epsilon_1 L }
e^{ \epsilon_2 L } = e^{ ( \epsilon_1 + \epsilon_2 )L } =
G( \epsilon_1 + \epsilon_2 ),
\end{equation}
$$ \vec{G} ( \epsilon ) \vec{G} (- \epsilon ) = e^{ \epsilon L } 
e^{ - \epsilon L } = e^0 = 1. $$
These ideas will be made rigorous in the section on Lie series.  For now, we
simply state the group properties: \\[.1in]
{\bf Properties.}
\begin{description}
     \item[A:] $ \vec{G} (0, \vec{v}) = \vec{v} $
     \item[B:] $ \vec{G} ( \epsilon_1, \vec{G} ( \epsilon_2, \vec{v} )) = 
\vec{G} ( \epsilon_1 + \epsilon_2, \vec{v} ) $
     \item[C:] $ \vec{G} ( \epsilon, \vec{G} (- \epsilon, \vec{v} )) = 
\vec{G} (- \epsilon, \vec{G} ( \epsilon, \vec{v} ))= \vec{v} $
\end{description}
Note that property C follows from A and B by choosing  $ \epsilon_1 =
\epsilon , \epsilon_2 = - \epsilon $  or  $ \epsilon_1 = - \epsilon ,
\epsilon_2 = \epsilon $. \\[.1in]
{\bf Definition.}
A one parameter group of transformations is a set of transformations
$ G( \epsilon , \vec{v} ) $  that depends on one real parameter and satisfies
properties 1 and 2 above.

As we go along, we will illustrate each idea with two examples, translations
in one variable and rotations in two variables.  Later in this chapter we
will discuss a rather complete set of elementary examples. \\[.1in]
{\bf Example - Translations.}
Translations in one variable are given by
\begin{equation}
\xi = \xi ( \epsilon ) = G( \epsilon ,x) = x + \epsilon
\end{equation}
\begin{center}
	\begin{pspicture}(0,-2)(6,1)
		\psline[linewidth=0.03](0,0)(6,0)
		\psdots[dotsize=0.2](1,0)
		\psdots[dotsize=0.2](4,0)
		\rput[b](1,0.2){$x$}
		\psline[linewidth=0.02]{|-|}(1,0.8)(4,0.8)
		\rput[b](2.5,0.9){$\epsilon$}
		\rput[b](4,0.2){$\xi$}
		\rput(3,-1){Translations}
	\end{pspicture}
\end{center}

The group properties are satisfied:
$$ G(0,x) = x $$
$$ G( \epsilon_1 ,G( \epsilon_2 +x)) = G( \epsilon_1 ,x+ \epsilon_2 )
= x + \epsilon_2 + \epsilon_1 = G( \epsilon_1 + \epsilon_2 ,x) $$ \\[.1in]
{\bf Example - Rotations.}
Rotations in two variables are given by 
\begin{equation}
( \xi ( \epsilon ), \eta ( \epsilon )) = \vec{G} ( \epsilon ,x,y) = (G_1 
( \epsilon ,x,y), G_2 ( \epsilon ,x,y))
\end{equation}
where
\begin{equation}
\xi = \xi ( \epsilon ) = \cos( \epsilon )x - \sin( \epsilon )y, \eta = 
\eta ( \epsilon ) = \sin ( \epsilon )x + \cos( \epsilon )y
\end{equation}
\begin{center}
	\begin{pspicture}(0,-1)(6,4)
		\psline[linewidth=0.03]{->}(0,0)(6,0)
		\psline[linewidth=0.03]{->}(0,0)(0,4)
		\psdots[dotsize=0.2](5,0)
		\rput{30}(0,0){\psdots[dotsize=0.2](5,0)}
		\psarc[linewidth=0.04cm](0,0){5}{-10}{40}
		\rput(3,-1){Rotations}
		\rput[bl](5.2,0.2){$\bm v=(x,y)$}
		\rput[bl](4.6,2.6){$\bm v(\epsilon)=  \bm G( \epsilon , \bm v)=(\xi,\eta)$}
	\end{pspicture}
\end{center}

The group properties are satisfied:
\begin{align}
\nonumber
\vec{G} (0,x,y) & = (\cos(0)x - \sin(0)y, \sin(0)x + \cos(0)y ) = (x,y)\,. \\
\nonumber
\vec{G} ( \epsilon_1 , \vec{G} ( \epsilon_2 ,(x,y)) & = \vec{G} 
( \epsilon_1 , G_1 ( \epsilon_2 , x,y),G_2 ( \epsilon_2 ,x,y) \\
\nonumber
& = \vec{G} ( \epsilon_1 , \cos( \epsilon_2 )x - \sin( \epsilon_2 )y, 
\sin( \epsilon_2 )x + \cos( \epsilon_2 )y) \\
\nonumber
& = ( \cos( \epsilon_1 )[ \cos( \epsilon_2 )x - \sin( \epsilon_2 )y] \\
\nonumber
&\quad\quad - \sin( \epsilon_1 )[ \sin( \epsilon_2 )x + \cos( \epsilon_2 )y], \\
\nonumber
& \quad \sin( \epsilon_1 )[ \cos( \epsilon_2 )x - \sin( \epsilon_2) y] \\
\nonumber
& \quad\quad + \cos( \epsilon_1 )[ \sin( \epsilon_2 )x + \cos( \epsilon_2 )y]) \\
\nonumber
& = ([ \cos( \epsilon_1 ) \cos( \epsilon_2 ) - \sin( \epsilon_1 ) 
\sin( \epsilon_2 )]x \\
\nonumber
&\quad\quad - [ \cos( \epsilon_1 ) \sin( \epsilon_2 ) + 
\sin( \epsilon_1 ) \cos ( \epsilon_2 )]y, \\
\nonumber
& \quad\quad [ \sin( \epsilon_1 ) \cos( \epsilon_2 ) + \cos( \epsilon_1 ) 
\sin( \epsilon_2 )]x \\
\nonumber
& \quad\quad+ [ \cos( \epsilon_1 ) \cos( \epsilon_2 ) 
- \sin( \epsilon_1 ) \sin( \epsilon_2 )]y) \\
\nonumber & = ( \cos( \epsilon_1 + \epsilon_2 )x - \sin( \epsilon_1 + \epsilon_2 )y, \\
\nonumber
&\quad\quad \sin( \epsilon_1 + \epsilon_2 )x + \cos( \epsilon_1 + \epsilon_2 )y) \\
\nonumber & = \vec{G} ( \epsilon_1 + \epsilon_2 ,x,y) \;.
\end{align}
By the way, the group properties are obvious from geometric considerations.

We now turn to the definition of the infinitesimal of a one parameter group of
transformations.  It is usual to think of the infinitesimal as the transformation
given by infinitely small changes in  $ \epsilon $.  Observe that if  $ \vec{v}$ 
is fixed and  $ \epsilon $  is allowed to vary, then  $ G ( \epsilon , \vec{v}) $ 
is a parametric curve in  $ R^n $  passing through  $ \vec{v} $  when
$ \epsilon = 0 $.

\begin{center}
	\begin{pspicture}(0,-1)(6,2)
		\psdots[dotsize=0.2](0.4,0.4)
		\psdots[dotsize=0.2](4,1.6)
		\pscurve[showpoints=false,dotsize=0.2,linewidth=0.03]{-}(3.3,0.5)(4,1.6)(0.4,0.4)
		\rput[bl](0.6,0){$\bm v=(x,y)$}
		\rput[bl](4.3,1.7){$\bm v(\epsilon)=  \bm G( \epsilon , \bm v)=(\xi,\eta)$}
		\rput(3,-1){One parameter curve}
	\end{pspicture}
\end{center}

\noindent {\bf Definition.}
The infinitesimal of a one parameter group is a vector field  $ \vec{T} ( \vec{v} ) $ 
given by the tangent vector to the curve  $ \vec{G} ( \epsilon , \vec{v} ) $  at
$ \epsilon = 0 $, that is
\begin{equation}
\vec{T} ( \vec{v} ) = \frac{ \partial}{\partial \epsilon } \vec{G} ( \epsilon , \vec{v} ) \mid_{ \epsilon =0 } .
\end{equation}

Because of the group property, the tangent vector to the curve  $ \vec{G}
( \epsilon , \vec{v} ) $  at any point  $ \epsilon_0 $  can be found in terms of the
infinitesimal vector field;
\begin{align} \label{G ODE}
\nonumber \frac{ \partial \vec{G} ( \epsilon , \vec{v}) }{ \partial \epsilon } 
\mid_{ \epsilon = \epsilon_0 } & = \frac{ \partial }{ \partial \epsilon } 
\vec{G} ( \epsilon + \epsilon_0 , \vec{v} ) \mid_{ \epsilon = 0 } \\
& = \frac{ \partial }{ \partial \epsilon } \vec{G} ( \epsilon , G( \epsilon_0 , 
\vec{v} )) \mid_{\epsilon =0 } \\
\nonumber & = \vec{T} ( \vec{G} ( \epsilon_0 , \vec{v} )).
\end{align}
One interpretation of this last formula is that the curve  $ \vec{G} ( \epsilon , \vec{v} ) $ 
is everywhere tangent to the direction field  $ \vec{T} ( \vec{v} ) $.

Another interpretation is that the formula is, in fact, a system of ordinary differential
equations for determining  $ \vec{G} ( \epsilon , \vec{v} ) $  from  
$ \vec{T}( \vec{v} ) $.
If we think of  $ \vec{v} $  and  $ \vec{T} $  as given and write  $ d/d \epsilon $ 
for  $ \partial / \partial \epsilon $  then \eqref{G ODE}
along with the fact that
$ \vec{G} (0, \vec{v} ) = \vec{v} $  gives
\begin{equation}
\frac{d}{ d \epsilon } \vec{G} ( \epsilon , \vec{v} ) = \vec{T} ( \vec{G} 
( \epsilon , \vec{v} )), \vec{G} (0, \vec{v} ) = \vec{v} .
\end{equation}
This is an initial value problem that determines  $ \vec{G} ( \epsilon , \vec{v}) $ 
in terms  $ \vec{T} ( \vec{v} ) $. \\[.1in]
{\bf Example - Translations - Continued. }
The translation group in one dimension is given by
$$ G( \epsilon ,x) = x + \epsilon \;.$$
The infinitesimal group is given by
\begin{equation}
T(x) = \frac{d}{ d \epsilon } G( \epsilon ,x) \mid_{ \epsilon =0 } = 1.
\end{equation}
The initial value problem for determining  $ G( \epsilon ,x) $  in terms of
$ T(x) $  is
\begin{equation} \label{ODEnotation}
\frac{d}{ d \epsilon } G( \epsilon ,x) = 1, G(0,x) = x.
\end{equation}
This notation for an ordinary differential is not the usual so we introduce
\begin{equation}
\xi = \xi ( \epsilon ) = G( \epsilon ,x)
\end{equation}
and then rewrite \eqref{ODEnotation} as
\begin{equation}
\frac{ d \xi }{ d \epsilon } = 1, \xi (0) = x.
\end{equation}
This can be integrated to
\begin{equation}
\xi = \epsilon + C
\end{equation}
and then the initial condition gives  $ C = x $  or
\begin{equation}
\xi = \epsilon + x.
\end{equation}
Thus we see that the infinitesimal group does determine the group! \\[.1in]
{\bf Example - Rotations - Continued. }
The rotation group in two dimensions is given by  
\begin{equation}
\vec{G} ( \epsilon ,x,y) = ( \cos( \epsilon )x - \sin( \epsilon )y, 
\sin( \epsilon )x + \cos( \epsilon )x).
\end{equation}
The infinitesimal group is given by
\begin{align}
\nonumber \vec{T} (x,y) & = \frac{d}{ d \epsilon } \vec{G} ( \epsilon ,x,y) \mid_{ \epsilon =0 } \\
\nonumber & = \frac{d}{ d \epsilon } ( \cos( \epsilon )x- \sin( \epsilon )y, \sin( \epsilon )x+ \cos ( \epsilon )y) \mid_{ \epsilon =0 } \\
& = (- \sin( \epsilon )x- \cos( \epsilon )y, \cos( \epsilon )x- \sin( \epsilon )y) \mid_{ \epsilon =0 } \\
\nonumber & = (-y,x).
\end{align}
We can now make a sketch of the infinitesimal group which is nothing more than
the direction field (with magnitude) given by  
$ \vec{T} (x,y) = (-y,x) $
\begin{center}
	\begin{pspicture}(0,-1)(6,6)
		\psline[linewidth=0.03]{->}(0,0)(6,0)
		\psline[linewidth=0.03]{->}(0,0)(0,5)
		\psline[linewidth=0.04]{->}(0,0)(5,0)
		\psdots[dotsize=0.2](5,0)
		\rput{30}(0,0){\psdots[dotsize=0.2](5,0)}
		\psarc[linewidth=0.04cm](0,0){5}{-10}{40}
		\rput(3,-1.5){Infinitesimal Rotations}
		\psline[linewidth=0.04]{->}(5,0)(5,5)
		\rput[bl](5.2,0.2){$\bm v=(x,y)$}
		\rput[bl](5.2,4.6){$\bm T(x,y)$}
	\end{pspicture}
\end{center}
\vspace{2in}

The initial value problem for determining  $ \vec{G} ( \epsilon ,x,y) $  in terms
of  $ \vec{T} (x,y) $  is easier to write down if we introduce
\begin{equation}
( \xi , \eta ) = ( \xi ( \epsilon ), \eta ( \epsilon )) = (G_1 ( \epsilon ,x,y),G_2 ( \epsilon ,x,y)).
\end{equation}
Then the initial value problem becomes
\begin{equation}
\frac{d}{ d \epsilon } ( \xi , \eta ) = T( \xi , \eta ) = (- \eta , \xi ),( \epsilon (0) , \eta (0)) = (x,y)
\end{equation}
or
\begin{equation}
\frac{ d \xi }{ d \epsilon } = - \eta , \frac{ d \eta }{ d \epsilon } = \xi ,
\xi (0) = x, \eta (0) = y.
\end{equation}
This simple initial value problem can be solved in many different ways.  We note
that
\begin{equation}
\frac{ d^2 \xi }{ d \epsilon^2 } + \xi = 0
\end{equation}
so that
\begin{equation}
\xi = A\cos( \epsilon ) + B\sin( \epsilon )
\end{equation}
and then
\begin{equation}
\eta = - \frac{ d \xi }{ d \epsilon } = -B\cos( \epsilon ) + A\sin( \epsilon )
\end{equation}
The initial conditions then give
\begin{equation}
A=x, B=y
\end{equation}
or
\begin{equation}
\xi = x \cos( \epsilon ) - y \sin( \epsilon ), \eta = x \sin( \epsilon )
+ y \cos( \epsilon ).
\end{equation}
Again we see that the infinitesimal group does determine the group.

There is an alternate method for determining the infinitesimal group that uses
power series.  The expansion of  $ \vec{G} ( \epsilon , \vec{v} ) $  at
$ \epsilon = 0 $  is given by
\begin{align}
\nonumber \vec{G} ( \epsilon , \vec{v} ) & = \vec{G} (0, \vec{v} ) + \frac{d}{ d \epsilon } \vec{G} ( \epsilon , \vec{v} ) \mid_{ \epsilon =0 } \epsilon + \frac{ d^2 }
{ d \epsilon^2 } \vec{G} ( \epsilon , \vec{v} ) \mid_{ \epsilon =0 } 
\frac{ \epsilon^2 }{2} + \cdots \\
\nonumber & = \vec{v} + \vec{T} ( \vec{v} ) \epsilon + \cdots
\end{align}
Thus  $ \vec{T} ( \vec{v} ) $  is the coefficient of  $ \epsilon $  in the power
series expansion of  $ \vec{G} ( \epsilon , \vec{v} ) $. \\[.1in]
{\bf Example - Rotations - Continued.}  As before
\begin{align}
G( \epsilon ,x,y) & = (\cos( \epsilon )x - \sin( \epsilon )y, \sin( \epsilon )
x + \cos( \epsilon )y) \\
\nonumber
& = ((1- \frac{ \epsilon^2 }{2} + \cdots )x - ( \epsilon - 
\frac{ \epsilon^3 }{6} + \cdots )y, \\
\nonumber
&\quad\quad ( \epsilon -` \frac{ \epsilon^3 }{6} + \cdots )x + (1- \frac{ \epsilon^2 }{2} + \cdots )y) \\
\nonumber & = (x,y) + (-y,x) \epsilon + (-x,-y) \frac{ \epsilon^2 }{2} + \cdots 
\end{align}
and consequently 
\begin{equation}
\vec{T} (x,y) = (-y,x).
\end{equation}

Here is a catalog of some elementary one dimensional examples that appear in
applications.
\begin{center}
\begin{tabular}{l|l|l}
$ \xi ( \epsilon ) = G( \epsilon ,x) $ & $ T(x) $ & common name \\
\hline
$ \xi ( \epsilon ) = x + \epsilon $ & $ T(x) = 1 $ & translations \\
$ \xi ( \epsilon ) = e^\epsilon x $ & $ T(x) = x $ & dilations \\
$ \xi ( \epsilon ) = x/(1- \epsilon x) $ & $ T(x) = x^2 $ & conformal maps
\end{tabular}
\end{center}
\centerline{Examples}
\vspace{.2in}
{\bf Example - Conformal.}  If
\begin{equation}
\xi ( \epsilon ) = x/(1- \epsilon x),
\end{equation}
then
\begin{equation}
T(x) = \frac{d}{ d \epsilon } \xi ( \epsilon ) \mid_{ \epsilon =0 } =
\frac{ x^2 }{ (1- \epsilon x )^2 } \mid_{ \epsilon =0 } = x^2 .
\end{equation}
Alternately, the power series for  $ 1/(1- x) $  is given by
\begin{equation}
\frac{1}{1-x} = 1 + x + O(x^2 )
\end{equation}
and consequently
\begin{equation}
\frac{x}{ 1- \epsilon x } = x(1+ \epsilon x +O( \epsilon x )^2 ) = x + \epsilon x^2 + O( \epsilon^2 ).
\end{equation}
As before,
\begin{equation}
T(x) = x^2 .
\end{equation}

On the other hand, if we know that  $ T(x) = x^2 $, then the group  $ \xi ( \epsilon ) = G( \epsilon ,x) $
satisfies the initial value problem
\begin{equation}
\frac{ d \xi }{ d \epsilon } = \xi^2 , \xi (0) = x.
\end{equation}
This initial value problem can be solved by separating variables and the
result is
\begin{equation}
\xi ( \epsilon ) = \frac{x}{ 1- \epsilon x } .
\end{equation} \\[.1in]
{\bf Exercise.}  Repeat the above calculation for the remainder of the previous table.

In two or more dimensions it is helpful to use column vector and matrix
notation.  To see how to do this we redo the rotation example. \\[.1in]
{\bf Example - Rotation - Continued.}  We rewrite the rotation group in the form
\begin{equation}
\left[ \begin{array}{c} 
\xi ( \epsilon ) \\
\eta ( \epsilon ) 
\end{array} \right] = \left[ \begin{array}{c}
\xi \\
\eta
\end{array} \right] = \left[ \begin{array}{ll}
\cos( \epsilon ) & - \sin( \epsilon ) \\
\sin( \epsilon ) & \cos( \epsilon )
\end{array} \right] \left[ \begin{array}{c}
x \\
y
\end{array} \right] .
\end{equation}
The infinitesimal group is given by
\begin{align}
\nonumber \frac{d}{ d \epsilon } \left[ \begin{array}{c}
\xi \\
\eta
\end{array} \right] 
\mid_{ \epsilon =0 } & =
\left[ \begin{array}{ll}
- \sin( \epsilon ) & - \cos( \epsilon ) \\
- \cos( \epsilon ) & - \sin( \epsilon )
\end{array} \right] 
\left[ \begin{array}{c}
x \\
y
\end{array} \right] 
\mid_{ \epsilon =0 } \\
& = \left[ \begin{array}{rr}
0 & -1 \\
1 & 0
\end{array} \right] \left[ \begin{array}{c}
x \\
y
\end{array} \right] \\
\nonumber
& = \left[ \begin{array}{c}
-y \\
x
\end{array} \right] .
\end{align}
The differential equations for determining the rotation group in terms of the
infinitesimal rotation group are then
\begin{align}
\frac{d}{ d \epsilon } \left[ \begin{array}{c}
\xi \\ \eta \end{array}
\right] & = \left[ \begin{array}{c}
\xi^\prime \\ \eta^\prime \end{array}
\right] = \left[ \begin{array}{c}
-\eta \\ \xi \end{array}
\right] = \left[ \begin{array}{rr}
0 & -1 \\ 1 & 0 \end{array}
\right] \left[ \begin{array}{c}
\xi \\ \eta \end{array}
\right] , \\
\nonumber
\left[ \begin{array}{c}
\xi (0) \\ \eta (0) \end{array}
\right] & = \left[ \begin{array}{c}
x \\ y \end{array}
\right] .
\end{align}
These are the same differential equations that occurred previously in this example.
Note that the solution of this {\it constant coefficient linear} system of
differential equations can be written using {\it matrix exponentials} as
\begin{equation}
\left[ \begin{array}{c} 
\xi \\ \eta \end{array}
\right] = e^{ \left[ \begin{array}{rr}
0 & -1 \\
1 & 0 \end{array}
\right] \epsilon } \left[ \begin{array}{c}
x \\ y \end{array}
\right] .
\end{equation}
We will discuss the meaning of this notation shortly.

Let us now turn to some examples in  $ R^n $.  If
$ \vec{a} = (a_1, \ldots , a_n ) $
is any given vector, then the translations in the direction  $ \vec{a} $  are
written
\begin{equation}
\vec{G} ( \epsilon , \vec{v} ) = \vec{v} + \epsilon \vec{a} .
\end{equation}
{\bf Exercise.} Check that this is a group of transformations with infinitesimal
\begin{equation}
\vec{T} ( \vec{v} ) = \vec{a} .
\end{equation}
Make a sketch of this group and its infinitesimal!

If  $ A $  is any  $ n \times n $  real matrix then  $ A $  will generate a
transformation group on  $ R^n $  that is analogous to the rotation 
group.  This analogy is based on matrix exponentials [9, Chapter 10].
The definition of the exponential of a matrix is a direct generalization 
of the power series definition of an exponential of a number. \\[.1in]
{\bf Definition.}  If  $ A $  is a given matrix, then
\begin{equation}
e^{ \epsilon A } = \sum_{n=0}^{\infty} \epsilon^n \frac{ A^n }{n!} = I + 
\epsilon A + \frac{ \epsilon^2 A }{2} + \cdots
\end{equation}
where  $ I $  is the identity matrix. \\[.1in]
{\bf Note.}
The parameter  $ \epsilon $  may be omitted (choose  $ \epsilon =1 $) in this
definition.  However, we will find this parameter useful.

It is a standard result that this series converges for all  $ A $  and
$ \epsilon $  and that the entries in the matrix
\begin{equation}
M( \epsilon ) = \exp( \epsilon A ) = e^{ \epsilon A }
\end{equation}
are analytic functions of  $ \epsilon $ and the entries of  $ A $.  Clearly
\begin{equation}
e^{0A} = I.
\end{equation}
An argument involving only rearranging summations will show that
\begin{equation}
e^{ \epsilon_1 A } e^{ \epsilon_2 A } = e^{ ( \epsilon_1 + \epsilon_2 )A }
\end{equation}
and then this identity implies that
\begin{equation}
e^{ -\epsilon A } e^{ \epsilon A } = I = e^{ \epsilon A } e^{ -\epsilon A } .
\end{equation}
Note that
\begin{equation}
Ae^{ \epsilon A } = e^{ \epsilon A } A
\end{equation}
and that
\begin{equation}
\frac{d}{ d \epsilon } e^{ \epsilon A } = \sum_{n=1}^{\infty} \epsilon^{n-1} 
\frac{ A^n }{(n-1)!} = A \sum_{m=0}^{\infty} \epsilon^m 
\frac{ A^m }{m!} = Ae^{ \epsilon A } ,
\end{equation}
that is, the matrix function  $ M( \epsilon ) = \exp( \epsilon A ) $  is
a one parameter group of matrices and is a solution of the initial value problem
\begin{equation} \label{Matrix DE}
\frac{d}{ d \epsilon } M( \epsilon ) = AM( \epsilon ) , M(0) = I .
\end{equation}
Frequently  $ M( \epsilon ) $  is called a fundamental solution matrix for the
differential equation \eqref{Matrix DE}. \\[.1in]
{\bf Exercise.}  Suppose that a matrix function  $ M( \epsilon ) $  satisfies
\begin{equation}
M(0) = I, M( \epsilon_1 )M( \epsilon_ 2 ) = M( \epsilon_1 + \epsilon_2 ).
\end{equation}
Then show that
\begin{equation}
M( \epsilon ) = e^{ \epsilon A }
\end{equation}
where
\begin{equation}
A = \frac{d}{ d \epsilon } M( \epsilon ) \mid_{ \epsilon =0 } .
\end{equation}

It is now possible to generate many groups of transformations on
$ R^n $. \\[.1in]
{\bf Proposition.}
If  $ A $  is any real  $ n \times n $  matrix, and  $ M( \epsilon ) = \exp(
\epsilon A ) $, then
\begin{equation}
\vec{\nu} = \vec{\nu} ( \epsilon ) = M( \epsilon ) \vec{v} 
\end{equation}
is a transformation group on  $ R^n $  with infinitesimal
\begin{equation}
\vec{T} ( \vec{v} ) = A \vec{v} .
\end{equation}
{\bf Exercise.}  Prove this proposition. \\[.1in]
{\bf Exercise.}  Show that the matrix
\begin{equation}
\left[ \begin{array}{cc}
a & 0 \\
0 & b
\end{array} \right]
\end{equation}
generates the group
\begin{equation}
\left[ \begin{array}{c}
\xi \\ \eta
\end{array} \right] = \left[ \begin{array}{c}
e^{ \epsilon a } x \\ e^{ \epsilon b } y
\end{array} \right]
\end{equation}
which is called a dilation group.  Note that a similar result holds for any
diagonal matrix.

One useful way to evaluate matrix exponentials is based on diagonalization
results.  Thus suppose that  $ S $  is a matrix who's columns are linearly
independent eigenvectors of  $ A $.  Then 
\begin{equation}
D=S^{-1} AS
\end{equation}
is a diagonal matrix.  The power series definition of the exponential then
gives
\begin{equation}
e^{ \epsilon A } = e^{ \epsilon SDS^{-1} } = Se^{ \epsilon D } S^{-1}
\end{equation}
and, as we saw above, $ \exp( \epsilon D ) $  is easy to compute
\begin{equation}
e^{ \epsilon \left( \begin{array}{ccc}
\lambda_1 & 0 & \cdots \\
0 & \lambda_2 & \cdots \\
\cdots & \cdots & \cdots
\end{array} \right) } = \left( \begin{array}{ccc}
e^{ \epsilon \lambda_1 } & 0 & \cdots \\
0 & e^{ \epsilon \lambda_2 } & \cdots \\
\cdots & \cdots & \cdots
\end{array} \right) .
\end{equation}
{\bf Exercise.}  Show that the matrix
\begin{equation}
\left[ \begin{array}{rr}
0 & 1 \\
1 & 0
\end{array} \right]
\end{equation}
generates the group
\begin{equation}
\left[ \begin{array}{c}
\xi \\ \eta
\end{array} \right] = 
\left[ \begin{array}{ll}
\cosh( \epsilon ) & \sinh( \epsilon ) \\
\sinh( \epsilon ) & \cosh( \epsilon )
\end{array} \right] 
\left[ \begin{array}{c}
x \\ y
\end{array} \right] .
\end{equation}
{\bf Exercise.}  Use the diagonalization procedure to show that
\begin{equation}
\left[ \begin{array}{rr}
0 & 1 \\
-1 & 0
\end{array} \right]
\end{equation}
generates the rotation group.  Note that  $ D $  and  $ S $  may be complex
matrices.

\section{Action on Functions and Lie Series.}

If  $ \vec{G} ( \epsilon , \vec{v} ) $  is a transformation group on
$ R^n $,  then  $ \vec{G} $  has a natural action on functions,
written  $ G( \epsilon ,f) $,  and given by
\begin{equation} \label{Function Action}
G( \epsilon ,f)( \vec{v} ) = f( \vec{G} ( \epsilon , \vec{v} )) \;.
\end{equation}
It is an abuse of notation to use the letter  $ G $  to label both the
transformation group and the action of the group on functions.  However,
these objects are so closely related that we will come to identify them and
then the notation is appropriate.  Note that $ G $
is not bold face
in the action on functions formula, \eqref{Function Action}.
Some authors put a 
minus sign in the definition of the action on functions
\begin{equation}
G( \epsilon ,f)( \vec{v} ) = f( \vec{G} (-\epsilon , \vec{v} )) \;.
\end{equation}
We will {\it not} use the minus sign. \\[.1in]
{\bf Example - Translations.}  If  $ f $  is a function of the real 
variable  $ x $, $ y=f(x) $,  and
$ G( \epsilon ,x) = x + \epsilon $  is the translation group, then
\begin{equation}
G ( \epsilon ,f)(x) = f(x+ \epsilon ) \;.
\end{equation}
\begin{center}
	\begin{pspicture}(0,0)(9,3)
		\psgrid[gridcolor=green,subgridcolor=yellow,gridlabels=0]
		\pscurve[showpoints=false,dotsize=0.2,linewidth=0.03]{-}(4,1.6)(3.3,0.5)(0.4,0.4)
		\pscurve[showpoints=false,dotsize=0.2,linewidth=0.03]{-}(8,1.6)(7.3,0.5)(4.4,0.4)
		\rput[bl](1.5,0.6){$f(x+\epsilon)$}
		\rput[bl](5.5,0.6){$f(x)$}
	\end{pspicture}
\end{center}
\centerline{Translation of Functions}\ \\[12pt]

Thus we see that right translation of points corresponds to left translation
of functions.  If we had included the minus sign in the definition of  $ G $,
then right translation of points would have corresponded to right translation 
of functions.

We now compute the infinitesimal transformation group of $ G( \epsilon ,f) $.
\\[.1in]
{\bf Proposition.}  If  $ \vec{G} ( \epsilon , \vec{v} ) $  is a one parameter transformation group
on  $ R^n $  with infinitesimal  $ \vec{T} ( \vec{v} ) $, $ f( \vec{v} ) $ 
is a function on  $ R^n $  and  $ G( \epsilon ,f) $  is defined by
\begin{equation}
G( \epsilon ,f) ( \vec{v} ) = f( \vec{G} ( \epsilon , \vec{v} )),
\end{equation}
then the infinitesimal  $ T (f) $  of  $ G( \epsilon ,f) $  is given by
\begin{equation}
T(f) ( \vec{v} ) = \vec{T} ( \vec{v} ) \cdot \vec{\nabla} f(v)
\end{equation}
where
\begin{equation}
\vec{\nabla} = \left( \frac{ \partial }{ \partial x_1 } , \cdots ,
\frac{ \partial }{ \partial x_n } \right)
\end{equation}
{\bf Proof.}  The infinitesimal is given by
\begin{equation}
T(f)( \vec{v} ) = \frac{d}{ d \epsilon } G( \epsilon ,f)( \vec{v} ) 
\mid_{ \epsilon =0 } = \frac{d}{ d \epsilon } f( \vec{G} ( \epsilon , 
\vec{v})) \mid_{ \epsilon =0 } \;.
\end{equation}
The chain rule then gives
\begin{align}
\nonumber T(f)( \vec{v} ) & = \sum_{k=1}^{n} \frac{ \partial f }{ \partial x_k } 
( \vec{G} ( \epsilon , \vec{v} )) \frac{ dG_k }{ d \epsilon } 
( \epsilon , \vec{v} ) \mid_{ \epsilon =0 } \\
& = \sum_{k=1}^{n} \frac{ \partial f }{ \partial x_k } ( \vec{v} )T_k 
( \vec{v} ) \\
\nonumber & = \vec{T} ( \vec{v} ) \cdot \vec{\nabla} f( \vec{v} )\;.
\end{align}
This result provides us with an identification 
of vector fields with first order partial differential operators,
\begin{equation}
\vec{T} ( \vec{v} ) < - >  \vec{T} ( \vec{v} ) \cdot \vec{\nabla} f = T(f)\;.
\end{equation}
This identification is used extensively in the literature and in this
manuscript.

In our work we will meet situations where we are given a vector field
$ \vec{T} ( \vec{v} ) $, or equivalently a first order partial differential
operator  $ \vec{T} ( \vec{v} ) \cdot \vec{\nabla} $  and we will want to find the
group action associated with the vector field or operator.  A useful
representation of group action is given by the exponential of the
differential operator
\begin{equation}
e^{ \epsilon \vec{T} ( \vec{v} ) \cdot \vec{\nabla} } \;.
\end{equation}
Such exponentials are called {\it Lie series}.  Lie series have been 
studied extensively [10, 101, 188, 189].  For the convenience of the reader
we now list the properties of Lie series.  The proofs of the first 6 properties
are not difficult and proofs of all properties are included in the cited
literature.  After we state the properties we will give some examples.

A Lie series is an exponential
\begin{equation}
e^{tL} = \sum_{k=0}^{\infty} \frac{ t^k L^k }{k!}
\end{equation}
of a first order differential operator
\begin{equation}
L = \vec{T} ( \vec{v} ) \cdot \vec{\nabla} = \sum_{i=1}^{n} T_i ( \vec{v} )
\frac{ \partial}{ \partial x_i }
\end{equation}
in the  $ n $  variables  $ \vec{v} = (x_1, \cdots ,x_n ) $. A Lie series
operates on scaler functions  $ g( \vec{v} ) $  that are analytic in the
$ n $  variables  $ \vec{v} = (x_1, \cdots , x_n ) $.
The action of the Lie series on a function  $ g( \vec{v} ) $, analytic near
$ \vec{v} = 0 $, is given by
\begin{equation}
e^{ \epsilon L } g( \vec{v} ) = \sum_{n=0}^{\infty} \frac{ \epsilon^k L^k }{k!}
g( \vec{v} ) = \sum_{n=0}^{\infty} \frac{ \epsilon^k }{k!} \left( 
\sum_{i=1}^{n} f_i ( \vec{v} ) \frac{\partial}{ \partial x_i } \right)^k 
g( \vec{v} )\;.
\end{equation}
\centerline{\bf PROPERTIES}

Assume that  $ f( \vec{v} ), g( \vec{v} ) $  and  $ h( \vec{v} ) $  are analytic functions near
$  \vec{v} = 0 $, that  $ a $  and  $ b $  are real constants, that
$ c( \epsilon ) $
is an analytic real valued function of $ \epsilon $ and that $ L $ is as in
1.81. \\[.1in]
1)  {\bf Convergence.}
\begin{equation}
e^{ \epsilon L } g( \vec{v} )
\end{equation}
is a well defined analytic function of $ \vec{v} $ and $ \epsilon $ 
for  $ \vec{v} $  and  $ \epsilon $  small.
This is a simple version of the Cauchy-Kowalewski theorem. \\[.1in]
2)  {\bf Time Derivative.}
\begin{equation}
\frac{d}{dt} e^{ c( \epsilon )L } = c^\prime( \epsilon )L e^{ c( \epsilon )L } 
= c^{ c( \epsilon )L } c^\prime ( \epsilon )L
\vspace{.1in}
\end{equation}
3)  {\bf Linearity.}
\begin{equation}
e^{ \epsilon L } (ag+bh) = ae^{ \epsilon L } g + be^{ \epsilon L } h
\vspace{.1in}
\end{equation}
4)  {\bf Product Preservation.}
\begin{equation}
e^{ \epsilon L } (gh)=(e^{ \epsilon L } g)(e^{ \epsilon L } h)
\end{equation}
Lie Series act on vector valued functions  $ \vec{f} ( \vec{v} )=(f_1 
( \vec{v} ), \ldots $, by acting on each component,
\begin{equation}
e^{ \epsilon L } \vec{f} ( \vec{v} ) = (e^{ \epsilon L } f_1 ( \vec{v} ),
\ldots ,)\;. 
\vspace{.1in}
\end{equation}
5)  {\bf Composition.}
\begin{equation}
e^{ \epsilon L } g( \vec{v} ) = g(e^{\epsilon L }  \vec{v} ),
\vspace{.1in}
\end{equation}
6)  {\bf Differential Equation Property.}

If
\begin{equation}
\vec{v} ( \epsilon ) = e^{ \epsilon L } \vec{v} 
\end{equation}

then
\begin{equation}
\vec{v}^\prime ( \epsilon ) = f( \vec{v} ( \epsilon )), \vec{v} (0) = \vec{v}\;.
\end{equation}
We now suppose that  $ P $  is another first order differential operator
and define successive commutators by:
$$ [L,\cdot]^0 P=P $$
\begin{equation}
[L,\cdot]^1 P=LP-PL
\end{equation}
$$ [L, \cdot ]^n P=[L, \cdot ]^{n-1} [L, P], n \geq 1 $$ \\[.1in]
7)  {\bf Similarity.}
\begin{equation}
e^{ \epsilon L } Pe^{ -\epsilon L } = e^{ \epsilon [L, \cdot ] } P = 
\sum_{k=0}^{\infty} \frac{ \epsilon^k }{k!} [L, \cdot ]^k P
\vspace{.1in}
\end{equation}
8)  {\bf Function Multiplier.}
\begin{equation}
e^{ \epsilon L } ge^{ -\epsilon L } h = (e^{ \epsilon L } g)h
\end{equation}
Although we do not yet have a use for the next property we list it because it is very
useful in other contexts, has important Lie algebraic applications and is
certainly the most interesting elementary fact about Lie series.
\\[.1in]
9)  {\bf Noncommuting Exponential Identities.}
\begin{align}
\nonumber e^{ \epsilon (L+P) } & =  e^{ \epsilon L } e^{ \epsilon P } e^{ \epsilon^2 
L_2 } e^{ \epsilon^3 L_3 } e^{ \epsilon^4 L_4 } e^{ \epsilon^5 L_5 } \cdots
\\
& =  \cdots e^{ \epsilon^5 L_5 } e^{ -\epsilon^4 L_4 } e^{ \epsilon^3 L_3 } 
e^{ -\epsilon^2 L_2 } e^\epsilon P e^\epsilon L \\
\nonumber e^{ \epsilon L } e^{ \epsilon P } & =  e^{ \epsilon L+ \epsilon P+ 
\epsilon^2 W_2 + \epsilon^3 W_3 + \cdots }
\end{align}
where
\begin{equation}
L_2 =- \frac{1}{2} [L,P]  \quad , \quad W_2 = \frac{1}{2} [L,P],
\end{equation}
and so forth.  Here each $ L_k $ and $ W_k $ are $ k $-fold commutators
of $ L $ and $ P $. \\[.1in]
{\bf Example - Translations - Continued.}

The infinitesimal action of translations is given by the differential operator
\begin{equation}
L= \frac{d}{dx}
\end{equation}
and consequently the one parameter group of translations should be given by
\begin{equation}
e^{ \epsilon \frac{d}{dx} }\;.
\end{equation}
The Composition Property gives
\begin{equation}
e^{ \epsilon \frac{d}{dx} } f(x) = f(e^{ \epsilon \frac{d}{dx} } x)
\end{equation}
so we need only compute
\begin{equation}
e^{ \epsilon \frac{d}{dx} } x = \sum_{k=0}^{\infty} \frac{\epsilon^k}{k!}
\left( \frac{d}{dx} \right)^k x = x + \epsilon
\end{equation}
as was desired. \\[.1in]
{\bf Exercise - Dilations - Continued.}

As in the above example, show that
\begin{equation}
e^{ \epsilon x \frac{d}{dx} } f(x) = f( e^\epsilon x ) \;.
\vspace{.1in}
\end{equation}
{\bf Exercise - Conformal - Continued.}

Show that
\begin{equation}
e^{ \epsilon x^2 \frac{d}{dx} } f(x) = f \left( \frac{x}{1-\epsilon x } 
\right) \;.
\vspace{.1in}
\end{equation}
{\bf Example - Rotations - Continued.}

The vector field that is the infinitesimal of the rotation group is
\begin{equation}
\vec{T} (x,y) = (-y,x)
\end{equation}
so the infinitesimal action on functions is given by
\begin{equation}
L = -y \frac{\partial}{\partial x} + x \frac{\partial}{\partial y} \;,
\end{equation}
that is,
\begin{equation}
(Lf)(x,y) = -y \frac{\partial f }{\partial x } (x,y) + x \frac{\partial f }
{ \partial y } (x,y) \;.
\end{equation}
Consequently the one parameter group of rotations is given by
\begin{equation}
e^{ \epsilon L } f(x,y) = f(e^{ \epsilon L } x,e^{ \epsilon L } y)
\end{equation}
so we need only compute
\begin{equation}
e^{ \epsilon L } x , \quad e^{ \epsilon L } y \;.
\end{equation}
The power series gives
\begin{align}
\nonumber e^{ \epsilon L } x & = \sum_{k=0}^{\infty} \frac{ \epsilon^k }{k!}
\left( -y \frac{\partial}{\partial x } + x \frac{\partial}{\partial y } 
\right)^k x
\\
& = x + \epsilon (-y) + \frac{ \epsilon^2 }{2} (-x) + \frac{ \epsilon^3 }{3!}
y + \cdots \\
\nonumber & = \cos( \epsilon )x - \sin( \epsilon )y \;.
\end{align}
Similarly
\begin{equation}
e^{ \epsilon L } y = \sin( \epsilon )x + \cos( \epsilon )y \;.
\vspace{.1in}
\end{equation}
{\bf Exercise.}  If  $ A $  is any  $ n \times n $  matrix, $ A=(a_{ij} ) $  and
\begin{equation}
L= \vec{v} A \vec{\nabla} = \sum_{i,j=1}^{n} x_i a_{ij} \frac{\partial}
{\partial x_j } \;,
\end{equation}
then show that
$$ e^{ \epsilon L } f( \vec{v} ) = f(e^{ \epsilon A } \vec{v} ) \;. $$
In this sense Lie series generalize the matrix exponential.

It is the author's opinion that the Lie series formalism is an invaluable tool for
understanding transformation groups.  However, the power series method of
evaluating Lie series is not very powerful.  Thus, when trying to evaluate
a complicated Lie series it is better to convert the problem to a system
of ordinary differential equations using the Differential Equation Property and
then apply classical ordinary differential equation techniques to the 
problem. \\[.1in]
{\bf Exercise.}  Use the Differential Equation Property to redo some of the previous examples.

\section{Action on Curves and Surfaces.}

This section will bring us substantially closer to the purpose of this
manuscript; the study of differential equations.  The solutions of a
differential equation or system of differential equations can be interpreted
as a curve, surface or in general, a hyper-surface in some Euclidean space.  We
want to know how transformation groups on the Euclidean space act on such
curves and surfaces.  To do this we need to expand the notation of the
previous sections.

Before we study the general case let us look at the situation in
$ R^2 $.
\begin{center}
	\begin{pspicture}(0,0)(9,2)
		\psgrid[gridcolor=green,subgridcolor=yellow,,gridlabels=0]
		\pscurve[showpoints=false,dotsize=0.2,linewidth=0.03]{-}(4,1.6)(3.3,0.5)(0.4,0.4)
		\pscurve[showpoints=false,dotsize=0.2,linewidth=0.03]{-}(8,2.1)(7.3,1)(4.4,0.9)
		\rput(1.95,0.7){$f(x+\epsilon,y+\eta)$}
		\rput(5.91,1.2){$f(x,y)$}
	\end{pspicture}
\end{center}
\centerline{Motion of a Curve in  $ R^2 $}
Now let
\begin{equation}
\vec{G} ( \epsilon ,x,y) = ( \xi ( \epsilon ,x,y)\;, \eta ( \epsilon ,x,y))
\end{equation}
be a group of transformations with infinitesimal
\begin{equation}
\vec{T} (x,y) = (r(x,y)\;, s(x,y)) \;.
\end{equation}
If  $ y=f(x) $  is a curve, then  $ y=f( \epsilon ,x) $  is to be a
curve whose graph is the image of the graph of  $ y=f(x) $  under the 
action of the group element  $ G( \epsilon ,x,y) $.  Under the action of
$ G(-\epsilon ,x,y) $  the points  $ (x,f( \epsilon ,x)) $  go into the points
$  ( \xi (-\epsilon ,x,f( \epsilon ,x)), \eta (-\epsilon ,x,f( \epsilon ,x))) $ 
which must lie in the graph of  $ y=f(x) $, that is,
\begin{equation} \label{Implicit Action}
\eta (-\epsilon ,x,f( \epsilon ,x)) = f( \xi (-\epsilon ,x,f( \epsilon ,x)))
\end{equation}
If  $ f( \epsilon ,x) $  is replaced by  $ y $, then \eqref{Implicit Action}
becomes
\begin{equation}
\eta (-\epsilon ,x,y) = f( \xi (-\epsilon ,x,y))
\end{equation}
which can be solved for  $ y=f( \epsilon ,x) $.  Thus  \eqref{Implicit Action}
which gives an implicit definition of  $ f( \epsilon ,x) $.
In some elementary cases \eqref{Implicit Action} can be solved for  $ f( \epsilon ,x) $  
but, in general, this cannot be done.  However, it is possible to compute the
{\it explicit} infinitesimal action!

The computation of the infinitesimal action of the group on curves requires
some formulas.  Recall that
\begin{equation}
\begin{array}{ll}
\nonumber \xi (0,x,y)=x \;, & \eta (0,x,y)=y\;, \\
\xi_\epsilon (0,x,y)=r(x,y) & \eta_\epsilon (0,x,y)=s(x,y)\;,
\end{array}
\end{equation}
and consequently
\begin{equation}
\xi_y (0,x,y) = 0 \quad, \quad \eta_y (0,x,y) = 1 \;.
\end{equation}
Differentiate  \eqref{Implicit Action} with respects to  $ \epsilon $  and set  
$ \epsilon = 0 $.   Then 
\begin{equation}
-s(x,f(x)) + f_\epsilon (0,x) = -f^\prime (x)r(x,f(x)) \;.
\end{equation}
Solving for  $ f_\epsilon $  gives
\begin{equation}
f_\epsilon (0,x) = -r(x,f(x)) \frac{df(x)}{dx} + s(x,f(x))
\end{equation}
which is the infinitesimal action.  Set
\begin{equation}
L(f)(x) = -r(x,f(x)) \frac{df(x)}{dx} + s(x,f(x))
\end{equation}
and note that  $ L $  is a nonlinear differential operator that gives the
infinitesimal group action.

To study the general case let  $ n $  and  $ m $  be two positive integers.
We will now consider transformation groups on  $ R^{n+m} $  where
the points in  $ R^{n+m} $  are labeled with two vectors
$ \vec{u} = (u_1, \ldots , u_m ) $  and  $ \vec{x} = (x_1, \ldots , x_n ) $.
The variables  $ \vec{u} $  will be considered {\it dependent} while the
variables  $ \vec{x} $  will be considered {\it independent}, that is, we will
be considering surfaces (hyper-surfaces) of the form
\begin{equation}
\vec{u} = \vec{f} ( \vec{x} ) \;.
\end{equation}
\begin{pspicture}(-3,-3)(3,4)
\pstThreeDCoor[%
  xMin=-4,%
  xMax=4,%
  yMin=-4,%
  yMax=4,%
  zMax=3,%
  arrows=<->%
]
\psplotThreeD[%
  plotstyle=line,%
  linecolor=blue,%
  yPlotpoints=40,%
  xPlotpoints=30,%
  linewidth=0.5pt,%
  hiddenLine=true%
](-2,2)(-2,2)
{
  x 100 mul cos y 100 mul sin sub
}
\psplotThreeD[%
  plotstyle=line,%
  linecolor=red,%
  yPlotpoints=40,%
  xPlotpoints=30,%
  linewidth=0.5pt,%
  hiddenLine=true%
](3,7)(3,7)
{
  x 5 sub 100 mul cos y 5 sub 100 mul sin sub 5 add 
}
\rput[tl](1,2.5){$\color{blue}u = \cos(100x_1) - \sin(100x_2)$}
\rput[tl](0,3.5){$\color{red}u = \cos(100(x_1-5)) - \sin(100(x_2-5))+5$}
\end{pspicture}

\centerline{Motion of a Surface in  $ R^{n+m} $}\ \\[12pt]

A transformation group on  $ R^{n+m} $  will be written
$ \vec{G} ( \epsilon , \vec{x} , \vec{u} ) $.
Geometrically it is clear that the transformation group will move the surface
given by the function  $ \vec{f} $  into a new surface given by a function
$ \vec{f} ( \epsilon , \vec{x} ) $.  Actually there may be exceptional points where the
new surface is vertical, but such singularities will not give us any difficulty.
First we write
\begin{equation}
\vec{G} ( \epsilon , \vec{x} , \vec{u} ) = ( \vec{G}_1 ( \epsilon , \vec{x} , 
\vec{u} ), \vec{G}_2 ( \epsilon , \vec{x} , \vec{u} ))
\end{equation}
where  $ \vec{G}_1 $  gives the  $ \vec{x} $  components of 
$ \vec{G} $  and  $ \vec{G}_2 $  gives the  $ \vec{u} $  components.  We also
write the infinitesimal of  $ \vec{G} $  in a similar way;
\begin{equation}
\vec{T} ( \vec{x} , \vec{u} ) = ( \vec{T}_1 ( \vec{x} , \vec{u} )\;, 
\vec{T}_2 ( \vec{x} , \vec{u} )) \;.
\end{equation}
The differential operator corresponding to  $ \vec{T} $  is then
\begin{equation}
\vec{T}_1 ( \vec{x} , \vec{u} ) \cdot \vec{\nabla}_x + \vec{T}_2 ( \vec{x} , 
\vec{u} ) \vec{\nabla}_u
\end{equation}
where
\begin{equation}
\vec{\nabla}_x = \left( \frac{\partial}{\partial x_1 },\ldots , 
\frac{\partial}{\partial x_n} \right) , \qquad 
\vec{\nabla}_u = \left( \frac{\partial}{\partial u_1}, \ldots , 
\frac{\partial}{\partial u_m} \right) \;.
\end{equation}
Now if the point  $ (\vec{x} , \vec{u} ) $  is on the surface given by
$ \vec{u} = \vec{f} ( \epsilon , \vec{x} ) $  then the point  
$ ( \vec{G}_1 (-\epsilon , \vec{x} , \vec{f} ( \vec{x} ) ) \;,
\vec{G}_2 ( -\epsilon , \vec{x} , \vec{f} ( \vec{x} ) )) $  is on the 
surface given by $ \vec{u} = \vec{f} ( \vec{x} ) $, that is,
\begin{equation}
\vec{G}_2 (-\epsilon , \vec{x} , \vec{u} ) = \vec{f} ( \vec{G}_1 (-\epsilon , 
\vec{x} , \vec{u} )) \;.
\end{equation}
If this equation is solved for  $ \vec{u} $  as a function of  $ \epsilon $  and
$ \vec{x} $, then the result defines  $ \vec{u} = \vec{f} ( \epsilon , 
\vec{x} ) $.  The implicit
function theorem guarantees that such a solution exists for  $ \epsilon $  
sufficiently small and  $ \vec{x} $  in some small set.  Thus  $ \vec{u} = 
\vec{f} ( \epsilon , \vec{x} ) $ is implicitly defined by
\begin{equation}\label{Implicitly Defined}
G_2 (-\epsilon , \vec{x} , \vec{f} ( \epsilon , \vec{x} )) = \vec{f} 
( \vec{G}_1 (-\epsilon , \vec{x} , \vec{f} ( \epsilon , \vec{x} ))) \;.
\end{equation}
Also, the group action on the function  $ \vec{u} = \vec{f} ( \vec{x} ) $  
is then written
\begin{equation}
\vec{G} ( \epsilon , \vec{f} )(x) = \vec{f} ( \epsilon , \vec{x} ) \;.
\end{equation}

The infinitesimal action on surfaces is
\begin{equation}
\vec{L} ( \vec{f} )( \vec{x} ) = \frac{d}{d \epsilon } \vec{G} ( \epsilon , 
\vec{f} ) ( \vec{x} ) \mid_{ \epsilon =0 } = \frac{d}{d \epsilon }  
f ( \epsilon , x) \mid_{ \epsilon =0 } \;.
\end{equation}
This is computed by differentiating \eqref{Implicitly Defined}
 implicitly with
respects to  $ \epsilon $.  The derivative of the left hand side of
\eqref{Implicitly Defined} at $ \epsilon =0 $  is
\begin{align} \label{Result}
& \frac{d}{d \epsilon } \vec{G}_2 ( -\epsilon , \vec{x} , \vec{f} 
( \epsilon , \vec{x} )) \mid_{ \epsilon =0 }   = \\
& \nonumber - \vec{T}_2 ( \vec{G}_2 (-\epsilon , \vec{x} , \vec{f} ( \epsilon , \vec{x} 
))) \mid_{ \epsilon =0 }
+ \vec{\nabla}_u \vec{G}_2 ( -\epsilon , \vec{x} , \vec{f} ( \epsilon , 
\vec{x} )) \cdot \frac{d}{d \epsilon } \vec{f} ( \epsilon , \vec{x} ) 
\mid_{ \epsilon =0 } \;.
\end{align}
Recall that  $ G_2 (0, \vec{x} , \vec{u} ) = \vec{u} $  so that
$ \vec{\nabla}_u G_2 (0, \vec{x} , \vec{u} ) $  is an identity matrix.  Thus the
right-hand side of \eqref{Result}  becomes
\begin{equation} \label{Simplified Result}
- \vec{T}_2 ( \vec{x} , \vec{f} ( \vec{x} )) + \vec{L} ( \vec{f} )
( \vec{x} ) \;.
\end{equation}
The derivative of the right-hand side of \eqref{Implicitly Defined} at $ \epsilon =0 $ is
\begin{align} \label{RHS}
& \vec{\nabla}_x \vec{f} ( \vec{G}_1 (-\epsilon , \vec{x} , \vec{f} 
(\epsilon , \vec{x} ))) = \\
 \nonumber & \{- \vec{T}_1 ( \vec{G}_1 (-\epsilon , \vec{x} , \vec{f} 
(\epsilon , \vec{x} )))
+ \vec{\nabla}_u \vec{G}_1 ( -\epsilon , \vec{x} , \vec{f} ( \epsilon , 
\vec{x} )) \frac{d}{d \epsilon } \vec{f} ( \epsilon , \vec{x} )\} 
\mid_{ \epsilon =0 } \;.
\end{align}
However, $ G_1 (0, \vec{x} , \vec{u} ) = \vec{x} $  and  $ \vec{\nabla}_u
\vec{x} =0 $  so  \eqref{RHS} becomes  
\begin{equation} \label{Simplified}
- \vec{T}_1 ( \vec{x} , \vec{f} ( \vec{x} )) \cdot \vec{\nabla}_x 
\vec{f} ( \vec{x} ) \;.
\end{equation}
Combining \eqref{Simplified Result} and \eqref{Simplified}  gives
\begin{equation}
- \vec{T}_2 ( \vec{x} , \vec{f} ( \vec{x} )) + \vec{L} ( \vec{f} )( \vec{x} )
= -\vec{T}_1( \vec{x} , \vec{f} ( \vec{x} )) \cdot \vec{\nabla}_x \vec{f} 
( \vec{x} )
\end{equation}
or
\begin{equation}
\vec{L} ( \vec{f} )( \vec{x} ) = - \vec{T}_1 ( \vec{x} , \vec{f} ( \vec{x} )) 
\vec{\nabla}_x \vec{f} ( \vec{x} ) + \vec{T}_2 ( \vec{x} , \vec{f} 
( \vec{x} )) \;.
\end{equation}
We summarize this. \\[.1in]
{\bf Proposition.}

If  $ ( \vec{T}_1 ( \vec{x} , \vec{u} ), \vec{T}_2 ( \vec{x} , \vec{u} )) $  
is a
vector field on  $ R^{n+m} $, the action of the group generated by the
vector field on functions  $ \vec{u} = \vec{f} ( \vec{x} ) $  has an 
infinitesimal given by
\begin{equation}
\vec{L} ( \vec{f} )( \vec{x} ) = - \vec{T}_1 ( \vec{x} , \vec{f} ( \vec{x} )) 
\cdot \vec{\nabla}_x \vec{f} ( \vec{x} ) + \vec{T}_2 ( \vec{x} , \vec{f} 
( \vec{x} )) \;.
\end{equation}
Note that  $ \vec{L} $  is a {\it nonlinear} (quasi-linear) first order 
differential operator.  This correspondence is easy to remember because
\begin{equation}
( \vec{T}_1 \cdot \vec{\nabla}_x +
\vec{T}_2 \cdot \nabla_u )( \vec{u} - \vec{f} ( \vec{x} ))
= \vec{T}_2 - \vec{T}_1 \cdot \vec{\nabla}_x \vec{f} ( \vec{x} ) \;.
\end{equation}

\section{Invariants and Canonical Coordinates.}

An important technique that we will use to study problems involves making
a change of coordinates so that some group becomes particularly simple.
Such transformations are found by computing functions that are invariant
under the group action.  Once the coordinate transformation is found, then
one approach to studying the given problem is to transform everything to the
new coordinates. 
The problem of transforming differential equations and operators to new
coordinate systems which involve changes in both the dependent and
independent variables is another example of a straight forward but tedious
algebraic procedure.  Consequently, we have provided a symbol manipulation
program to do such computations.
We return to the notation of the first two section to begin
our discussion.

A function  $ f $  mapping  $ R^n $  into  $ R $  is invariant under  $ G $  if it satisfies
\begin{equation}
f( \vec{G} ( \epsilon , \vec{v} ))=f( \vec{v} ) \;,
\end{equation}
that is, 
\begin{equation} \label{Invariant 1}
G( \epsilon ,f)( \vec{x} ) = f( \vec{v} ) 
\end{equation}
for all  $ \epsilon $.  If we differentiate this equation with respects to
$ \epsilon $  and set  $ \epsilon = 0 $, we obtain
\begin{equation} \label{Invariant 2}
\vec{L} f( \vec{v} )=0
\end{equation}
where  $ \vec{L} $  is the infinitesimal of  $ G $.  Conversely, if  
$ Lf( \vec{v} )=0 $, then
\begin{equation}
G( \epsilon ,f)( \vec{v} )=e^{ \epsilon L } f( \vec{v} ))=f( e^{ \epsilon L } 
\vec{v} )= e^{ \epsilon L } f( \vec{v} )=f( \vec{v} )
\end{equation}
so that \eqref{Invariant 2} and \eqref{Invariant 1} are equivalent.

In general a group will have more than one invariant.  If  $ f_i $, 
$ 1 \leq i \leq m $, are invariants, then they are said to be independent if
the Jacobian matrix
\begin{equation} \label{Jacobian}
\left( \frac{ \partial f_i }{ \partial x_j } \right) \qquad 
1 \leq i \leq m \;,\; 1 \leq j \leq n
\end{equation}
has maximal rank.  Because the rank of the matrix \eqref{Jacobian}
must be less than or 
equal to  $ n $, a transformation group on  $ R^n $  can have no more than
$ n $  independent invariants.  If a group has  $ n $  independent invariants,
then set
\begin{equation}
\xi_i = f_i ( \vec{v} )
\end{equation}
Because the invariants are independent, the $ \xi_i $  will serve as 
coordinates on  $ R^n $.  If
$ g( \vec{v} ) $  is any function, $ \vec{\nu} = ( \xi_1, \ldots , \xi_n ) $,
and  $ \tilde{g} ( \vec{\nu} ) = g ( \vec{v} ) $, then
\begin{equation}
e^{ \epsilon L } g( \vec{v} ) = e^{ \epsilon L } \tilde{g} ( \vec{\nu} )
= \tilde{g} ( e^{ \epsilon L } \vec{\nu} ) = \tilde{g} ( \vec{\nu} ) = 
g( \vec{v} ) \;.
\end{equation}
Thus  $ e^{ \epsilon L } $  is the identity group.  Consequently any nontrivial 
one parameter group on  $ R^n $  can have at most  $ n-1 $  independent
invariants.

We can now prove the following useful and elementary theorem. \\[.1in]
{\bf Theorem.}  Any nontrivial one parameter transformation group on  
$ R^n $  has exactly  $ n-1 $  independent invariant functions.
\\[.1in]
{\bf Proof.}  We already know that the group can have at most  $ n-1 $  
invariants.  Also $ f( \vec{v} ) $  is an invariant
if and only if  $ Lf( \vec{v} ) = 0 $.  This is just a
first order linear partial differential equation for  $ f( \vec{v} ) $.  The
standard approach to solving this type of problem is the method of
characteristics [6].  Thus, if
\begin{equation}
L = \vec{T} \cdot \vec{\nabla} = \sum T_i ( \vec{v} ) \frac{ \partial }{
\partial x_i} \;,
\end{equation}
then the characteristic equations are
\begin{equation} \label{Characteristics}
\frac{ dx_1 }{ T_1 ( \vec{v} ) } = \frac{ dx_2 }{ T_2 ( \vec{v} ) } = 
\cdots = \frac{ dx_n }{ T_n ( \vec{v} ) } \;.
\end{equation}
If we set all these equations equal to  $ d \epsilon  $, then we see this 
is just the system of equations given by the Differential Equation property 
listed in Section 1.3.  It is well known from the theory of first order 
partial differential equations that \eqref{Characteristics} has
$ n-1 $  integrals which is just another way of saying invariants of the group.
This can be seen by solving \eqref{Characteristics} for the  $ n $  function
$ x_i = f_i ( \epsilon ) $, using one of the equations to eliminate
$ \epsilon $, and then noting that the remaining  $ n-1 $  functions are
invariants (or integrals).  Of course these integrals are the usual integrals 
of the autonomous system \eqref{Characteristics}. \\[.1in]
{\bf Example - Rotations - Continued.}  If
\begin{equation}
L = -y \frac{ \partial }{ \partial x } + x \frac{ \partial }{ \partial y } \;,
\end{equation}
then the characteristic equations are
\begin{equation}
\frac{dx}{-y} = \frac{dy}{x} \;.
\end{equation}
Thus
\begin{equation}
x\; dx + y\; dy = 0
\end{equation}
and an integration gives
\begin{equation}
\frac{x^2}{2} + \frac{y^2}{2} = c \;.
\end{equation}
Clearly, the function
\begin{equation}
f(x,y) = x^2 + y^2
\end{equation}
is an invariant of the rotation group. \\[.1in]
{\bf Exercise - Dilations - Continued.}  The infinitesimal dilation group is
\begin{equation}
L = x \frac{ \partial }{ \partial x } + y \frac{ \partial}{ \partial y } \;.
\end{equation}
Show that  $ f(x,y)=y/x $  is an invariant of this group.

The next item that we will work on is the problem of giving a canonical
representation of every one parameter transformation group.  What we will
show is that under an appropriate change of coordinates, every one parameter
transformation group is isomorphic to a group of translations.  As in many
other branches of mathematics, this canonical representation will allow us
to easily predict the outcome of many computations.  It is the author's
opinion that it is hard to over-rate the usefulness of such canonical
representation theorems. \\[.2in]
{\bf Theorem.}  If  $ \vec{G} ( \epsilon , \vec{v} ) $  is a transformation 
group on  $ R^n $,
then there exists a change of coordinates  $ \vec{\nu} = \vec{C} ( \vec{v} ) $  such that
in the new coordinates  $ \vec{G} ( \epsilon , \vec{v} ) $  becomes
\begin{equation}
\vec{G} ( \epsilon , \vec{\nu} ) = ( \xi_1 + \epsilon , \xi_2 , \ldots
, \xi_n )
\end{equation}
\vspace{.2in}
{\bf Proof.}  If  $ L $  is the infinitesimal operator of  $ \vec{G} $, then
the theory of first order partial differential says that it is possible to
choose  $ \xi_1 = \xi_1 ( \vec{v} ) $  such that
\begin{equation}
L \xi_1 = 1 \;.
\end{equation}
Next, choose  $ \xi_i = \xi_i ( \vec{v} ) $  to be any  $ n-1 $  independent
invariants of  $ G( \epsilon , \vec{\nu} ) $.   Set  $ \vec{C} ( \vec{v} ) = 
( \xi_1 ( \vec{v} ) , \ldots , \xi_n ( \vec{v} )) $.  If the Jacobian of this 
transformation is zero, then the vectors
$ \nabla \xi_i = ( \partial \xi_i / \partial \xi_1 , \ldots , 
\partial \xi_i / \partial \xi_n ) $,
$ i = 1 , \ldots , n $  must be linearly dependent.  It was assumed that
$ \xi_2 , \ldots , \xi_n $  are linearly independent; consequently
\begin{equation}
\vec{\nabla} \xi_1 = \sum_{i=2}^n \alpha_i \vec{\nabla} \xi_i
\end{equation}
for some scalers  $ \alpha_i $.  Now
\begin{equation}
L \xi_1 = T( \vec{v} ) \cdot \vec{\nabla} \xi_1
= \sum_{i=1}^n \alpha_i T(v) \cdot \nabla \xi_i
= \sum_{i=1}^n \alpha_i L( \xi_i ) = 0 \;.
\end{equation}
However, it was assumed that  $ L( \xi_1 ) = 1 $  so the assumption of linear
dependence lead to a contradiction.  The linear independence implies that the
Jacobian of the transformation is nonzero.

We have by definition
\begin{equation}
G( \epsilon , \vec{\nu} ) = e^{ \epsilon L } \vec{\nu}
= (e^{ \epsilon L } \xi_1 , \ldots , e^{ \epsilon L } \xi_n )
= ( \xi_1 + \epsilon , \xi_2 , \ldots , \xi_n ) \;.
\end{equation}

This canonical representation theorem immediately gives a well known canonical
representation for first order differential operators. \\[.2in]
{\bf Theorem.}  If  $ L $  is a first order linear differential operator
\begin{equation}
L = \vec{T} \cdot \vec{\nabla} \;,\; T(v) = (T_1 ( \vec{v} ) , \ldots , 
T_n ( \vec{v} )) \;,
\end{equation}
then there exists a change of coordinates $ \vec{\nu} = \vec{C} ( \vec{v} ) $  
such that  $ L $  transforms into  $ \tilde{L} $  and
\begin{equation}
\tilde{L} = \frac{ \partial}{ \partial \xi_1 } \;.
\end{equation} \\[.2in]
{\bf Proof.}  Under the change of coordinates given in the previous theorem, 
$ L $  goes into the
infinitesimal generator of the translation group which is just  $ \partial /
\partial \xi_1 $. \\[.2in]
{\bf Example - Rotations.}  Previously we showed that
\begin{equation}
x^2 + y^2 = c
\end{equation}
is an invariant of the rotation group.  Consequently all invariants are given
by
\begin{equation}
g(x,y) = G(x^2 + y^2 )
\end{equation}
for any function  $ G $  of one variable.  Next we need to solve
\begin{equation}
Lf = 1 \;.
\end{equation}
We guess that a particular solution of this equation is
\begin{equation}
f = \arctan \left( \frac{y}{x} \right) \;.
\end{equation}
Consequently the general solution is given by
\begin{equation}
f(x,y) = F(x^2 + y^2 ) + \arctan \left( \frac{y}{x} \right)
\end{equation}
for any function  $ F $  of one variable.

Thus any change of variables
\begin{equation}
\xi = F(x^2 + y^2 ) + \arctan \left( \frac{y}{x} \right)
\;,\; \eta = G(x^2 + y^2 )
\end{equation}
with nonzero Jacobian will give us a set of canonical coordinates.  In fact the
Jacobian of such a transformation is
\begin{equation}
J = -2G^\prime
\end{equation}
so that if  $ G^\prime = -1/2 $  the transformation will have unit Jacobian.  If
the coordinates are to be orthogonal, it is easy to check that
$ F prime = 0 $  independent of the choice of  $ G $.  Thus under these two
constraints we have
\begin{equation}
\xi = c_1 + \arctan \left( \frac{y}{x} \right)
\;,\; \eta = c_2 - \frac{ (x^2 + y^2 ) }{2}
\end{equation}
which bears an obvious relationship to polar coordinates and is essentially
the action-angle coordinates for the harmonic oscillator given in
Hamiltonian mechanics. \\[.2in]
{\bf Exercise - Dilations.}  Recall that the infinitesimal group is given by
\begin{equation}
L = x \frac{ \partial}{ \partial x } + y \frac{ \partial}{ \partial y }
\end{equation}
Show that canonical coordinates are given by
\begin{equation}
\xi = F \left( \frac{y}{x} \right) + \ln \mid x \mid
\;,\; \eta = G \left( \frac{y}{x} \right) \;.
\end{equation}
An interesting special transformation is
\begin{equation}
\xi = \ln \sqrt{x^2 + y^2} \;,\;
\eta = \arctan \left( \frac{y}{x} \right) \;.
\end{equation}
\vspace{.2in}
{\bf Exercise.}  Find canonical coordinates for the operator
\begin{equation}
x \frac{d}{dx} \;,\; x^2 \frac{d}{dx} \;.
\end{equation}

Now that we know how to find canonical coordinates, we need to know how to transform
differential equations to the new coordinates.  Before we turn to the general
case let us do the two variable problem.  Thus assume that we have an
infinitesimal group of the form
\begin{equation} \label{Infinitesimal}
L = a(x,y) \frac{\partial}{\partial x} + b(x,y) \frac{\partial}{\partial y}
\end{equation}
and that we have found two functions  $ \xi = \xi (x,y) $  and
$ \eta = \eta (x,y) $  such that
\begin{equation} \label{Origional}
L( \xi ) = 1 \;,\; L( \eta ) = 0 \;.
\end{equation}
The transformation that we will use is
\begin{equation}
\xi = \xi (x,y) \;,\; \eta = \eta (x,y) \;.
\end{equation}
Our theory shows that this transformation is invertible and we write the inverse
as
\begin{equation}
x = x( \xi , \eta ) \;,\; y = y( \xi , \eta ) \;.
\end{equation}

Let us first verify that  $ L $  is transformed into translation.  Let
$ f(x,y) $  be any function and  $ \tilde{f} ( \xi , \eta ) $  be defined 
so that
\begin{equation}
\tilde{f} ( \xi (x,y) , \eta (x,y)) = f(x,y) \;.
\end{equation}
Then the chain rule gives
\begin{align}
\frac{\partial f}{\partial x} & = \frac{\partial \tilde{f}}{\partial \xi }
\frac{\partial \xi}{\partial x} + \frac{\partial \tilde{f}}{\partial \eta }
\frac{\partial \eta}{\partial x} \;,\; \\
\nonumber
\frac{\partial f}{\partial y} 
& =
\frac{\partial \tilde{f}}{\partial \xi} \frac{\partial \xi}{\partial y} +
\frac{\partial \tilde{f}}{\partial \eta} \frac{\partial \eta}{\partial y}
\end{align}
and consequently
\begin{align}
L(f) & = a \frac{\partial f}{\partial x} + b \frac{\partial f}{\partial y} \\
\nonumber & =
(a \frac{\partial \xi}{\partial x} + b \frac{\partial \xi}{\partial y} )
\frac{\partial \tilde{f}}{\partial \xi} + (a \frac{\partial \eta}{\partial x}
+ b \frac{\partial \eta}{\partial y} ) \frac{\partial \tilde{f}}{\partial \xi} \\
\nonumber
& = \frac{\partial \tilde{f}}{\partial \xi} = \tilde{L} (f)
\end{align}
as was desired.

Now let us consider what happens to a curve  $ y = f(x) $  in the canonical
coordinates.  First, the curve goes into a curve  $ \eta = g(\xi) $  determined
by solving the equation
\begin{equation}
y(\xi ,\eta ) = f(x(\xi ,\eta ))
\end{equation}
for  $ \eta $  as a function of  $ \xi $ , that is, $ g $  satisfies
\begin{equation}
y( \xi , g( \xi )) = f(x( \xi ,g( \xi ))) \;.
\end{equation}

Now compute what happens to the slope of the curve  $ y = f(x) $.
The use of differentials makes this computation transparent.  Thus
\begin{align}
d \xi & = \frac{\partial \xi}{\partial x} dx + \frac{\partial \xi}{\partial
y} dy  = \xi_x dx + \xi_y dy \;,\;  \\
\nonumber
d \eta & = \frac{\partial \eta}{\partial x} dx + \frac{\partial \eta}
{\partial y} dy = \eta_x dx + \eta_y dy
\end{align}
and consequently
\begin{equation} \label{Stuff}
\eta^\prime = \frac{d \eta}{d \xi} = \frac{\eta_x dx + \eta_y dy}
{\xi_x dx + \xi_y dy} = \frac{\eta_x + \eta_y y^\prime}{\xi_x + \xi_y
y^\prime} \;.
\end{equation}
The quantities  $ \eta_x , \eta_y , \xi_x $  and  $ \xi_y $
still depend on  $ x $  and  $ y $  so the inverse transformation should
be used to remove  $ x $  and  $ y $  in favor of  $ \xi , \eta $.
Solving \eqref{Stuff} for  $ y^\prime $  gives
\begin{equation}
y^\prime = - \frac{\eta_x - \xi_x \eta^\prime}{\eta_y - \xi_y \eta^\prime} \;.
\end{equation}

The action of the infinitesimal group \eqref{Infinitesimal}
on the curve $ y = f(x) $  is 
given by the operator
\begin{align}
\nonumber S(f) & =  -af^\prime + b = -ay^\prime + b = \\
\nonumber & =  \frac{a(\eta_x - \xi_x \eta^\prime ) + b(\eta_y - \xi_y
\eta^\prime) }{\eta_y - \xi_y \eta^\prime} \\
& =  \frac{(a\eta_x + b \eta_y ) - (a \xi_x + b \xi_y ) \eta^\prime}
{\eta_y - \xi_y \eta^\prime} \\
\nonumber & =  \frac{\eta^\prime}{\xi_y \eta^\prime - \eta_y} \;.
\end{align}

This result seems a bit surprising.  To see why we say this we summarize our
results to this point.  We started with an operator
\begin{equation}
L = a \frac{\partial}{\partial x} + b \frac{\partial}{\partial y} \;.
\end{equation}
The action of this operator on curves is given by  $ y=f(x) $
\begin{equation}
S(f) = -af^\prime + b
\end{equation}
These two operators transform into
\begin{equation}
\tilde{L} = \frac{\partial}{\partial \xi} \;,\; \tilde{S} ( \tilde{f} ) = 
\frac{\tilde{f}^\prime}{\xi_y \tilde{f}^\prime - \eta_y} \;.
\end{equation}
It seems natural to have expected that  $ \tilde{S} \tilde{f} = \tilde{L} 
\tilde{f} $.    
However, for this to happen the transformation must be particularly simple.
What is needed is
\begin{equation}
\xi_y = 0 \;,\; \eta_y = 1 \;.
\end{equation}
The original equations \eqref{Origional} for  $ \xi $  and
$ \eta $  then become
\begin{equation} \label{Simple}
a \xi_x = 1 \;,\; a \eta_x - b = 0 \;.
\end{equation}
If the transformation is to have nonzero Jacobean then it must be the case that
\begin{equation}
\xi_x \neq 0 \;.
\end{equation}
Differentiating \eqref{Simple}  gives  $ a_y \xi_x = 0 $, that is  $ a_y = 0 $.  
Again differentiating \eqref{Simple}  gives  $ b_y = 0 $.  For such a coordinate 
system
\begin{equation}
\xi = \xi (x) \;,\; \eta = - \mu (x)y + \nu (x)
\end{equation}
for some functions  $ \xi (x) \;\; \mu (x) \;,\; \nu (x) $.

Let us now turn to the general case of transformations on  $ R^{n+m} $ 
acting on surfaces  $ \vec{u} = \vec{f} ( \vec{x} ) $  where  
$ \vec{u} = ( u_1, \ldots , u_m ) $  and  $ \vec{x} = (x_1, \ldots , 
x_n ) $.  As before, the infinitesimal transformation group will be written
\begin{equation}
L = \vec{T}_1 ( \vec{x} , \vec{u} ) \cdot \vec{\nabla}_x + \vec{T}_2 
( \vec{x} , \vec{u} ) \cdot \nabla_u \;.
\end{equation}
Assume that we have found functions  $ \xi_i ( \vec{x} , \vec{u} ) $,
$ 1 \leq i \leq n $, $ \eta_i ( \vec{x} , \vec{u} ) $, $ 1 \leq i \leq m $ 
such that
\begin{align}
\nonumber L \xi_1 & =  1 \;, \\
L \xi_i & =  0 \;, \qquad 2 \leq i \leq n \;, \\
\nonumber L \eta_i & =  0 \;, \qquad 1 \leq i \leq m \;.
\end{align}
As before, the transformation
\begin{equation}
\vec{\xi} = (\xi_1 ( \vec{x} , \vec{u} ), \ldots , \xi_n ( \vec{x} , \vec{u} ))
\;,\; \vec{\eta} = (\eta_1 ( \vec{x} , \vec{u} ), \ldots , \eta_m ( \vec{x} , 
\vec{u} )) \;,
\end{equation}
has nonzero Jacobean.  Under the transformation,  $ L $  is transformed into
$ \tilde{L} = \partial / \partial \xi_1 $  as was shown in the previously.

What does the infinitesimal action on surfaces 
\begin{equation}
\vec{S} ( \vec{u} )( \vec{x} ) = - \vec{T}_1 ( \vec{x} , \vec{u} ) 
\vec{\nabla}_x \vec{U} + \vec{T}_2 ( \vec{x} , \vec{u} )
\end{equation}
transform into?  Again, the use of differentials make the computations
simple.  The chain rule gives
\begin{equation} \label{dxi deta}
d \vec{\xi} = \vec{\nabla}_x \vec{\xi} d \vec{x} + \vec{\nabla}_u \vec{\xi} 
d \vec{u} \;,\; d \vec{\eta} = \vec{\nabla}_x \vec{\eta} d \vec{x} + 
\vec{\nabla}_n \vec{\eta} d \vec{u} \;.
\end{equation}
We assume that  $ \vec{u} = \vec{f} ( \vec{x} ) $  is transformed into
$ \vec{\eta} = g( \vec{\xi} ) $.  The differentials of these equations give
\begin{equation} \label{du}
d \vec{u} = \vec{\nabla}_x \vec{f} d \vec{x} \;,\; d \vec{\eta} = 
\vec{\nabla}_\xi \vec{g} d \vec{\xi} \;.
\end{equation}
Combining \eqref{dxi deta}  and \eqref{du}  gives
\begin{equation}
d \vec{\eta} = \vec{\nabla}_\xi \vec{g} d \xi = \vec{\nabla}_\xi \vec{g} \cdot 
( \vec{\nabla}_x \vec{\xi} + \vec{\nabla}_u \vec{\xi} \vec{\nabla}_x \vec{f} )
d \vec{x} = \vec{\nabla}_x \vec{\eta} + \vec{\nabla}_u \vec{\xi} 
\vec{\nabla}_x \vec{f} )dx \;.
\end{equation}
Consequently
\begin{equation}
\vec{\nabla}_\xi \vec{g} \vec{\nabla}_x \vec{\xi} + \vec{\nabla}_\xi \vec{g} 
\vec{\nabla}_u \vec{\xi} \vec{\nabla}_x \vec{f} = \vec{\nabla}_x \vec{\eta} 
+ \vec{\nabla}_u \vec{\eta} \vec{\nabla}_x \vec{f}
\end{equation}
or
\begin{equation}
\nabla_x \vec{f} = - ( \vec{\nabla}_u \vec{\eta} - \vec{\nabla}_\xi \vec{g} 
\vec{\nabla}_u \vec{\xi} )^{-1} ( \nabla_x \eta - \vec{\nabla}_\xi \vec{g} 
\vec{\nabla}_x \vec{\xi} )
\end{equation}
which is the desired formula.

We conclude this section with the important but trivial observation that
symmetries are preserved under changes of coordinates, that is, if a given
system of differential equations has a symmetry and then a change of
coordinates will
send the symmetry into a symmetry of the transformed system.  This also
means that the symmetry method does not depend on the coordinate system
that is used to describe the given system of differential equations.

\section{Transformation Symmetries}

The notion of a symmetry can be associated with any problem and simply means
a mapping of the solution of the problem into solutions of the problem.  In
this section the problems we will consider will be systems of ordinary
or partial differential equations {\it without} boundary or initial conditions 
and the mappings will be one parameter groups of transformations.
Such symmetries are called point symmetries to distinguish them from
from jet symmetries that will be introduced later.

We will consider a situation where  $ \vec{u} = (u_1, \ldots , u_m ) $  are
dependent variables and  $ \vec{x} = (x_1, \ldots , x_n ) $  are the
independent variables.  We will write the system
of differential equations in the operator form
\begin{equation}
\vec{F} ( \vec{f} ) = (F_1 ( \vec{f} ), \ldots , \vec{F}_m ( \vec{f} ))
\end{equation}
where we are thinking of  $ \vec{u} $  being a function  $ \vec{x} $,
\begin{equation}
\vec{u} = \vec{f} ( \vec{x} ) \;.
\end{equation}
Here each  $ F_j $  is an expression in  $ \vec{x} $, $ \vec{f} $  and finite
number of the derivatives
\begin{equation}
\frac{\partial^k f_j}{\partial x_i^k} \;.
\end{equation}
The function  $ \vec{f} ( \vec{x} ) $  is a solution of the system of equations
provided
\begin{equation}
\vec{F} ( \vec{f} ) = 0 \;.
\end{equation}

Next let  $ \vec{G} (\epsilon , \vec{x} , \vec{u} ) $  be a one parameter group of
transformations on  $ R^{n+m} $  and  $ \vec{G} (\epsilon , \vec{f} ) $  be
the corresponding action surfaces  $ \vec{u} = \vec{f} ( \vec{x} ) $.  The
infinitesimal of the transformation group can be written
\begin{equation}
S = \vec{s} ( \vec{x} , \vec{u} ) \vec{\nabla}_x + \vec{r} ( \vec{x} , \vec{u}) 
\vec{\nabla}_{\vec{u}}
\end{equation}
and consequently the infinitesimal action on surfaces is given by
\begin{equation}
\vec{S} (f) = (S_1 (f), \ldots , S_j (f))
\end{equation}
where
\begin{equation}
S_j ( \vec{f} ) = - \vec{s} ( \vec{x} , \vec{f} ) \vec{\nabla}_x f_j + r_j 
( \vec{x} , \vec{f} ) \;.
\end{equation}

The requirement that  $ \vec{G} (\epsilon , \vec{f} ) $  sends solutions of
$ \vec{F} ( \vec{f} ) = 0 $  into solutions of the same problem can be written
\begin{equation} \label{Group Invariance}
\vec{F} ( \vec{f} ) = 0  = > \vec{F} ( \vec{G} (\epsilon , \vec{f} )) = 0
\;.
\end{equation}
This is called the {\it group invariance condition}.
The power of the infinitesimal method comes from differentiating the right hand
side of the group invariance condition \eqref{Group Invariance}
with respects to  $ \epsilon $  and
setting  $ \epsilon = 0 $.  Before we do this we need a definition.  \\[.2in]
{\bf Definition.}  The directional derivative of the functional  
$ \vec{F} ( \vec{f} ) $  at the point $ \vec{f} $  in the direction  
$ \vec{g} $  is given by
\begin{equation}
(D_{\vec{f}} \vec{F} )( \vec{g} ) = \frac{d}{ d \epsilon } \vec{F} 
( \vec{f} + \epsilon \vec{g} ) \mid_{\epsilon =0 } \;.
\end{equation}
This derivative is sometimes called a Frechet or Gataux derivative, especially
when a notion of convergence is used in the definition.

To differentiate the expression
\begin{equation}
\vec{F} ( \vec{G} (\epsilon , \vec{f} ))
\end{equation}
we use a power series expansion to write
\begin{equation}
\vec{G} (\epsilon , \vec{f} ) \cong \vec{f} + \epsilon S(f)
\end{equation}
and then compute
\begin{equation}
\frac{d}{d \epsilon} \vec{F} ( \vec{G} (\epsilon , \vec{f} )) 
\mid_{\epsilon =0 } \cong \frac{d}{d \epsilon} \vec{F} ( \vec{f} + \epsilon 
\vec{S} ( \vec{f} )) \mid_{\epsilon =0} = (D_{\vec{f}} \vec{F} )( \vec{S} 
( \vec{f} )) \;.
\end{equation} \\[.2in]
{\bf Definition.}  The infinitesimal group
\begin{equation}
L = \vec{s} ( \vec{x} , \vec{u} ) \vec{\nabla}_{ \vec{x} } + \vec{r} 
( \vec{x} , \vec{u} ) \vec{\nabla}_{ \vec{u} }
\end{equation}
is an infinitesimal symmetry of
\begin{equation}
\vec{F} ( \vec{f} ) = 0
\end{equation}
provided the corresponding infinitesimal action on surfaces  $ S(f) $  satisfies
\begin{equation}
\vec{F} ( \vec{f} ) = 0  = > (D_f ( \vec{F} )( \vec{S} ( \vec{f} )) = 0 \;.
\end{equation}

This form of condition is not convenient for computations.  However, the
convenient form depends on the particular problem being studied
so we postpone further derivations until Chapter II and III.
Before we turn to these applications, we note a few useful properties of the
directional derivative. \\[.2in]
{\bf Proposition.}  The directional derivative  $ (D_{ \vec{f} } \vec{F} )
( \vec{g} ) $  is linear in  $ \vec{g} $. \\[.2in]
{\bf Proof.}  Set  $ \alpha = \epsilon a $, $ \beta = \epsilon b $  and then 
compute
\begin{align}
(D_{ \vec{f} } \vec{F} )( a \vec{g} + b \vec{h} ) & = 
\frac{d}{d \epsilon} F( \vec{f} + \epsilon (a \vec{g} + b \vec{h} )) 
\mid_{\epsilon =0} \\
\nonumber
& = \frac{d \alpha}{d \epsilon} \frac{\partial}{\partial \alpha} \vec{F} 
( \vec{f} + \alpha \vec{g} + \beta \vec{h} ) \mid_{\epsilon =0} \\
\nonumber
& \quad\quad + \frac{d \beta}{d \epsilon} \frac{\partial}{\partial \beta} \vec{F} 
( \vec{f} + \alpha \vec{g} + \beta \vec{h} ) \mid_{\epsilon =0 } \\
\nonumber & = a(D_{ \vec{f} } \vec{F} )(g) + b(D_{ \vec{f} } \vec{F} )(h)
\;.
\end{align}
\vspace{.2in}
{\bf Proposition.}  If  $ \vec{F} ( \vec{f} ) $  is linear in  $ \vec{f} $ , 
then
\begin{equation}
(D_{ \vec{f} } \vec{F} )(g) = \vec{F} ( \vec{g} ) \;.
\end{equation}
\vspace{.2in}
{\bf Proof.}  Compute
\begin{equation}
(D_{ \vec{f} } \vec{F} )( \vec{g} ) = \frac{d}{d \epsilon} \vec{F} 
( \vec{f} + \epsilon \vec{g} ) \mid_{\epsilon =0}
= \frac{d}{d \epsilon} \{ F( \vec{f} ) + \epsilon \vec{F} ( \vec{g} ) \} 
\mid_{\epsilon =0} = \vec{F} ( \vec{g} ) \;.
\end{equation}

\newpage \clearpage
\setcounter{equation}{0}
\chapter{ORDINARY DIFFERENTIAL EQUATIONS}

\section{Introduction}

We now apply the material developed in Chapter I to systems of
ordinary differential equations.
In this section we will describe a general system of differential
equations because this is what our VAXIMA programs work with. The 
reader who is new to this material will find that a light reading of
this section followed by a more careful reading of the next few sections
will make this material more understandable. The next few sections
do not explicitly depend on this section!  
Let  $ \vec{y} = (y_1, \ldots , y_m ) $ 
be the dependent variables and  $ t $  be the independent variable.  We will
be interested in higher order nonlinear systems of the form
\begin{equation} \label{Nonlinear}
\frac{d^{p_i} y_i}{dt^{p_i}} = H_i ( \vec{y} ) \;,\; 1 \leq i \leq m \;.
\end{equation}
Here the  $ p_i $  are positive integers, $ p_i > 0 $,
and  $ H_i ( \vec{y} ) $  is an operator function of  $ \vec{y} $ 
and the derivatives of  $ \vec{y} $  that are of lower order than the 
derivatives on the left hand side of the equation
\eqref{Nonlinear}. Thus
\begin{equation}
H_i ( \vec{y} ) = h_i \left( t,y_1, \ldots , y_m, \ldots ,
\frac{d^{k_1} y_1}{dt^{k_1}} \;, \ldots ,\; 
\frac{d^{k_m} y_m}{dt^{k_m}}, \ldots , \right)
\end{equation}
where  $ k_j < p_j $ and  $ h_j $  is an analytic function of its arguments.
More precisely, let
\begin{equation}
p = \sum_{j=1}^{m} p_j
\end{equation}
and then introduce the  $ p-m $  variables
\begin{equation}
v_j^k \;,\; 1 \leq j \leq m \;,\; 1 \leq k < p_j \;.
\end{equation}
We think of  $ v_j^k $  being short hand for $ d^k y_j /dt^k $.
Then the  $ h_i $  are analytic functions of the  $ p+1 $  variables
$ (t, \vec{y} , v_j^k ) $, that is,
\begin{equation}
h_i = h_i (t, y_1, \ldots , y_m, \ldots , v_j^k, \ldots ) \;.
\end{equation}
To apply the results of Section 6 of Chapter 1, let
\begin{equation}
\vec{f} (t) = (f_1 (t), \ldots , f_m (t))
\end{equation}
and then introduce the operator
\begin{equation}
\vec{F} ( \vec{f} ) = (F_1 ( \vec{f} ), \ldots , \vec{F}_m ( \vec{f} ))
\end{equation}
where
\begin{equation}
F_j ( \vec{f} ) = \frac{d^{p_j}}{dt^{p_j}} f_j - H_j ( \vec{f} ) \;.
\end{equation}
The derivative of  $ \vec{F} $  in the direction  $ \vec{g} $  is given by
\begin{equation}
(D_{ \vec{f} } \vec{F} )( \vec{g} ) = ((D_{ \vec{f} } F_1 )( \vec{g} ),
\ldots , (D_{ \vec{f} } F_m )( \vec{g} ))
\end{equation}
where
\begin{align}
(D_{ \vec{f} } F_j )( \vec{g} ) & = \frac{d}{d \epsilon} \left\{ 
\frac{d^{p_j} (f_j + \epsilon g_j )}{dt^{p_j}} - H_j ( \vec{f} + \epsilon 
\vec{g}) \right\}_{\epsilon =0} \\
\nonumber
& = \frac{d^{p_j} g_j}{dt^{p_j}} - (D_{ \vec{f} } H_j )( \vec{g} ) \;.
\end{align}
Next,
\begin{align}
\nonumber  (D_{ \vec{f} } H_j )( \vec{g} ) & = \frac{d}{d \epsilon} h_j 
\left( t, f_1 + \epsilon g_1, \ldots , f_m + \epsilon g_m, \ldots , 
\frac{d^k (p_j + \epsilon g_j ) }{dt^k }, \ldots \right) 
\mid_{\epsilon =0} \\
& = \frac{\partial h_j}{\partial y_1} \cdot g_1 + \ldots +
\frac{\partial h_j}{\partial y_m} \cdot g_m + \ldots +
\frac{\partial h_j}{\partial v_j^k} \cdot \frac{d^k g_j}{dt^k} + \ldots
\;.
\end{align}

The condition of infinitesimal invariance (see Section 6 of Chapter 1) is
\begin{equation}
\vec{F} ( \vec{f} ) = 0 = > (D_{ \vec{f} } F ) ( \vec{S} ( \vec{f} )) = 0
\;.
\end{equation}
Recall that  $ S $  has the form
\begin{align}
\nonumber \vec{S} ( \vec{f} ) & = (S_1 (f),\ldots , S_m (f)),
\text{ where }\\
S_j ( \vec{f} ) & = - \vec{S} (t, \vec{f} ) \cdot \vec{\nabla}_x f_j + 
r_j (t, \vec{f} ) \;,\; 1 \leq j \leq m \;,
\end{align}
where $ s ( t , \vec{y} ) $ and $ r_j (t, \vec{y} ), 1 \leq j \leq m $
are functions that are to be determined.
The way this condition is used is to replace all derivatives of the form
\begin{equation}
\frac{d^{p_{i}+k} f_i}{dt^{p_{i}+k}}
\end{equation}
that occur in
\begin{equation}
(D_{ \vec{f} } \vec{F} )( \vec{S} ( \vec{f} )) = 0
\end{equation}
by the right hand side (or appropriate derivative thereof)
of one of the differential equations \eqref{Nonlinear}.
We call the resulting expression  $ \vec{E} $.  The expression
$ \vec{E} $  now contains only derivatives of  $ f_i $  of order less than
$ p_i $.  At this point the  $ f_i $  are still restricted to
solutions of the system of differential equations which are not normally known.
However, the existence theorem for the initial value problem says that it is
always possible to find a solution of the differential equation satisfying
the initial conditions
\begin{equation}
\frac{d^k y_j}{dt^k} (t) = v_j^k \;,\; 1\leq j \leq m \;,\;
0 \leq k < p_j \;,
\end{equation}
where  $ v_j^0 = y_i $  and the  $ v_j^k $, $ 1 \leq j \leq m $, 
$ 0 \leq k < p_j $  are arbitrary. Consequently
the derivatives  $ d^k f_j /dt^k $  in the expression
$ \vec{E} $  may be replaced by the variables  $ v_j^k $.  Thus
$ \vec{E} $  becomes an expression of the form
\begin{equation} \label{Expression}
\vec{E} (t, y_1, \ldots , y_m , v_1^1, \ldots , v_m^1, \ldots , v_j^k,
\ldots ) = 0
\end{equation}
and the equality holds for all values of 
$ (t,y_1, \ldots , y_m, v_1^1, \ldots , v_m^1, \ldots , v_j^k, \ldots ) $.
This expression also depends on the coefficients of the unknown infinitesimal
symmetry operators.  It is this last expression \eqref{Expression} 
that is solved for the coefficients of the infinitesimal transformation.

\section{One First Order Equation}

It is important to realize that the case of a single first order differential
equation is very simple compared to other cases.  Thus this section does not
provide good insight into what will happen with second order equations, systems 
of equations or partial differential equations.  However, the calculations are
simple and thus this is a nice place to start, other authors have discussed
this case [4, Section 1.9].

The differential equation under consideration is to be written in the form
\begin{equation} \label{First Order}
y^\prime = a(x,y)
\end{equation}
where  $ y^\prime = dy/dx $.
The solutions of differential equations of this form are curves
$ y = f(x) $  in  $ R^2 $  and consequently we will need to consider
infinitesimal groups on  $ R^2 $,
\begin{equation}
L = r(x,y) \frac{\partial}{\partial x} + s(x,y) \frac{\partial}{\partial y} \;.
\end{equation}
The corresponding infinitesimal action on curves  $ y=f(x)  $  is given by
(see Chapter 1, Section 4)
\begin{equation}
S(f) = -r(x,f)f^\prime + s(x,f)
\end{equation}

The directional derivative of the operator form of the equation,
\eqref{First Order}
\begin{equation}
F(f) = f^\prime - a(x,f) \;,
\end{equation}
in the direction  $ g $  is
\begin{align}
\nonumber (D_f F)(g) & = \frac{d}{d \epsilon} F(f + \epsilon g) 
\mid_{\epsilon =0} \\
 & = \frac{d}{d \epsilon} \{ f^\prime + \epsilon g^\prime - a(x,f + \epsilon g)
\} \mid_{\epsilon =0} \\
\nonumber & = g^\prime - a_y (x,f)g \;.
\end{align}
The invariance condition (see Chapter 1, Section 6) 
\begin{equation}
F(f) = 0 \Rightarrow (D_f F)(S(f)) = 0 \;,
\end{equation}
then becomes
\begin{equation}
F(f) = 0 \Rightarrow \frac{d}{dx} S(f) - a_y (x,f)S(f) = 0 \;,
\end{equation}
that is (we now suppress function arguments),
\begin{equation} \label{fprimeprime}
f^\prime = a(x,f) \Rightarrow -rf^{\prime \prime} - r_x f^\prime - r_y 
(f^\prime )^2 - a_y (-rf^\prime + s) = 0 \;.
\end{equation}

The next step is to use the differential equation
\eqref{First Order}
to eliminate  
$ f^\prime $  and $ f^{\prime \prime} $  from \eqref{fprimeprime}.  The chain rule applied 
to  $ f^\prime = a(x,f) $  gives
\begin{equation}
f^{\prime \prime} = a_x + a_y f^\prime = a_x + aa_y
\end{equation}
and then this converts \eqref{fprimeprime} to
\begin{equation}
-r(a_x + aa_y ) - r_x a - r_y a^2 + s_x + s_y a + raa_y + sa_y = 0 \;.
\end{equation}
or
\begin{equation} \label{Every Solution}
-ra_x - r_x a - r_y a^2 + s_x + s _a + sa_y = 0 \;.
\end{equation}
This equation \eqref{Every Solution} is to hold for every solution  $ y=f(x) $  of the given
differential equation \ref{First Order}.

\eqref{Every Solution}

However, the existence theorem for the initial value problem for \ref{First Order}
says that for every  $ x_0 $  and  $ y_0 $  there exists a solution
$ y=f(x) $  of the differential equation with  $ y_0 = f(x_0 ) $.  
Consequently, \eqref{Every Solution} must hold for all  $ x_0 $  and  $ y_0 $.  If we
relabel  $ x_0 $  and  $ y_0 $  by  $ x $  and  $ y $,  then \eqref{Every Solution}
must hold for all  $ x $  and  $ y $, that is,
\begin{align} \label{Invariance}
& -r(x,y)a_x (x,y) - r_x (x,y)a(x,y) - r_y (x,y) a^2 (x,y) + \\
& \quad \quad 
\nonumber s_x (x,y) + s_y (x,y)a(x,y) + s(x,y)a_y (x,y) \equiv 0 \;.
\end{align}

Before we start looking at examples, let us look for a moment at the converse of our
standard problem.  The standard problem is, of course, given a differential
equation find the groups that leave the solution space of the equation
invariant.  The converse is, given a group find the equations whose solution
space is left invariant by the group.  To illustrate this we consider
translation groups in either  $ x $  or  $ y $.

Suppose a differential equation is invariant under the group of translations
in the  $ y $  variable.  In this case, the infinitesimal group is given by
(see Chapter 1, Section 2)
\begin{equation}
\frac{\partial}{\partial y} \;,
\end{equation}
that is,
\begin{equation}
r(x,y) = 0 \;,\; s(x,y) = 1
\end{equation}
and invariance condition \eqref{Invariance} becomes
\begin{equation}
a_y (x,y) = 0 \;,
\end{equation}
that is, $ a=a(x) $  and then the differential equation becomes
\begin{equation}
y^\prime = a(x) \;.
\end{equation}
The differential equation is a simple integration problem with the solution
\begin{equation}
y = \int a(x)dx \;.
\end{equation}

On the other hand, suppose that a differential equation \ref{First Order} is invariant 
under
the group of translations in the x variable.  In this case, the infinitesimal
group is given by
\begin{equation}
\frac{\partial}{\partial x}
\end{equation}
or
\begin{equation} \label{Translation r s}
r(x,y) = 1 \;,\; s(x,y) = 0
\end{equation}
and invariance condition \eqref{Invariance} becomes
\begin{equation}
a_x (x,y) = 0 \;.
\end{equation}
Thus the differential equation becomes
\begin{equation}
y^\prime = a(y)
\end{equation}
which is a special case of a separable equation.  An implicit solution is 
given by integration,
\begin{equation}
\int \frac{dy}{a(y)} = x \;.
\end{equation}
This illustrates why we hope that transforming the differential equation
to canonical coordinates for one of its invariance
groups will result in a solution or simplification 
of the given differential equation.

Before we check this out, let us study the simplest example we can imagine.
Thus we will choose  $ a \equiv 0 $  and study the equation
\begin{equation}\label{Trivial}
y^\prime = 0 \;.
\end{equation}
The invariance condition \eqref{Invariance} becomes
\begin{equation}
s_x = 0 \;.
\end{equation}
This then implies that any infinitesimal with
\begin{equation}
r = r(x,y) \;,\; s = s(y) \;,
\end{equation}
yields a group that leaves the equation \eqref{Trivial} invariant.
Written in operator notation, the infinitesimal transformations have the
form
\begin{equation}
L = r(x,y) \frac{\partial}{\partial x} + s(y) \frac{\partial}{\partial y} \;,
\end{equation}
Consequently this ``simplest" differential equation has an
infinite dimensional invariance group.

Although this example is very simple we have set up a program,
ode\_sym\_1, to do the calculation.  This program is in the file examples.v.
We include the program listing here to illustrate the ease with which the 
programs can be used.  For more details on the meaning of the program see 
the chapter on programs and the MACSYMA Manual [29].
The next two sections also include example programs and this chapter
concludes with a more interesting example. 
\newpage
\centerline{Program ode\_sym\_1}
\begin{verbatim}
ode_sym_1() := block(
/* This program computes the symmetries of the simplest first order
   ordinary differential equation in one variable. */ 

/* The veryverbose mode will allow the user to see some of the inner
   working of the program. */ 
	verbose : true,
	veryverbose : true, 

/* The flag num_diff is used in more complicated examples. */
	num_dif:0, 

/* Set the dependent and independent variables. */
	dep : [y],
	indep : [x], 

/* Define the differential equation. Note the use of the noun form 
   of the diff operator. */
	diffeqn : ['diff(y,x) = 0], 

/* Now load and execute the symmetries program. Note that the
   program doitall attempts a more complete solution than symmetry
   but is not appropriate for such a simple example. */
	load(symmetry),
	symmetry(),
	end_ode_sym_1)$
\end{verbatim}

The one parameter transformation group is obtained by exponentiating
the infinitesimal,
\begin{equation}
\left( 
\begin{array}{c}
x (\epsilon) \\ y (\epsilon) 
\end{array}
\right) 
= e^{\epsilon \left( r(x,y) \frac{\partial}{\partial x} + s(y) 
\frac{\partial}{\partial y} \right) } 
\left( 
\begin{array}{c}
x \\ y 
\end{array}
\right) \;.
\end{equation}
The Differential Equation property for Lie series says that
\begin{equation} \label{DE Property}
\dot{x} = r(x,y) \;,\; \dot{y} = s(y) \;.
\end{equation}
Consequently, if we know  $ s $  and  $ r $, then  $ y $  and then  $ x $ 
can be found by integration.

Note that the solutions of the differential equation are all horizontal
lines.  The transformation group can be thought of as follows.
Starting at a point  $ (x,y) $  first move some distance in  $ y $  that is
independent of  $ x $.  Next, move some distance in the  $ x $  direction.
Thus, if two points are on some horizontal line, then both points will
end up on the same horizontal line.  Clearly, any such transformation
sends solutions of the differential equation \eqref{Trivial} into a
solution of the differential equation.

The next example cannot be done using our programs because of certain
difficulties with the differentiation routines in MACSYMA [35].  Once
the differentiation routine is fixed it would be a simple matter to extend
our programs to handle this type of calculation.  The examples concerns finding
a group of transformations that leaves the solution space of each of a
{\it class} of differential equations invariant.  This procedure is not 
general and works here because single first order equations have so many 
symmetries. \\[.2in] 
{\bf Example.}  The equation
\begin{equation} \label{Scale Invariant}
y^\prime = a \left( \frac{y}{x} \right)
\end{equation}
for all functions  $ a $  of a single variable is a common example in many
ordinary differential equation texts.  The infinitesimal invariance
condition \eqref{Invariance} for this equation is
\begin{equation}
s_x + s_y a - r_x a - r_y a^2 + \frac{r_y a^\prime}{x^2} - 
\frac{sa^\prime}{x} = 0 \;.
\end{equation}

The calculation that follows is an excellent example of the techniques used
to find infinitesimal symmetries.  Because  $ a $  is arbitrary we must have
\begin{align} \label{Over}
\nonumber s_x & = 0\; ,& r_y & = 0 \;, \\
    s_y - r_x & = 0\; ,& yr - xs & = 0 \;.
\end{align}
This is an {\it over determined} system of equations for  $ r $  and  
$ s $, that is, there are 4 equations and 2 unknowns.
Differentiating the equation  $ s_y - r_x $  with respects to  $ x $ 
and  $ y $  gives
\begin{equation}
r_{xx} = s_{xy} = 0 \;,\; s_{yy} = r_{xy} = 0 \;.
\end{equation}
The condition  $ r_y = 0 $  implies that  $ r = r(x) $  and then the
condition  $ r_{xx} = 0 $  implies that  $ r = c_1 x + d_1 $  for
some constants  $ c_1 $  and  $ d_1 $.  Similarly,
$ s = c_2 y + d_2 $.  Now the condition \eqref{Over}, $ yr - xs = 0 $, becomes
\begin{equation}
c_1 xy + d_1 y - c_2 xy + d_2 x = 0 \;.
\end{equation}
This must hold for all  $ x $  and  $ y $  so  $ d_1 = 0 $,
$ d_2 = 0 $  and  $ c_1 = c_2 $.  The value of  $ c_1 $
$ (c_1 \neq 0) $   doesn't affect the results because the symmetries
form a linear space, so
\begin{equation}
L = x \frac{\partial}{\partial x} + y \frac{\partial}{\partial y} \;.
\end{equation}
This is the infinitesimal generator of the dilation or scaling group
(see Chapter 1, Section 2)
\begin{equation}
\xi = e^\epsilon x \;,\; \eta = e^\epsilon y \;.
\end{equation}
This is the {\it only} symmetry of this {\it class} of equations.

We now point out that it is obvious that scalings are a symmetry of this
class of equations although it is not completely obvious that this is the
{\it only} symmetry.  Set
\begin{equation}
\lambda = e^\epsilon
\end{equation}
so that our notation agrees with that found many places in the literature.
Then choose
\begin{equation}
\xi = \lambda x \;,\; \eta = \lambda y
\end{equation}
so that
\begin{equation}
d \xi = \lambda dx \;,\; d \eta = \lambda dy \;.
\end{equation}
Consequently
\begin{equation}
\frac{\eta}{\xi} = \frac{y}{x} \;,\; \frac{d \eta}{d \xi} = \frac{dy}{dx} \;.
\end{equation}
In the  $ ( \xi , \eta ) $  coordinates the differential equation \eqref{Scale Invariant}
becomes
\begin{equation}
\eta^\prime = \frac{d \eta}{d \xi} = a \left( \frac{\eta}{\xi} \right)
\end{equation}
which is the same equation as the original equation \eqref{Scale Invariant}.

Let us now transform everything in this example to canonical coordinates 
(see Chapter 1, Section 5) for the dilation group.
This requires the computation of functions  $ \xi = \xi (x,y) $,
$ \eta = \eta (x,y) $  such that
\begin{equation}
L \xi = 0 \;,\; L \eta = 1
\end{equation}
where
\begin{equation}
L = x \frac{\partial}{\partial x} + y \frac{\partial}{\partial y} \;.
\end{equation}
The characteristic equation for  $ L $  is
\begin{equation}
\frac{dx}{x} = \frac{dy}{y}
\end{equation}
which has
\begin{equation}
\frac{y}{x} = k
\end{equation}
as an integral.  Consequently
\begin{equation}
\xi = h_1 \left( \frac{y}{x} \right)
\end{equation}
for any function  $ h_1 $  of a single variable.  To find  $ \eta $  we try  
$ \eta = \eta (y) $  and then  $ L \eta = 1 $  becomes
\begin{equation}
y \frac{d \eta}{dy} = 1
\end{equation}
or  $ \eta = \ln \mid y \mid + c $.  Consequently
\begin{equation}
\eta = h_2 \left( \frac{y}{x} \right) + \ln \mid y \mid
\end{equation}
for any function  $ h_2 $  of a single variable.

We now choose (because things come out nice)
\begin{align}
\nonumber \eta & = \ln \mid y \mid \;,&  y & = e^\eta \;, \\
\xi & = \frac{y}{x}  \;, &	x & = \frac{e^\eta}{\xi} \;.
\end{align}
The differential equation \eqref{Scale Invariant} becomes
\begin{equation}
\frac{\xi^2 \eta^\prime}{\xi \eta^\prime -1} = f(\xi)
\end{equation}
or
\begin{equation}
\eta^\prime = \frac{1}{\xi f(\xi)- \xi^2 } \;.
\end{equation}
This equation can now be solved by integration,
\begin{equation}
\eta = \int \frac{d \xi}{\xi f( \xi )- \xi^2 } \;.
\end{equation}
\vspace{.2in}
{\bf Exercise.}  Show that another choice for coordinates is
\begin{equation}
\eta = \frac{y}{x} + \ln \mid x \mid \;,\; \xi = \frac{y}{x}
\end{equation}
and in these coordinates the differential equation becomes
\begin{equation}
\eta^\prime = 1 + \frac{1}{f( \xi )- \xi} \;.
\end{equation}

By the way, the usual transformation applied to this equation is
\begin{equation}
\eta = \frac{y}{x} \;,\; \xi = x
\end{equation}
which then gives rise to the separable equation
\begin{equation}
\xi \eta^\prime = f( \eta ) - \eta \;.
\end{equation}
Because our methods force the equation to become an integral we would not find
this transformation. \\[.2in]
{\bf Exercise.}  Show that the infinitesimal invariance condition for the 
class of equations
\begin{equation}
y^\prime = a(x-y) \;,
\end{equation}
for any function  $ a $  of one variable reduces to
\begin{align}
\nonumber s_x & = 0 \; , & r + s & = 0 \;, \\
          r_y & = 0 \; , & r_x - s_y & = 0 \;.
\end{align}
Consequently the symmetry operators are given by
\begin{equation}
L = \frac{\partial}{\partial x} - \frac{\partial}{\partial y} \;.
\end{equation}
Canonical coordinates are given by
\begin{equation}
\eta = \frac{y-x}{2} \;,\; \xi = \frac{y+x}{2} \;,
\end{equation}
and in these coordinates the differential equation reduces to the integral
\begin{equation}
\frac{d \eta}{d \xi} = \frac{f(2 \xi )-1}{f( 2 \xi )+1} \;.
\end{equation}

The symmetry properties of a first order equation can be derived from the
fact that symmetry groups are invariant under a change of variables.
The equation we are studying is
\begin{equation} \label{General ODE}
y^\prime = a(x,y) \;.
\end{equation}
We now construct a change of variables.
\begin{align}
\nonumber \xi & = \xi (x,y) \; , &	x & = x(\xi , \eta) \;, \\
         \eta & = \eta (x,y) \; , &	y & = y(\xi , \eta) \;,
\end{align}
that will transform \eqref{General ODE} into the equation  $ y^\prime = 0 $.
It is important to note that we are setting things
up so that $ y $  and  $ \eta $  are dependent variables while  $ x $  and
$ \xi $  are independent variables.  Now
\begin{equation}
dy = y_\xi d \xi + y_\eta d \eta \;,\; dx = x_\xi d \xi + y_\eta d \eta
\end{equation}
and consequently
\begin{equation}
y^\prime = \frac{dy}{dx} = \frac{y_\xi d \xi + y_\eta d \eta }
{x_\xi d \xi + x_\eta d r\eta }
= \frac{y_\xi + y_\eta \eta^\prime}{x_\xi + x_\eta \eta^\prime } \;.
\end{equation}
Now solve for $ \eta^\prime $,
\begin{equation}
\eta^\prime = - \frac{x_\xi y^\prime -y_\xi}{x_\eta y^\prime -y_\eta} \;.
\end{equation}
Thus, in the  $ ( \xi , \eta ) $  coordinates the differential equation
\eqref{General ODE} becomes
\begin{equation}\label{Condition}
\eta^\prime = - \frac{x_\xi a -y_\xi}{x_\eta a-y_\eta} \;.
\end{equation}

If
\begin{equation}
x_\xi f - y_\xi = 0 \;,
\end{equation}
that is,
\begin{equation}
x_\xi ( \xi , \eta )f(x( \xi , \eta ),y( \xi , \eta )) - y_\xi ( \xi , \eta )
= 0 \;,
\end{equation}
then the differential equation becomes
\begin{equation}
\eta^\prime = 0 \;.
\end{equation}

The above condition can be written
\begin{equation}
\frac{y_\xi}{x_\xi} = f(x( \xi , \eta ),y( \xi , \eta )) \;.
\end{equation}
This equation says that if  $ \eta $  is fixed and $ \xi $  varies, then
$ (x( \xi , \eta ),y( \xi , \eta )) $  is a parametric curve which is a solution
of the differential equation  $ y^\prime = f(x,y) $.  Another way to 
describe this curve is that it is a level curve of  $ \eta $,
\begin{equation}
\eta (x,y) = {\rm const.}
\end{equation}
Thus  $ \eta (x,y) $  must be an integral of the equation \eqref{General ODE}.   
On the other hand, if  $ F(x,y) $  is an integral of \eqref{General ODE}, then we can choose
$ \eta = F(x,y) $  and  $ \xi = \xi (x,y) $  as new coordinates, where
$ \xi (x,y) $  is fairly arbitrary, and in this coordinate system the
differential equation \eqref{General ODE} becomes
\begin{equation}
\eta^\prime = 0 \;.
\end{equation}

Previously we found the infinitesimal symmetries and one parameter groups 
of symmetries of  $ y^\prime = 0 $.    However, if we want
the equation  $ y^\prime = 0 $  to transform into  $ \eta^\prime = 0 $, 
then equation \eqref{Condition} tells us that we must have  $ y_\xi = 0 $. 
Thus any transformation of the form
\begin{equation}
x = x( \xi , \eta ) \;,\; y = y( \eta ) \;,
\end{equation}
is a symmetry of  $ y^\prime = 0 $.  Interchanging the roles of  $ (x,y) $  and
$ ( \xi , \eta ) $  and then solving for  $ x $  and  $ y $  gives
\begin{equation}
\eta = \eta (y) \;,\; \xi = \xi (x,y)
\end{equation}
which is the analog of the previously derived condition \eqref{DE Property} on the 
symmetries.
Because symmetry groups are preserved under changes of variables, the
equation  $ y^\prime = a(x,y) $  must have infinitely many symmetries.

\section{One Second Order Equation}

The case of a second order equation is fairly typical of the general case of
higher order equations and partial differential equations.  In
particular, the result that the infinitesimal symmetries form a finite
dimensional space is important and typical.  We will consider an equation of the
form
\begin{equation} \label{Second Order}
y^{\prime \prime} = a(x,y,y^\prime ) \;.
\end{equation}
Of course, $ y^\prime = dy/dx $  and  $ y^{ \prime \prime } = dy^\prime
/dx $  and  $ a $  is an arbitrary function of three variables  
$ a = a(x,y,v) $.

The operator form of the differential equation is
\begin{equation}
F(f) = f^{ \prime \prime } - a(x,f,f^\prime )
\end{equation}
and the derivative of  $ F $  in the direction  $ g $  is
\begin{align}
\nonumber (D_f F)(g) & = \frac{d}{d \epsilon} F(f + \epsilon g) 
\mid_{\epsilon =0} \\
& = \frac{d}{d \epsilon} (f + \epsilon g)^{ \prime \prime } - a(x,f + \epsilon 
g, (f + \epsilon g)^\prime ) \mid_{\epsilon =0} \\
\nonumber & = g^{\prime \prime} - a_y ( x,f,f^\prime )g - a_v (x,f,f^\prime
)g^\prime
\end{align}
The infinitesimal symmetries have the same form as in the first order
equation case,
\begin{equation}
L = r(x,y) \frac{\partial}{\partial x} + s(x,y) \frac{\partial}{\partial y} \;.
\end{equation}
The corresponding infinitesimal action on curves is given by
\begin{equation}
S(f) = -r(x,f)f^\prime + s(x,f) \;.
\end{equation}

Now, the invariance condition (see Chapter 1, Section 6) becomes
\begin{equation}
F(f) = 0 = > (D_f F)(S(f)) = 0 \;,
\end{equation}
that is,
\begin{align}\label{Begin}
f^{\prime \prime} & - a(x,f^\prime ,f^{\prime \prime}) = 0 = > \\ 
\nonumber
& (-r(x,f)f^\prime + s(x,f))^{\prime \prime} - a_y (x,f,f^\prime )(-r(x,f)
f^\prime + s(x,f)) \\
& \nonumber - a_v (x,f,f^\prime )(-r(x,f)f^\prime + s(x,f))^\prime = 0 \;.
\end{align}
Now
\begin{equation}
(-rf^\prime + s)^\prime = -r_x f^\prime - r_y (f^\prime)^2 -
rf^{\prime \prime} + s_x + s_y f^\prime
\end{equation}
and
\begin{align}
 \nonumber (-rf^\prime + s)^{\prime \prime} & = -r_{xx} f^\prime - 2r_{xy} 
(f^\prime)^2 - 2r_x f^{\prime \prime} - r_{yy} (f^\prime)^3 - 3r_y f^\prime
f^{\prime \prime} \\
& \quad - rf^{\prime \prime \prime} 
 + s_{xx} + 2s_{xy} f^\prime + s_{yy} (f^\prime)^2 \;.
\end{align}
The differential equation \eqref{Second Order} gives $ f^{\prime \prime} = a(x,f,f^\prime) $
and differentiating this gives
\begin{equation} \label{End}
f^{\prime \prime \prime} = a_x (x,f,f^\prime) + a_y (x,f,f^\prime)f^\prime
+ a_v (x,f,f^\prime)f^{\prime \prime} \;.
\end{equation}
Combining \eqref{Begin} through \eqref{End}) gives
\begin{align} \label{Big Equation}
& a_v (-s_x -s_y y^\prime + r_x y^\prime + r_y (y^\prime)^2 ) + \\ 
\nonumber
& a_y (-s-ry^\prime) + a_x (-r)+ a(s_y -2r_x -3r_y f^\prime) + \\
\nonumber
& s_{xx} - r_{xx} y^\prime + 2s_{xy} y^\prime - 2r_{xy} (y^\prime)^2 
+ s_{yy} (y^\prime)^2 - r_{yy} (y^\prime)^3 = 0 \;.
\end{align}
This equation holds for all  $ x $  and all solutions  $ f(x) $  of the
differential equation \eqref{Second Order}. 

The fact that the initial value problem
\begin{equation}
f^{\prime \prime} = a(x,f,f^\prime) \;,\; f(x) = y \;,\; f^\prime (x) = v
\;,
\end{equation}
always has a solution allows us to replace  $ f $  by  $ y $  and  
$ f^\prime $  by $ v $  in \eqref{Big Equation}, 
\begin{align} \label{Really Big}
& a_v (x,y,v)(-s_x (x,y) - s_y (x,y)v + r_x (x,y)v^2) \\
& \nonumber + a_y (x,y,v)(-s(x,y) - r(x,y)v) + a_x (x,y)(-r(x,y)) \\
& \nonumber + a (x,y,v) (s_y (x,y) - 2 r_x (x,y) - 3 r_y (x,y) v ) + s_{xx} (x,y)
\\
& \nonumber
- r_{xx} (x,y)v + 2s_xy (x,y)v - 2r_{xy} (x,y)v^2 
+ s_{yy} (x,y)v^2 - r_{yy} (x,y)v^3 = 0
\end{align}
Equation \eqref{Really Big} must hold for all  $ (x,y,v) $.
It is also helpful to collect the terms in the following way,
\begin{align} \label{Nice Big}
\nonumber
& s_{xx} + 2vs_{xy} + v^2 s_{yy} - vr_{xx} - 2v^2 r_{xy} - v^3 r_{yy} \\
\nonumber
&\quad + r_x (va_v -2a) + r_y (v^2 a_v -3va) + r(-va_y -a_x) \\
&\quad + s_x (-a_v) + s_y (-va_v + a) + s(-a_y) = 0 \;.
\end{align}

Thus we see that)  \eqref{Nice Big}
is linear in  $ a $.  Consequently, the inverse problem of finding an
equation that is invariant under a given group becomes a problem of
solving a first order linear partial differential equation for
$ a = a(x,y,v) $.  The existence theory for this type of equation tells
us that there are infinitely many solutions.  However, when the differential
equation is given, that is, $ a = a(x,y,v) $  is given, then \eqref{Nice Big}
is a single second order linear homogeneous partial differential equation for
determining  $ r(x,y) $  and  $ s(x,y) $.  Because  $ r(x,y) $  and
$ s(x,y) $  do not depend on  $ v $, this equation implies much more.  If
$ a(x,y,v) $  is expanded as a power series in  $ v $  and this is
substituted into \eqref{Nice Big}, then the coefficients of the powers of  $ v $  
must be zero.  This will
produce an infinite set of linear homogeneous second order (at most)
partial differential equations for determining  $ r(x,y) $  and  $ s(x,y) $.
Because the explicit powers of  $ v $ in \eqref{Nice Big}
range between  0  and  3  and the second order terms do depend on  $ a $,
this system must contain at least 4 independent equations for  $ r $  and
$ s $.  Consequently the system is overdetermined implying that there is
at most a finite number of symmetries [56, 18]. \\[.2in]
{\bf Example.}  Let us look at the simplest possible example,
\begin{equation}
y^{\prime \prime} = 0 \;.
\end{equation}
Thus  $ a \equiv 0 $  and equation \eqref{Nice Big} becomes
\begin{equation} \label{Simple Big}
-vr_{xx} - 2v^2 r_{xy} - v^3 r_{yy} + s_{xx} + 2vs_{xy} + v^2 s_{yy} = 0
\;.
\end{equation}
This is a polynomial in  $ v $  so its coefficients must be zero,
\begin{align} \label{Coefficients}
\nonumber
r_{yy} & = 0 \;,            & -2r_{xy} + s_{yy} & = 0 \\
-r_{xx} + 2s_{xy} & = 0 \;, & s_{xx} & = 0 \;.
\end{align}

This last system of equations \eqref{Coefficients} is an {\it over determined}
system of equations for  $ r $  and  $ s $, that is, there are 2 unknowns
and 4 equations.  The following discussion illustrates an important technique
for solving such equations.

Differentiate the second equation in \eqref{Coefficients} with respects to  $ y $,
\begin{equation}
s_{yyy} = 2r_{xyy} = 0 \;,
\end{equation}
and the third equation in  with respects to  $ x $,
\begin{equation}
r_{xxx} = 2s_{xxy} = 0 \;.
\end{equation}
The technique for solving these equations is discussed in Appendix A.
The equations
\begin{equation}
r_{yy} = 0 \;,\; r_{xxx} = 0
\end{equation}
imply that
\begin{equation} \label{Imply 1}
r = c_1 x^2 y + c_2 xy + c_3 y + c_4 x^2 + c_5 x + c_6
\end{equation}
where  $ c_i $  are constants.  The conditions
\begin{equation}
s_{xx} = 0 \;,\; s_{yyy} = 0
\end{equation}
imply that
\begin{equation} \label{Imply 2}
s = d_1 xy^2 + d_2 xy + d_3 x + d_4 y^2 + d_5 y + d_ 6 \;,
\end{equation}
where the  $ d_i $  are constants.  Plugging \eqref{Imply 2} and \eqref{Imply 1}
into the middle two equations in \eqref{Coefficients} gives
\begin{equation}
-2(2c_1 x + c_2 ) + (2d_1 x + 2d_4 ) = 0 \;,\; -(2c_1 y+c_4 ) + 2(2d_1 
y+d_2 ) = 0 \;.
\end{equation}
This must hold for all  $ x $  and  $ y $  and consequently
\begin{equation}
c_1 = c_4 = d_1 = d_4 = 0 \;,
\end{equation}
and then
\begin{equation}
L = (c_2 xy+c_3 y +c_5 x +c_6 ) \frac{\partial}{\partial x } + 
(d_2 xy+d_3 x+d_5 y+d_6 ) \frac{\partial}{\partial y } \;.
\end{equation}
Thus the infinitesimal symmetries of the equation  $ y^{\prime \prime}=0 $  
form an 8
dimensional linear space called the projective algebra.
In the case of partial differential equations we will find similar algebras.
The operators in this algebra are easy to exponentiate using Lie series, see
[4, Section 1.7] for the details of the exponentiation using classical
methods.

We have included in our example programs a program, ode\_sym\_2, to do the
above computation.  This program is contained in the file examples.v.
The program uses a slightly different notation where the
symmetry operator has the form
\begin{equation}
a_2 y^\prime + a_1
\end{equation}
and consequently
\begin{equation}
a_2 = -r \;,\; a_1 = s \;.
\end{equation}
This is an example that anyone wishing to use the programs should run.
Note that in this elementary example the program will produce the
symmetries with no help from the user. The program uses a slightly
different notation for the various functions, however, comparing
the program output with our discussion will make all conventions
transparent.
For a more complete discussion of the programming details see
the chapter of programs and the last section of this chapter. 
\newpage
\centerline{Program ode\_sym\_2}

\begin{verbatim}
ode_sym_2() := block( 
/* This program computes the symmetries of the simplest second
   order ordinary differential equation in one variable. */ 

/* The veryverbose mode will allow the user to see some of the inner
   working of the program. */ 
	verbose : true, 
	veryverbose : true, 

/* The flag num_diff is used to limit the number of differentiations
   made in attempting a solution of the equation list. */ 
	num_dif:3, 

/* Set the dependent and independent variables. */ 
	dep : [y], 
	indep : [x], 

/* Define the differential equation. Note the use of the noun form 
   of the diff operator. */ 
	diffeqn : ['diff(y,x,2) = 0], 

/* Now load and execute the program doitall. The program doitall attempts
   a more complete solution than the program symmetry. */ 
	load(doitall), 
	doitall(), 
	end_ode_sym_2)$ 
\end{verbatim}

\section{Two First Order Equations}

We will write the system of equations in the form
\begin{equation}
x^\prime = a(t,x,y) \;,\; y^\prime = b(t,x,y)
\end{equation}
where  $ x^\prime = dx/dt $  and  $ y^\prime = dy/dt $.  The operator form 
of these equations is given by
\begin{equation}
\vec{F} (f,g) = (f^\prime -a(t,f,g) \;,\; g^\prime - b(t,f,g))
\end{equation}
where  $ f=f(t) $, $ g=g(t) $.  The derivative of  $ \vec{F} $  in the direction
$ (u,v) $  is
\begin{align}
 (D_{f,g} \vec{F} )(u,v) &= \frac{d}{ d \epsilon } \vec{F} (f+ \epsilon u ,
g+ \epsilon v) \mid_{\epsilon =0 } \\
& \nonumber = ( u^\prime -a_x (t,f,g)u - a_y (t,f,g)v, \\
\nonumber
& \quad \quad v^\prime  - (a_x (t,f,g)u - b_y (t,f,g)v) \;.
\end{align}
The infinitesimal symmetries have the form
\begin{equation}
L = q(t,x,y) \frac{\partial}{\partial t } + r(t,x,y) \frac{\partial}
{\partial x } + s(t,x,y) \frac{\partial}{\partial y }
\end{equation}
and the corresponding action on curves  $ (f(t),g(t)) $  is given by
\begin{equation}
\vec{S} (f,g) = (-qf^\prime +r,-qg^\prime +s) \;.
\end{equation}

The invariance condition (see Chapter 1, Section 6) becomes
\begin{align} 
\nonumber
& \vec{F} (f,g) = 0  \Rightarrow \\
\nonumber
& (D_{f,g} \vec{F} )( \vec{L} (f,g))  =  \\
& \quad\quad ((-qf^\prime +r)^\prime - (-qf^\prime +r)a_x - (-qg^\prime+s)a_y + \\
\nonumber
& \quad\quad (-qg^\prime +s)^\prime - (qf^\prime +r)b_x - (qg^\prime +s)b_y)=0
\;,
\end{align}
that is,
\begin{align} \label{Start List}
&(-qf^\prime +r)^\prime - (-qf^\prime +r)a_x - (-qg^\prime +s)a_y =0 \;, \\
& \nonumber (-qg^\prime +s)^\prime -(-qf^\prime +r)b_x -(-qg^\prime +s)b_y =0
\;.
\end{align}
The system of differential equations gives
\begin{equation} 
f^\prime = a(t,f,g) \;,\; g^\prime = b(t,f,g)
\end{equation}
and differentiating with respects to  $ t $  gives
\begin{align}\label{Second Derivative}
f^{\prime \prime} & = a_t + a_x f^\prime + a_y g^\prime = a_t + aa_x + ba_y \;,
\\
\nonumber g^{\prime \prime} & = b_t + b_x f^\prime + b_y g^\prime = b_t + ab_x 
+ bb_y \;.
\end{align}
Combining \eqref{Start List}  through 
\eqref{Second Derivative} gives
\begin{align} \label{Final Result}
\nonumber aq_t + a^2 q_x + abq_y + a_t q + a_x r + a_y s - r_t - 
ar_x - br_y & = 0 \;, \\
bq_t + abq_x + b^2 q_y + b_t q + b_x r + b_y s - s_t - as_x - bs_y & = 0
\;.
\end{align}
Here  $ a $, $ b $, $ q $, $ r $, and  $ s $  are all functions of  
$ (t,x,y) $.  Consequently,
\eqref{Final Result}
is a system of 2 first order linear partial differential equations for
determining  $ q $, $ r $, and  $ t $.  Thus the system \eqref{Final Result}
is under determined and will have infinitely many solutions. \\[.2in]
{\bf Example.}  Again, we look at the simplest possible example,
\begin{equation} \label{Simple ODEs}
x^\prime = 0 \;,\; y^\prime = 0 \;.
\end{equation}
The invariance condition \eqref{Final Result} becomes
\begin{equation} \label{No Name}
r_t = 0 \;,\; s_t = 0 \;.
\end{equation}
The solutions of \eqref{Simple ODEs}
are lines in  $ (t,x,y) $ space that are parallel to the $ t $ axis.  The
exponential of an infinitesimal satisfying
\eqref{No Name} 
will move any point  $ (t,x,y) $  to some point  $ ( \tilde{t} , \tilde{x} ,
\tilde{y} ) $  where  $ \tilde{x} $  and  $ \tilde{y} $  are independent of
$ t $  and then move this point parallel to the  $ t $  axis, see the
Differential Equation Property in Section 3 of Chapter 1 and Section 2 of
this chapter.  Such a transformation clearly sends solutions into solutions.
The following program will do this example.

Again, we wrote a VAXIMA program ode\_sym\_11,
contained in the file examples.v, that will do this example.  As before,
a comparison of the discussion in this section with the output
of the program will make the notation clear.
For more details see the chapter on programming and the next section
of this chapter.
\newpage
\centerline{Program ode\_sym\_11} 

\begin{verbatim}
ode_sym_11() := block( 
/* This program finds the symmetries of the simplest system of
   two first order ordinary differential equations. */ 

/* The veryverbose mode will allow the user to see some of the inner
   working of the program. */ 
	verbose : true, 
	veryverbose : true, 

/* The flag num_diff is used in more complicated examples. */ 
	num_dif:0, 

/* Set the dependent and independent variables. */ 
	dep : [x,y], 
	indep : [t], 

/* Define the differential equation. Note the use of the noun form 
   of the diff operator. */ 
	diffeqn : ['diff(x,t) = 0 , 'diff(y,t) = 0], 

/* Now load and execute the program symmetry. Note that the program
   doitall attempts a more complete solution than symmetry but is not
   appropriate for such a simple example. */ 
	load(symmetry), 
	symmetry(), 
	end_ode_sym_11)$ 
\end{verbatim}
\vspace{.2in}
{\bf Example.}  The equation  $ y^{\prime \prime} = 0 $  can be converted 
to the equivalent system
\begin{equation}
x^\prime = 0 \;,\; y^\prime = x \;.
\end{equation}
We will now apply the results of this section to this system and then compare 
this to the results on second order equations.
Thus  $ a(t,x,y)=0 $  and  $ b(t,x,y)=x $  and the determining equations
\eqref{Final Result} become
\begin{align} \label{New System}
\nonumber r_t + xr_y & = 0 \\
xq_t + x^2 q_y + r - s_t - xs_y & = 0
\end{align}
The method of characteristics will allow us to construct infinitely many
solutions to \eqref{New System} .
To compare this result to that obtained for second order equations,
eliminate  $ r $  from \eqref{New System} 
because  $ r $  determines the change in  $ x $  which corresponds to the
derivative of  $ y $  and this was not included in the second order case.
Differentiate the last equation in \eqref{New System} 
with respects to  $ t $  and  $ y $,
\begin{align} \label{Diff Of}
\nonumber xq_{tt} + x^2 q_{ty} + r_t - s_{tt} - xs_{ty} & = 0 \;, \\
xq_{ty} + x^2 q_{yy} + r_y - s_{ty} - xs_{yy} & = 0 \;.
\end{align}
Now solve \eqref{Diff Of}
for  $ r_t $  and  $ r_y $  and plug into the first of \eqref{New System}
\begin{equation} \label{More Stuff}
-s_{tt} + (q_{tt} +2s_{ty} )x + (2 \, q_{ty} -s_{yy} )x^2 + q_{yy} x^3 = 0
\;.
\end{equation}
If we assume that  $ s $  and  $ q $  are independent of  $ x $, then
\eqref{More Stuff}
yields 4 equations that are the same as those obtained for
$ y^{\prime \prime} = 0 $  in the previous section.  The equation for  $ r $ in
\eqref{New System} is the standard extension of a transformation to derivatives,
see Chapter 4. \\[.2in]
{\bf Exercise.}  An autonomous system of equations
\begin{equation} \label{General System}
x^\prime = a(x,y) \;,\; y^\prime = b(x,y)
\end{equation}
can be reduced to the single first order equation
\begin{equation} \label{Ratio}
\frac{dy}{dx} = \frac{b(x,y)}{a(x,y)} \;.
\end{equation}
Apply the theory in this section to the system
 \eqref{General System}
and the results on first order equations to \eqref{Ratio}
and then make a comparison!
\clearpage

\section{The Toda Lattice}

Here we wish to provide the user with a nontrivial example that will
illustrate the power of our symbol manipulation programs.
We decided on the Toda lattice equations because of the 
current interest in this system.
This problem is sufficiently complex so as to be unpleasant to do
by hand but not so complex that it produces a hard to understand
example.
It is known that the Toda lattice has a nontrivial symmetry.
However, this symmetry depends on the
derivative of the solutions of the differential equations
(momenta) and consequently is not a point symmetry. Thus our
programs will not find this symmetry.  On the other hand,
the Toda lattice is an autonomous system and consequently
time translations will be a symmetry of the system. Thus
our programs must produce this symmetry. The theory of 
Hamiltonian systems implies that there no other Hamiltonian
symmetries (symmetries that are canonical transformations).
We show, in fact, that translations are the only point symmetries
of the Toda lattice.

The remainder of this section consists of a severely edited VAXIMA output.
To remain in the domain of point symmetries we use the two
second order equations that model the Toda lattice.
To give the reader some idea of the size of this problem, it is worth
noting that the file that contained all of the output from the VAXIMA session
where we calculated the symmetries contained 5096 lines and the problem
required 95 cpu minutes to run on a VAX11/780.
The program can be found in the file examples.v.
We will return to this example in a later section.

This section will not be very understandable with out some prior 
knowledge of MACSYMA [29]. The chapter on programs can be used
to look up a description of our programs. As in our other examples
we build a file that contains the information necessary to do the 
computation and then load this file into VAXIMA. \\[.2in]
\centerline{VAXIMA Output}
\begin{verbatim}
We begin by loading a file.
(c2) load(toda);
Batching the file toda.v

(c3) ode_toda() := block( 
/* The Toda lattice. */

/* The veryverbose mode will allow the user to see some of the
   inner working of the program. */ 
        verbose:true, 
        veryverbose : true,

/* The flag num_diff is used to control the equation solver. */ 
        num_dif:2,

/* Set the dependent and independent variables. */ 
        dep:[x,y], 
        indep:[t],
\end{verbatim}

\begin{verbatim}
/* The Toda lattice is a Hamiltonion system so we use 
   that notation. */ 
        H:(exp(2*y+2*sqrt(3)*x)+exp(2*y-2*sqrt(3)*x) 
                                +exp(-4*y))/24 - 1/8,

/* Define the differential equation. */ 
        diffeqn:[ 
                'diff(x,t,2) = -diff(H,x), 
                'diff(y,t,2) = -diff(H,y)   ],

/* Now load and execute the program doitall. */ 
        load(doitall), 
        doitall(), 
        end_toda)$

Batching done.

(c5) ode_toda();

Putting the differential equations in standard form.  
      2             2 y - 2 sqrt(3) x            2 y + 2 sqrt(3) x
     d x   sqrt(3)%e                    sqrt(3)%e
(e8) --- = -------------------------- - -------------------------- 
       2              12                           12
     dt

        2        2 y + 2 sqrt(3) x     2 y - 2 sqrt(3) x     - 4 y
       d y     %e                    %e                    %e
(e9)   --- = - ------------------- - ------------------- + ------- 
         2             12                    12               6
       dt

Creating the symmetry operators. 
The coefficients a1, a2 and 
a3 are unknown functions of the variables (x,y,t).

                                        dx
                                l  = a3 -- + a1 
                                 1      dt

                                        dy
                                l  = a3 -- + a2 
                                 2      dt

When the veryverbose flag is true the program 
prints a running commentary. Here are some 
of the comments. 
Computing the determining equations.  
Eliminating the time derivatives.  
Collecting the determining equations. 
Cleaning up a list.  
The length of the list is 2  . 
Calculating the coefficients of a polynomial. 
Preparing the equation list for printing.  
The number of equations is  18 .  
Ordering a list of length 18  .  
\end{verbatim}

\begin{verbatim}
                                    2
                                   d a1
(e13)                              ---- = 0 
                                     2
                                   dy

                                    2
                                   d a3
(e14)                              ---- = 0 
                                     2
                                   dy

                                    2
                                   d a3
(e15)                              ----- = 0 
                                   dx dy

                                    2
                                   d a3
(e16)                              ---- = 0 
                                     2
                                   dx

                                    2
                                   d a2
(e17)                              ---- = 0 
                                     2
                                   dx
\end{verbatim}

\begin{verbatim}
                                2         2
                               d a3      d a1
(e18)                        2 ----- + 2 ----- = 0 
                               dt dy     dx dy

                                 2       2
                                d a3    d a1
(e19)                         2 ----- + ---- = 0 
                                dt dx     2
                                        dx

                                 2       2
                                d a3    d a2
(e20)                         2 ----- + ---- = 0 
                                dt dy     2
                                        dy

                                2         2
                               d a3      d a2
(e21)                        2 ----- + 2 ----- = 0 
                               dt dx     dx dy
\end{verbatim}

\begin{verbatim}
               da3  2y + 2sqrt(3) x          da3  2y - 2sqrt(3) x
        sqrt(3)---%e                  sqrt(3)---%e
               dy                            dy
(e22) - --------------------------- + --------------------------- 
                     6                            6

                                                   2
                                                  d a1
                                              + 2 ----- = 0
                                                  dt dy

The remaining equations in the equation list become progressively more
complicated. Here are the functional dependencies. 

              [a1(x, y, t), a2(x, y, t), a3(x, y, t), x(t), y(t)]

What follows are comments printed (veryverbose:true) 
as the program does its work. 
Starting the solution of the equation list. 
Collecting all one term equations.  
Length of list is  15  .  
Ordering a list of length 5  .  
Solving all one term equations.  
Fixing up a one term equation.  

                      2
                     d a1
Attempting to solve  ----  = 0 . 
                       2
                     dy

The program continues to comment as it solves the first five
equations in the equation list. The program then pauses to print
the equation list.
Preparing the equation list for printing. 
The number of equations is  10 . 
Cleaning up a list. 
The length of the list is 10  . 
Ordering a list of length 10  . 
\end{verbatim}

\begin{verbatim}
                                 da8     da5
(e30)                          2 --- + 2 --- = 0
                                 dt      dx

                                da12     da10
(e31)                         2 ---- + 2 ---- = 0
                                 dy       dt

                   2 y + 2 sqrt(3) x              2 y - 2 sqrt(3) x
        sqrt(3)a8%e                    sqrt(3)a8%e
(e32) - ---------------------------- + ----------------------------
                    6                              6

                                                    da5
                                                + 2 --- = 0
                                                    dt

                           2        2
                          d a5     d a4     da10
(e33)                     ---- y + ---- + 2 ---- = 0
                            2        2       dt
                          dx       dx
\end{verbatim}

\begin{verbatim}
Again, the remaining equations in the equation list become
progressively more complicated so we do not print them.
Now the symmetry operators are beginning to contain some
information.

                          dx                  dx      dx
(e40)             l  = a8 -- y + a5 y + a10 x -- + a9 -- + a4
                   1      dt                  dt      dt


                           dy         dy      dy
(e41)            l  = a8 y -- + a10 x -- + a9 -- + a12 x + a11
                  2        dt         dt      dt

[a1(x, y, t), a2(x, y, t), a3(x, y, t), a4(x, t), a5(x, t),
  a6(x, t), a7(x, t), a8(t), a9(t), a10(t), a11(t, y), a12(t, y),
       x(t), y(t)]
\end{verbatim}

\begin{verbatim}
The next strategy for solving the equation list is to
differentiate some of the simpler equations in the list and
look for more one term equations.
Differentiating the equation list for the first time. 
Differentiating a list of length 10  . 
Collecting all one term equations. 
Length of list is  16  . 
Ordering a list of length 2  . 
Solving all one term equations. 
Fixing up a one term equation. 

                      2
                     d a5
Attempting to solve  ----  = 0 . 
                       2
                     dx
\end{verbatim}

\begin{verbatim}
The program goes on this way until it has differentiated the
equation list 3 times and solved all one term equations. We note
that the third differentiation does not produce any new solvable
equations.
Preparing the equation list for printing. 
The number of equations is  26 . 
Cleaning up a list. 
The length of the list is 26  . 
Ordering a list of length 26  . 

                                 da8
(e74)                          2 --- + 2 a14 = 0
                                 dt

                                        da10
(e75)                         2 a16 + 2 ---- = 0
                                         dt

As before the equation list goes on with the equations becoming
more and more complicated.  The symmetry operators are now taking
on a distinctly interesting form.

              dx                             dx
(e94) l  = a8 -- y + a14 x y + a13 y + a10 x --
       1      dt                             dt

                              dx        2
                         + a9 -- + a19 x + a18 x + a17
                              dt

                                                   

                dy         dy      dy        2
(e95) l  = a8 y -- + a10 x -- + a9 -- + a22 y 
       2        dt         dt      dt

                         + a16 x y + a21 y + a15 x + a20

Here are the functional dependencies.

[a1(x, y, t), a2(x, y, t), a3(x, y, t), a4(x, t), a5(x, t),
    a6(x, t), a7(x, t), a8(t), a9(t), a10(t), a11(t, y), a12(t, y),
      a13(t), a14(t), a15(t), a16(t), a17(t), a18(t), a19(t),
         a20(t), a21(t), a22(t), x(t), y(t)]

The symmetry program now gives up!

(d95)                              end_toda
\end{verbatim}

\begin{verbatim}
Looking back at the equation list we see that the first four
equations can be solved for one of the unknown functions.
We now do this interactively.

(c99) globalsolve:true$

(c102) linsolve(eqnlist[1],a14);

                                          da8
(d102)                           [a14 = - ---]
                                          dt

We proceed to solve the first 4 equations in the equation list.
Next plug the solutions back into the equation list to see if 
there are any more simple equations.

(c113) eqnlist:cleanup(eqnlist);

Cleaning up a list. 
The length of the list is 20  . 
Now lets see what we have. Note that the equation list consists
of expression that are to be set equal to zero.
\end{verbatim}

\begin{verbatim}
(c114) first(eqnlist);
                    2 y + 2 sqrt(3) x              2 y - 2 sqrt(3) x
         sqrt(3)a8%e                    sqrt(3)a8%e
(d114) - ---------------------------- + ----------------------------
                     6                              6

                                             2
                                            d a8       da13
                                        - 2 ---- x + 2 ----
                                              2         dt
                                            dt

The functions a8 and a13 depend only on the variable t. Consequently
differentiating the previous equation with respects to x or y will
kill some of the terms in the equation.

(c116) diff(d114,x);

                                                              2
              2 y + 2 sqrt(3) x        2 y - 2 sqrt(3) x     d a8
(d116) - a8 %e                  - a8 %e                  - 2 ----
                                                               2
                                                             dt

(c117) diff(%,x);

                   2 y - 2 sqrt(3) x               2 y + 2 sqrt(3) x
(d117) 2sqrt(3)a8%e                  - 2sqrt(3)a8%e
\end{verbatim}

\begin{verbatim}
The last equation impliest that:
(c119) a8:0$

We apply the same trick to the second equation in the equation
list.  This reduces the first two equation in the equation list
to one term equations so we take care of them.

(c129) alloneterm(eqnlist);
Collecting all one term equations. 
Length of list is  6  . 
Ordering a list of length 2  . 
Solving all one term equations. 
Fixing up a one term equation. 

                     da13
Attempting to solve  ----  = 0 . 
                      dt

The second one term equation is done in the same way.
We now start to manipulate the third equation in the equation list
in hopes of finding a simple equation. At first, the output is
messy so we do not display it.
(c136) expand(eqnlist[3]*exp(-2*y));
(c137) diff(%,y);
(c138) diff(%,x);
(c139) diff(%,y);

                                   2
                                  d a18   - 2 y
(d139)                          4 ----- %e
                                     2
                                   dt
\end{verbatim}

\begin{verbatim}
(c141) %*exp(2*y);
(c143) alloneterm([%]);
Collecting all one term equations. 
Length of list is  1  . 
Ordering a list of length 1  . 
Solving all one term equations. 
Fixing up a one term equation. 

                      2
                     d a18
Attempting to solve  -----  = 0 . 
                        2
                      dt
\end{verbatim}

\begin{verbatim}
This should have simplified the first equation in the equation list.

(c146) first(eqnlist);

                                   2
                                  d a9
(d146)                            ---- + 2 k4
                                    2
                                  dt

Some of our programs depend on recognizing the constants that are 
introduced so we use the function newc(k) to add the constant to
the integral of the previous equation. Note that a9 depends only
on t so that it is not necessary to add an arbitrary function of
x and y.

(c148) integrate(%,t)+newc(k);


                                        da9
(d148)                         2 k4 t + --- + k5
                                        dt

(c149) integrate(%,t)+newc(k);

                                2
(d149)                      k4 t  + k5 t + a9 + k6

(c150) linsolve(%,a9);

                                      2
(d150)                    [a9 = - k4 t  - k5 t - k6]

We proceed in a fashion similar to the last few computations until
we have solved for all of the "a" functions. Now all that is left to
do is to solve for the "k" constants.  The equation list consists of
two complicated equations so we compute the first few terms of the
power series expansion of the equations.

(c189) subst(0,x,eqnlist);
(c190) subst(0,y,%);

                         2
(d190)             [k10 t  + k9 t + k8 + 2 k10, k12 t + k11]

(c191) k10:0$k9:0$k8:0$k12:0$k11:0$

(c201) diff(subst(0,x,eqnlist),y);
                        
                  - 6 y 
(d201) [k1 - k1 %e     ,
                        
                                                          - 6 y
                       - 6 y            - 6 y   11 k4 t %e
          - 4 k4 t y %e      + 2 k7 y %e      + ---------------
                                                       3

                               - 6 y
                          k7 %e               - 6 y   k4 t   k7
                        + ---------- + 2 k5 %e      + ---- - --]
                              6                        3     6
(c202) k1:0$
\end{verbatim}

\begin{verbatim}
We now just list the input lines that are used
to finish the computation.
(c203) subst(0,y,last(d201));
(c204) k4:0$
(c205) k5:0$
(c206) eqnlist:ev(eqnlist);
(c207) subst(0,x,diff(eqnlist,x));
(c208) k2:0$
(c209) k7:0$
(c210) eqnlist:ev(eqnlist);
(c211) k3:0$

The equation list is solved!
Now print out the results of the computation.

(c213) results();
Here are the only symmetries!

                                           dx
(e213)                           l  = - k6 --
                                  1        dt


                                           dy
(e214)                           l  = - k6 --
                                  2        dt

(d214)                               done
\end{verbatim}
\vspace{.1in}
When the symmetry operators are written as a transformation group on
$ R^3 $ labeled with the variables $ (x, y, t) $, they become
constant multiples of the single infinitesimal transformation
$$ \frac{\partial}{\partial t } \;. $$
As was mentioned at the beginning of this section, time translations are
the only point symmetries of the Toda lattice.
Thus we see that it is important to generalize the notion of symmetry.

\newpage \clearpage
\setcounter{equation}{0}
\chapter{PARTIAL DIFFERENTIAL EQUATIONS}

\section{Introduction}

The material in this section is the basis for our programs that compute
point symmetries.  As in Chapter 2, it is perhaps best to give this 
section a light reading and then turn to the next section where these
ideas are applied to the heat equation. It is not necessary to understand
this section to understand the next section. The third section uses our 
computer code to find the symmetries of Burger's equation.
The method
for partial differential equations is similar to the method for ordinary
differential equations. Let  $ \vec{u} = (u_1, \ldots , u_m ) $ 
be the dependent variables while  $ t $  and  $ \vec{x} = (x_1, \ldots ,
x_n ) $  are the independent variables.  The problems of interest are
nonlinear partial differential equations that can be written in the form
\begin{equation} \label{Nonlinear PDE}
\frac{d^{ p_i } u_i }{ dt^{ p_i } } = H_i ( \vec{u} ) \;,\;
1 \leq i \leq m \;,
\end{equation}
where  $ p_i $, are positive integers, $ p_i > 0 $, and  $ H_i ( \vec{u} ) $  
is operator notation for a function of  $ \vec{u} $  and derivatives of  
$ \vec{u} $  that are of lower
order than the derivatives on the left hand side of the equation
\eqref{Nonlinear PDE} .  Thus
\begin{equation}
H_i ( \vec{u} ) =
h_i (t, \vec{x} , \vec{u} , \frac{\partial u_1 }{\partial t } \;,
\frac{\partial u_1 }{\partial x_1 } \;,\; \ldots )
\end{equation}
where  $ h_i $  is a function of a finite number of variables.
More precisely, if we introduce the new variables
\begin{equation}
u_i^{ (j, \vec{k} ) } \;,\;
\vec{k} = ( k_1, \ldots , k_n ) \;,\;
1 \leq j < p_j \;,\; 1 \leq i \leq m \;,
\end{equation}
then $ h_i  $ should be an analytic function of its arguments,
\begin{equation}
h_i = h_i (t, x , \vec{u} , \ldots , u_i^{ (j, \vec{k} ) } , \ldots )
\;.
\end{equation}
There are no restrictions on the $ \vec{k} $ indicies.  Here we think of 
the superscripted  $ u $ variables as short hand for derivatives,
\begin{equation}
u_i^{ (j, \vec{k} ) } =
\frac{\partial^j }{\partial t^j }
\frac{\partial^{ \vec{k} } }{\partial \vec{x}^{ \vec{k} } }
u_i 
\end{equation}
where
\begin{equation}
\frac{\partial^{ \vec{k} } }{\partial \vec{x}^{ \vec{k} } } =
\frac{\partial^{ k_1 } }{\partial x_1^{ k_1 } } \ldots 
\frac{\partial^{ k_n } }{\partial x_n^{ k_n } } \;.
\end{equation}

To apply the results of Section 6 of Chapter I, let
\begin{equation}
\vec{f} (t, \vec{x} ) = (f_1 (t, \vec{x} ), \ldots , f_m (t, \vec{x} ))
\end{equation}
(here we think of  $ \vec{u} = \vec{f} (t, \vec{x} ) $  as solutions) and then
introduce the operator
\begin{equation}
\vec{F} ( \vec{f} ) = (F_1 ( \vec{f} ), \ldots , F_m ( \vec{f} ))
\end{equation}
where
\begin{equation}
F_j ( \vec{f} ) =
\frac{\partial^{ p_j } }{\partial t^{ p_j }} 
f_j - H_i ( \vec{f} ) \;.
\end{equation}
The derivative of  $ \vec{F} $  in the direction  $ \vec{g} $  is given by
\begin{equation}
(D_{ \vec{f} } \vec{F} )( \vec{g} ) = ((D_{ \vec{f} } F_1 )( \vec{g} ),
\ldots , (D_{ \vec{f} } F_m )( \vec{g} ))
\end{equation}
where
\begin{align}
(D_{ \vec{f} } F_j )( \vec{g} ) & =
\frac{d}{ d \epsilon } \left\{
\frac{ d^{ p_j } }{ dt^{ p_j } }
(f_j + \epsilon g )
- H_i ( f + \epsilon g ) \right\}_{\epsilon =0 } \\
\nonumber & = \frac{ d^{ p_j } }{ dt^{ p_j } } g_j
- ( D_{ \vec{f} } H_j )( \vec{g} ) \;.
\end{align}
Next,
\begin{align}
\nonumber (D_{ \vec{f} } H_i )( \vec{g} ) & =
\frac{d}{ d \epsilon } h_j (t, \vec{x} ,f_1 + \epsilon g_1, \ldots ,
\frac{\partial^j }{\partial t^j } 
\frac{\partial^{ \vec{k} } }{\partial \vec{x}^{ \vec{k} } }
(f_i + \epsilon g_i ), \ldots ) \mid_{\epsilon =0 } \\
& = \frac{\partial h_j }{\partial u_1 } g_1 + \ldots + 
\frac{\partial h_j }{\partial u_i^{ (j, \vec{k} ) } }
\frac{\partial^j }{\partial t^j } 
\frac{\partial^{ \vec{k} } }{\partial \vec{x}^{ \vec{k} } }
g_i + \ldots \;.
\end{align}

The condition of infinitesimal invariance is
\begin{equation}
\vec{F} ( \vec{f} ) = 0  = > (D_{ \vec{f} } \vec{F} ) ( \vec{S} ( \vec{f} ))
= 0 \;.
\end{equation}
The infinitesimal symmetries have the form
\begin{equation}
S = T \frac{\partial}{\partial t } +
\sum_{i=1}^{n} X_i \frac{\partial}{\partial x_i } 
+ \sum_{j=1}^{m} U_j \frac{\partial}{\partial u_j }
\end{equation}
where  $ T $, $ X_i $  and  $ U_j $  depend on  $ (t, x , \vec{u} ) $.
The infinitesimal action on surfaces is given by
\begin{equation}
\vec{S} ( \vec{f} ) = S_1 ( \vec{f} ), \ldots , S_m ( \vec{f} ) )
\end{equation}
where
\begin{equation} 
S_i ( \vec{f} ) = -T \frac{\partial f_i }{\partial t } -
\sum_{k=1}^{n} X_k \frac{\partial f_i }{\partial x_k }
+ U_i \;.
\end{equation}

The way this condition is used is to replace all derivative of the form
\begin{equation}
\frac{\partial^{ p_i +k } f_i }{ dt^{ p_i + k } } \;,\; k \geq 0 \;,
\end{equation}
that occur in  $ (D_{ \vec{f} } \vec{F} )( \vec{L} ( \vec{f} )) $
by the right hand side
(or an appropriate derivative thereof) of the differential equation
\eqref{Nonlinear PDE} .  The
resulting expression is called  $ \vec{E} $.  The expression  $ \vec{E} $  still
contains solutions of the given system
which need to be removed.

Now suppose that an arbitrary but finite set of values
\begin{equation}
u_i^{ (j, \vec{k} ) } \;,\; 1 \leq i \leq m \;,\; 0 \leq j \leq p_i \;,
\end{equation}
are given and that it is possible to find a solution  $ \vec{f} $  of the
differential equation \eqref{Nonlinear PDE}  that satisfies
\begin{equation}
\frac{\partial^j }{\partial t^j }
\frac{\partial^{ \vec{k} } }{\partial \vec{x}^{ \vec{k} } }
f_i = u_i^{ (j, \vec{k} ) } \mid_{t=0} \;.
\end{equation}
Note that this condition is considerably weaker than requiring that
\eqref{Nonlinear PDE}  have a well posed initial value problem. The
Cauchy-Kowalewski theorem
[6] can frequently be used to show that the system of partial
differential equations satisfy a considerable stronger condition.
Under this condition the equation
\begin{equation}
\vec{E} = 0
\end{equation}
must hold with  $ \vec{f} $  and all of its derivatives replaced by the right
hand side of \eqref{Nonlinear PDE} .  The resulting expression must be zero for all values of 
the variables
\begin{equation}
(t, \vec{x} , \vec{u}, \ldots , u_i^{ (j, \vec{k} ) } , \ldots ) \;.
\end{equation}
Here it is important to note the coefficients
of the infinitesimal symmetry do not depend on the variables
$ u_i^{ (j, \vec{k} ) }  $.
Consequently the power series expansion of the expression $ E $ in the 
$ u_i^{ (j, \vec{k} ) }  $
variables will not involve derivatives of the coefficients of the
symmetry operator and consequently each nontrivial coefficient in
the expansion will produce an equation for the coefficients of the
symmetry operator.  If, as is frequently the case, the expression  
$ \vec{E} $  is a polynomial in some of the variables 
$ u_i^{ (j, \vec{k} ) }  $ then, the coefficients of this polynomial must 
be zero.  When we generalize the notion of a symmetry this will no longer
be true and this fact will produce one of our major difficulties.

We now turn to some examples.

\section{The Heat Equation}

The heat equation is fairly typical of partial differential equations
that we would call very symmetric. Here we are discussing the heat 
equation in one space variable so we have two independent variables
$ ( x, t) $ and one dependent variable $ u $.  The heat equation is
then written
\begin{equation}
\frac{\partial u }{\partial t } = 
\frac{\partial^2 u }{\partial x^2 } \;.
\end{equation}
It is easy to see that the heat equation has at least 5 symmetries.
There are three translations, one each in $ x $, $ t $ and $ u $.
There are two scaling symmetries, any scaling in $ u $ and
scaling $ t $ with the square of a scaling in $ x $. Because
the heat equation is linear there is also and infinite symmetry group,
the addition of any solution to all solutions sends the solution space
into the solution space.
We will find [4] that the heat equation has two additional symmetries
which are usually referred to as a hidden symmetries.

If $ u = f(x,t) $ is a surface, then the operator form of the heat equation is
\begin{equation}
F(f) = f_t - f_{xx}
\end{equation}
where we have used subscript notation
\begin{equation}
f_t = \frac{\partial f }{\partial t } \;,\;
f_{xx} = \frac{\partial^2 f }{\partial x^2 } \;,
\end{equation}
for partial derivatives.
Because there is one dependent and two independent variables,
the infinitesimal symmetries act on surfaces and have the form
\begin{equation}
S(f) = A(x,t,f) f_t + B(x,t,f) f_x + C(x,t,f)
\end{equation}
where $ A(x,t,u) $, $ B(x,t,u) $ and $ C(x,t,u) $ are functions
that are to be determined.

Because the heat equation is linear, it is its own directional
derivative. Thus 
\begin{equation}
F(f + \epsilon g) = (f + \epsilon g)_t - (f + \epsilon g )_{xx} =
f_t - f_{xx}  + \epsilon ( g_t - g_{xx} )
\end{equation}
and consequently
\begin{equation}
(D_f F )(g) = \frac{d}{ d \epsilon } F ( f + \epsilon g) \mid_{\epsilon = 0 } 
= g_t - g_{xx} = F(g) \;.
\end{equation}
The invariance condition is then (Chapter 1 Section 6)
\begin{equation}
F(f) = 0 \Rightarrow (D_f F) ( S (f) ) = 0 \;.
\end{equation}
Thus, to find the conditions on $ A $, $ B $ and $ C $, we substitute 
$ f_{xx} $ for $ f_t $ in $ F (S(f)) = 0 $ to obtain
(here $ A $, $ B $ and $ C $ have arguments $ ( x , t, f ) $ )
\begin{align} \label{Heat Condition}
& A_u f_x f_{xxx} - 2 A_x f_{xxx} - A_{uu} f_x^2 f_{xx} - 2B_u f_x f_{xx} \\
& \nonumber - 2A_{ux} f_x f_{xx} - 2B_x f_{xx} - A_{xx} f_{xx} + A_t f_{xx}
\\
& \nonumber - B_{uu} f_x^3 - 2B_{ux} f_x^2 - C_{uu} f_x^2 - B_{xx} f_x \\
& \nonumber + B_t f_x - 2C_{ux} f_x - C_{xx} + C_t = 0 \;.
\end{align}
Because the initial value problem for the heat equation is well posed,
we may replace $ f $, and the derivatives of $ f $
in the previous expression by variables, say 
\begin{align}
\nonumber
& f \rightarrow u ,&  & f_x \rightarrow u^1 \;, \\
& f_{xx} \rightarrow u^{11} ,&  & f_{xxx} \rightarrow u^{111} \;.
\end{align}
Now the expression
\eqref{Heat Condition}
is and identity in the variables $ t $, $ x $, $ u $,
$ u^1 $, $ u^{11} $ and $ u^{111} $.
Because this expression is a polynomial in $ u^1 $, $ u^{11} $
and $ u^{111} $, the coefficients of this polynomial must be zero,
which yields the following set of equations:
\begin{align} \label{Heat System}
\nonumber
A_u = 0 \;,\; A_x = 0 \;,\; A_{uu} = 0 \;,\; B_{uu} & = 0 \;, \\
C_t - C_{xx} = 0 \;,\; B_u + A_{ux}  = 0 \;,\;
2B_{ux} - C_{uu} & = 0 \;, \\
\nonumber
2B_x - A_{xx} + A_t = 0 \;,\; 2C_{ux} - B_t + B_{xx} & = 0 \;.
\end{align}
Note that these equations are redundant. However, there are three
unknown and certainly more than three equations so they are over
determined. Thus we expect the solution space to be finite dimensional
and find that this is nearly true. See [56, 18] for some
theorems on this point.

We now solve these equations. We note that our computer programs
try to mimic, with some success, this method of solution.
The first two equations in \eqref{Heat System}  give
\begin{equation}
A = A (t) \;.
\end{equation}
We should introduce a new function here, as our computer codes do, but
this only makes a mess for humans to read. Plugging this back into
the equations yields:
\begin{align}
C_t - C_{xx} = 0 \;,\; B_u & = 0 \;, \\
\nonumber 2B_{ux} - C_{uu} = 0 \;,\; 2B_x + A_t & = 0 \;, \\
\nonumber 2C_{ux} - B_t + B_{xx} & = 0 \;.
\end{align}
As before, the second equation gives
\begin{equation}
B = B (x,t) \;.
\end{equation}
If the third equation is differentiated with respects to $ x $, then
\begin{equation}
B_{xx} = 0 \;.
\end{equation}
The equation list can now be written:
\begin{align} \label{Heat Simple}
\nonumber
B_{xx} = 0 \;,\; C_{uu} & = 0 \;, \\
A_t + 2B_x = 0 \;,\; B_t - 2C_{ux} & = 0 \;, \\
\nonumber
C_t - C_{xx} & = 0 \;.
\end{align}
Now differentiate the fourth equation in \eqref{Heat Simple}
twice with respects to $ x $, then differentiate the last equation in
\eqref{Heat Simple} twice, once with respects to $ u $ and $ x $ and once with respects to
$ u $ and $ t $, then differentiate the fourth equation \eqref{Heat Simple} with respects
$ t $ and finally differentiate the third equation \eqref{Heat Simple} twice with respects 
$ t $ twice.  A little algebra the gives the following equation list:
\begin{align} \label{Equation List}
A_{ttt} & = 0 \;, \\
\nonumber B_{tt} = 0 \;,\; B_{xx} & = 0 \;, \\
\nonumber C_{uxxx} = 0 \;,\; C_{uxt} = 0 \;,\; C_{utt} = 0 \;,\; 
C_{uu} & = 0 \;, \\
\nonumber A_t + B_x = 0 \;,\; C_t - C_{ux} = 0 \;,\; C_t - C_{xx} & = 0 \;.
\end{align}
Equations like the first seven equations in \eqref{Equation List}
occur frequently in symmetry calculation.
This gives:
\begin{align}
\nonumber
A & = c_1 t^2 + c_2 t + c_3 \;, \\
\nonumber
B & = c_4 xt + c_5 x + c_6 t + c_7 \;, \\
C & = R(x,t) u + h(x,t) \;, \\
\nonumber
R & = c_8 t + c_9 x^2 + c_{10} x + c_{11} \;, \\
\nonumber
& h_t - h_{xx}  = 0 \;.
\end{align}
If these results are plugged back into the equation list, then
some constraints on the constants are obtained. If these constraints
are solved and then the constants are relabeled, then the symmetry
operator
\begin{align} \label{Symmetries Heat}
\nonumber
S(f) & = ( 4 k_1 t^2 + 2 k_4 t + k_5 ) f_t \\
& + ( 4 k_1 tx + k_4 x + 2 k_2 t + k_3 ) f_x \\
\nonumber
& + ( k_1 x^2 + k_2 x + k_1 t + k_6 ) f + h
\end{align}
is obtained. Here $ k_1 $ through $ k_6 $ are arbitrary constants
while $ h $ is any solution of the heat equation.
The symmetry corresponding to $ h $ is obvious because the heat 
equation is linear.  The remaining symmetries can be translated to
the vector field notation (Chapter 1 Section 4) where they from 
a linear space. Corresponding to each constant in
 \eqref{Symmetries Heat}
is a basis element. Here is a list of the basis elements.
\begin{align}
\nonumber
& k_5 \rightarrow \frac{\partial}{\partial t } \;,\;
\text{time translation} \;, \\
\nonumber
& k_3 \rightarrow \frac{\partial}{\partial x } \;,\;
\text{\rm space translation} \;, \\
\nonumber
h =\; & k_7 \rightarrow \frac{\partial}{\partial u } \;,\;
\text{\rm add constant to solution} \;, \\
& k_6 \rightarrow u \frac{\partial}{\partial u } \;,\; 
\text{\rm rescale the solution} \;, \\
\nonumber
& k_4 \rightarrow 2t \frac{\partial}{\partial t } +
x \frac{\partial}{\partial x } \;,\; 
\text{\rm scale in time and space} \, \\
\nonumber
& k_2 \rightarrow 2t \frac{\partial}{\partial x }
- xu \frac{\partial}{\partial u } \;,\;
\text{\rm hidden symmetry} \; \\
\nonumber
& k_1 \rightarrow 4t^2 \frac{\partial}{\partial t }
+ tx \frac{\partial}{\partial x } 
- x^2 \frac{\partial}{\partial u } \;,\; 
\text{\rm hidden symmetry} \;,
\end{align}

In the file examples.v there are three programs, heat\_1\_1,
heat\_1\_2 and heat\_1\_3. The first integer in the name gives
the number of spatial dimensions in the example and the second
integer is a version number. The first two programs are essentially
the same, heat\_1\_2 prints fewer intermediate results than heat\_1\_1.
The printing of the intermediate results requires a 
substantial amount of time because, before the results are 
printed they are sorted and cleaned up for the convenience of the reader.
The program heat\_1\_3 uses the program symmetry rather than doitall
and consequently runs substantially faster than the other versions.
However, the equation list is not completely solved so the user 
would normally finish this interactively.
To help the user
understand the interactive use of our programs we have included
in heat\_1\_3 the commands that we used when we solved the equations
interactively.  Running this program then is the same as watching
the author go through an interactive session.
The program heat\_1\_2 uses about 23 cpu minutes on a VAX11/780 while
the program heat\_1\_3 uses about 11 cpu minutes.
Here are listings of heat\_1\_1 and heat\_1\_3. \\[.2in]
\newpage
\centerline{Program heat\_1\_1}

\begin{verbatim}
heat_1_1() := block(
/* The one dimensional heat equation. Use the very verbose mode so that some
   of the inner workings of the code can be seen. This slows down the 
   program substantially. */
	verbose : true,
	veryverbose : true,

/* The parameter num_dif is the maximum number of terms in an equation that is
   to be differentiated by listsolver. */
	num_dif:4,

/* Set the dependent and independent variables. */
	dep : [u],
	indep : [t,x],

/* Define the differential equation. */
	diffeqn : ['diff(u,t)-'diff(u,x,2) = 0],

/* Now load a execute the program. */
	load(doitall),
	doitall(),
	end_heat_1_1)$

heat_1_3() := block(
/* The one dimensional heat equation. This is the same as heat_1_1 except that
   some special tricks are used to make the code run faster. */
	verbose : true,
	veryverbose : false,
	dep : [u],
	indep : [t,x],
	diffeqn : ['diff(u,t)-'diff(u,x,2) = 0],

/* Not that we call symmetry and not doitall. */
	load(symmetry),
	symmetry(),

/* The following commands illustrate what a user might do in solving the
   determining equations for the one space dimension heat equation. */
	oneterm(diff(first(eqnlist),x)),
	results(),
	oneterm(diff(eqnlist[2],x,2)),
	results(),
	eqnlist:cons(diff(last(eqnlist),u),eqnlist),
	oneterm(diff(first(eqnlist),x,2)),
	results(),
	oneterm(diff(eqnlist[2],x)),
	results(),
	oneterm(diff(eqnlist[2],t)),
	results(),
	oneterm(diff(eqnlist[3],x,1,t,1)),
	results(),
	oneterm(diff(eqnlist[2],t,2)),
	results(),
	oneterm(diff(eqnlist[3],t)),
	results(),
	allnondiff(eqnlist),
	results(),
	end_heat_1_3)$
\end{verbatim}

\section{Burger's Equation}

Burger's equation is a nonlinear equation similar to the heat equation,
\begin{equation} \label{Burger}
\frac{\partial u }{\partial t } = \frac {\partial^2 u }{\partial x^2 } 
+ u \frac{\partial u }{\partial x } \;,
\end{equation}
where $ u = f(x,t) $.
There has been considerable interest in the symmetries of Burger's
equation including that of the author [102]. Because \eqref{Burger}
is nonlinear the calculation of the directional derivative of
its operator form,
\begin{equation}
F(f) = \frac{df}{dt} - \frac{ d^2 f }{ dx^2 } - f \frac{df}{dx} \;,
\end{equation}
is interesting. Thus,
\begin{align}
& \frac{d}{ d \epsilon } F ( f + \epsilon g ) \mid_{\epsilon = 0 }  =  \\
\nonumber 
& \frac{\partial}{\partial \epsilon } \left[
\frac{\partial ( f + \epsilon g ) }{\partial t } -
\frac{\partial^2 ( f + \epsilon g ) }{\partial x^2 } -
( f + \epsilon g ) \frac{\partial ( f + \epsilon g ) }{\partial x }
\right]_{\epsilon =0} = \\
\nonumber
& \frac{\partial g }{\partial t } -
\frac{\partial^2 g }{\partial x^2 } - g 
\frac{\partial f }{\partial x } - f  
\frac{\partial g }{\partial x } \;.
\end{align}

The following is a substantially edited listing of the output of
a VAXIMA run that computes the symmetries of the Burger's equation.  \\[.2in]
\centerline{VAXIMA Output}

\begin{verbatim}
(c3) load(burgers);
Batching the file burgers.v

(c4) burgers() := block(
/* The Burger's equation. */
        verbose : true,
        veryverbose : true,

/* The parameter num_dif is the maximum number of terms in an
   equation that is to be differentiated by listsolver. */
        num_dif:7,

/* Set the dependent and independent variables. */
        dep : [u],
        indep : [t,x],

/* Define the differential equation. */
        diffeqn : ['diff(u,t)-'diff(u,x,2)-u*'diff(u,x) = 0],

/* Now load a execute the program. */
        load(doitall),
        doitall(),
        end_burgers)$

\end{verbatim}

\begin{verbatim}
Batching done.
(d5)                               burgers.v

(c6) burgers();

Putting the differential equations in standard form. 
                                      2
                                du   d u     du
(e9)                            -- = --- + u --
                                dt     2     dx
                                     dx

Creating the symmetry operators. 

                                    du      du
                            l  = a3 -- + a2 -- + a1
                             1      dx      dt

Computing the determining equations. 
Eliminating the time derivatives. 
Collecting the determining equations. 
Calculating the coefficients of a polynomial. 
Preparing the equation list for printing. 
The number of equations is  9 . 
\end{verbatim}

\begin{verbatim}
                                    da2
(e12)                               --- = 0
                                    du

                                    2
                                   d a2
(e13)                              ---- = 0
                                     2
                                   du

                                    da2
(e14)                               --- = 0
                                    dx

                             2        2
                            d a2     d a3     da2
(e15)                     - ---- u - ---- - 2 --- = 0
                              2        2      du
                            du       du
\end{verbatim}

\begin{verbatim}
                                               2
                            da2       da3     d a2
(e16)                   - 2 --- u - 2 --- - 2 ----- = 0
                            du        du      du dx

                             da1     d a1   da1
(e17)                      - --- u - ---- + --- = 0
                             dx        2    dt
                                     dx

                       d a2        d a3      da2   d a1
(e18)              - 2 ----- u - 2 ----- - 2 --- - ---- = 0
                       du dx       du dx     dx      2

                                           2
                          da2       da3   d a2   da2
(e19)                 - 3 --- u - 2 --- - ---- + --- = 0
                          dx        dx      2    dt

        da2  2   da3     d a2     da2     d a3   da3     d a1
(e20) - --- u  - --- u - ---- u + --- u - ---- + --- - 2 -----
        dx       dx        2      dt        2    dt      du dx
                         dx               dx

                                                     - a1 = 0

                                    du      du
(e21)                       l  = a3 -- + a2 -- + a1
                             1      dx      dt

           [a1(u, t, x), a2(u, t, x), a3(u, t, x), u(t, x)]
\end{verbatim}

\begin{verbatim}
Starting the solution of the equation list. 
Collecting all one term equations. 
Length of list is  9  . 
Ordering a list of length 3  . 
Solving all one term equations. 
Fixing up a one term equation. 

                     da2
Attempting to solve  ---  = 0 . 
                     du

Fixing up a one term equation. 

                     da4
Attempting to solve  ---  = 0 . 
                     dx

The programs continues in this fashion and solves two more one
term equations.
Preparing the equation list for printing. 
The number of equations is  3 . 
\end{verbatim}

\begin{verbatim}
                                da5     da6
(e22)                           --- - 2 --- = 0
                                dt      dx

                       2                        2
             da8  2   d a8     da8     da7     d a7   da7
(e23)      - --- u  - ---- u + --- u - --- u - ---- + --- = 0
             dx         2      dt      dx        2    dt
                      dx                       dx

                                                 2
                   da6     da5       da8        d a6   da6
(e24)     - a8 u - --- u + --- u - 2 --- - a7 - ---- + --- = 0
                   dx      dt        dx           2    dt
                                                dx

                                du      du
(e25)                   l  = a6 -- + a5 -- + a8 u + a7
                         1      dx      dt

[a1(u, t, x), a2(u, t, x), a3(u, t, x), a4(x, t), a5(t),
                         a6(x, t), a7(x, t), a8(x, t), u(t, x)]
\end{verbatim}

\begin{verbatim}
Differentiating the equation list for the first time. 
Differentiating a list of length 3  . 
Collecting all one term equations. 
Length of list is  12  . 
Ordering a list of length 1  . 
Solving all one term equations. 
Fixing up a one term equation. 

                      2
                     d a6
Attempting to solve  ----  = 0 . 
                       2
                     dx

The program now differentiates the equation list a second
time and then finds enough one term equations to completely
solve for all of the "a" functions. In this process constants
are introduced and some of them are redundant so some of 
the constants are eliminated.

Collecting the non-differential equations. 
Collecting coefficients 
Calculating the coefficients of a polynomial. 
Collecting the constants to be solved for. 
Solving for the constants. 
Preparing the equation list for printing. 
The number of equations is  0 . 
\end{verbatim}

\begin{verbatim}
                 du        du          
(e37) l  = k10 t -- x + k2 -- x + k10 x
       1         dx        dx          

                           du       du        2 du
                    + k6 t -- + k11 -- + k10 t  --
                           dx       dx          dt

                             du      du
                    + 2 k2 t -- + k8 -- + k10 t u + k2 u + k6
                             dt      dt

(d37)                             end_burgers
The program has successfully found all of the symmetries of
Burger's equation. Let us now use VAXIMA interactively to
find a basis of the symmetry operators. d40 is the right
hand side of e37.

(c43) coeff(d40,k10);

                             du          2 du
(d43)                      t -- x + x + t  -- + t u
                             dx            dt

(c44) coeff(d40,k2);

                               du         du
(d44)                          -- x + 2 t -- + u
                               dx         dt

(c45) coeff(d40,k6);

                                     du
(d45)                              t -- + 1
                                     dx

(c46) coeff(d40,k11);

                                      du
(d46)                                 --
                                      dx

(c47) coeff(d40,k8);

                                      du
(d47)                                 --
                                      dt
\end{verbatim}

This, we hope, illustrates the convenience of being able to 
use VAXIMA interactively to make formulas more readable.

\newpage \clearpage
\setcounter{equation}{0}
\chapter{GENERALIZED SYMMETRIES}

\section{Introduction}

This chapter was not completed.

In this chapter
we will extend the notion of symmetry of a system of differential
equations from geometric transformations to transformations that depend on the
derivatives of the solution of the system.  We call the most general form of
these transformations, {\it jet} transformations.  The term {\it jet} is 
borrowed from
differential geometry.  Many authors call this transformation Backlund or
Lie-Backlund transformations.
We will indicate in what sense contact transformations are a special case of
jet transformations.

Because the notation becomes complicated in this theory, Section 2 begins
by studying the situation in the plane.  One thing to note is that there is
at least an implicit choice of dependent and independent variables in this
theory.  Thus in three dimensions there are two distinct kinds of 
jet transformation,
depending on whether there are two dependent and one independent or one
dependent and two independent variables.
To study jet transformations we must introduce infinite set of variables.
In this context, it is still no completely understood how to do this in
a rigorous fashion.  The assumption that the various functions introduced
depend on only a finite subset of the variables makes many of the objects
under consideration well defined. Because this is just the assumption
that makes computer programs practical, we will assume that all of the
functions introduced are analytic in a finite number of variables.
In Section 3 we will study jet transformations in $ m $ dependent and
$ n $ independent variables. In Section 4 we generalize the notion of symmetry.

\section{Two Dimensional Jet Transformations}

This section is devoted to motivating several definitions that are
important to the theory of jet transformations.
When the transformations are allowed to depend on derivatives,
there are several points where confusion arises. We will
clarify these points here so that we can do the general case more
easily.  One point is that the choice of dependent and independent
variables is important. Once this choice is made, then we are always 
implicitly assuming the dependent variables depend on the independent
variables and the derivatives under consideration are the derivatives of
the dependent variables with respects to the independent variables.
However, it is frequently convenient to think of the derivatives as
variables in their own right. We believe that this will become apparent
as we go through this section.

In the case of transformations in the plane, we assume  $ y $ 
is dependent and  $ x $  is independent,
and that these variables are being transformed to $ \xi $ and $ \eta $,
where $ \xi $ is independent and $ \eta $ is dependent.
In this case, the transformations that depend on one derivative have the form
\begin{equation}
\xi = f(x,y, \frac{dy}{dx} ) \;,\; \eta = g(x,y, \frac{dy}{dx} )\;,
\end{equation}
where
\begin{equation}
f = f(x,y,v) \;,\; g = g(x,y,v) \;,
\end{equation}
are given functions of three variables.
Here we are beginning to distinguish between the derivatives and variables
that will have derivatives substituted for them.
The transformation of the derivatives can be computed using the chain rule,
\begin{equation} \label{No y''}
\frac{d \eta}{d \xi} = 
\frac{g_x dx + g_y dy + g_v dy^\prime}{f_x dx + f_y dy + f_v dy^\prime}
= \frac{g_x + g_y y^\prime + g_v y^{\prime \prime}}{f_x + f_y y^\prime +
f_v y^{\prime \prime}} \;.
\end{equation}

If we wish to restrict ourselves to first order contact transformations,
then we will require that $ d \eta / d \xi $ not to depend on 
$ y^{\prime \prime} $.  This can be done by requiring that
\begin{equation}
\frac{\partial (d \eta /d \xi)}{\partial y^{\prime \prime} } = 0 \;,
\end{equation}
which means that we must have
\begin{equation}
g_v f_x = f_v g_x \qquad {\rm and} \qquad g_v f_y = g_y f_v \;.
\end{equation}
Now if $ d \eta /d \xi $  does not depend on  $ y^{\prime \prime} $  then 
choosing $ y^{\prime \prime} = 0 $ in \eqref{No y''} gives
\begin{equation}
\frac{d \eta}{d \xi} = \frac{g_x + g_y y^\prime}{f_x + g_y y^\prime} \;.
\end{equation}
The definition of an $ n $-th order contact transformation is the obvious 
generalization of the above ideas.

To study jet transformations we introduce the infinite set of variables
$ x $, $ y $ and $ \vec{v} $ where 
\begin{equation}
\vec{v} = (v_1 , v_2, \ldots ) \;.
\end{equation}
As before $ x $ is independent, $ y $ is dependent and $ v_i $
is thought of as standing for the $ i $-th derivative of $ y $ with 
respects to $ x $.
These variables will be transformed to $ \xi $, $ \eta $ and $ \vec{\nu} $
where $ \xi $ is independent, $ \eta $ is dependent and $ \nu_i $
is thought of as standing for the $ i $-th derivative of $ \eta $ with 
respects to $ \xi $.  Let 
\begin{equation}
f(x,y, \vec{v} ) \qquad {\rm and} \qquad g(x,y, \vec{v} ) 
\end{equation}
be functions that depend on a finite number of variables and are
analytic in these variables.  A {\it jet} transformation will have the form
\begin{equation} \label{dxi deta jet}
\xi = f(x,y, \frac{dy}{dx}, \frac{d^2 y}{dx^2}, \ldots ) \;,\;
\eta = g(x,y, \frac{dy}{dx}, \frac{d^2 y}{dx^2}, \ldots ) \;.
\end{equation}
Once the functions $ f $ and $ g $ are given, then the chain rule
determines the transformation of the derivatives.
The transformed derivatives will have the form
\begin{equation}
\frac{d^i \eta}{d \xi^i} = g_i (x,y, \frac{dy}{dx},  
\frac{d^2 y}{dx^2}, \ldots ) \;,\; 1 \leq i < \infty \;.
\end{equation}
where, again, the $ g_i $ depend on only a finite subset of variables
$ \xi $, $ \eta $ and $ \vec{\nu} $.
If we replace the derivatives by the variables that stand for them,
then the jet transformation will have the form
\begin{equation}
\nu_i = g_i (x,y, \vec{v} ) \;,\; 1 \leq i < \infty \;.
\end{equation}
Now repeated applications of the chain rule to \eqref{dxi deta jet}
and then replacing derivatives by the variables that stand for them
gives the following form for the $ g $ functions (here $ g_0 = g $),
\begin{equation}
g_{i+1} (x,y, \vec{v} ) =
\frac{ \frac{\partial g_i}{\partial x} + \frac{\partial g_i}{\partial y} 
v_1 + \sum_{k=1}^\infty \frac{\partial g_i}{\partial v_k} v_{k+1}}{ 
\frac{\partial f}{\partial x} + \frac{\partial f}{\partial y} 
v_1 + \sum_{k=1}^\infty \frac{\partial f}{\partial v_k} v_{k+1}}
\;,\; i \geq 0 \;.
\end{equation}

The previous discussion was meant to motivate the following definition
of a group of jet transformations. A group of jet transforms is obtained by 
simply letting the previous formulas depend on  $ \epsilon $
and then requiring the group axioms to hold.
A group of jet transformations are transformations that act on the
infinite set of variables
\begin{equation}
x,y,^\prime , \vec{v} 
\end{equation}
and the transformations have the form
\begin{align}
\nonumber \xi & = f(\epsilon ,x,y, \vec{v} ) \\
\eta & = g(\epsilon ,x,y, \vec{v} ) \\
\nonumber \nu_i & = g_i (\epsilon ,x,y, \vec{v} ) \;,\; 1 \leq i \leq \infty \;.
\end{align}
The group axioms in $ \epsilon $ must hold and if we set
\begin{equation}
f (\epsilon) = f (\epsilon , x,y, \vec{v} ) \;,\;
g_i (\epsilon) = g_i (\epsilon , x, y, \vec{v} )  \;,
\end{equation}
then the following contact conditions must be satisfied,
\begin{equation} \label{Contact Condition}
g (\epsilon)_{i+1} =
\frac{ \frac{\partial g_i (\epsilon) }{\partial x} + \frac{\partial g_i (\epsilon) }{\partial y} v_1 
+ \sum_{k=0}^\infty \frac{\partial g_k (\epsilon) }{\partial v_k} v_{k+1} }{ \frac{\partial f (\epsilon) }{\partial x} 
+ \frac{\partial f (\epsilon) }{\partial y} 
v_1 + \sum_{k=0}^\infty 
\frac{\partial f (\epsilon) }{\partial v_k} 
v_{k+1} } \;,\; i \geq 0 \;.
\end{equation}
To obtain the infinitesimal of a group of jet transformations,
the previous formulas are differentiated with respects to  $ \epsilon $
and then $ \epsilon $ is set equal to zero.  Before we do the calculations 
we define
\begin{align}
\nonumber
r(x,y, \vec{v} ) & = \frac{d}{d \epsilon} 
f(\epsilon ,x,y, \vec{v} ) \mid_{\epsilon =0 } \;, \\
s(x,y, \vec{v} ) & = \frac{d}{d \epsilon} 
g(\epsilon ,x,y, \vec{v} ) \mid_{\epsilon =0 } \;, \\
\nonumber
t_i (x,y, \vec{v} ) & = \frac{d}{d \epsilon} 
g_i (\epsilon ,x,y, \vec{v} ) \mid_{\epsilon =0 } \;.
\end{align}
The fact that a group is the identity when $ \epsilon = 0 $, implies that
\begin{equation}
f(0,x,y, \vec{v} ) = x,\;
g(0,x,y, \vec{v} ) = y,\;
\nu_i (0,x,y, \vec{v} ) = v_i \;.
\end{equation}
Now differentiating \eqref{Contact Condition}
with respects to $ \epsilon $ and setting $ \epsilon = 0 $ yields
(here $ t_0 = s $)
\begin{align} \label{Jet Transformation}
t_{i+1} (x,y, \vec{v} ) & = 
\frac{\partial t_i}{\partial x}
+ \frac{\partial t_i}{\partial y} 
v_1 + \sum_{k=1}^\infty
\frac{\partial t_k}{\partial v_k} v_{k+1} \\
\nonumber
& -v_{i+1} \left(
\frac{\partial r}{\partial x}
+ \frac{\partial r}{\partial y} 
v_1 + \sum_{k=1}^\infty
\frac{\partial r_k}{\partial v_k} 
v_{k+1} \right) \;,\; i \geq 0 \;.
\end{align}
An infinitesimal jet transformation is now a vector field 
\begin{equation}
\vec{T} (x, y, \vec{v} ) = (r (x, y, \vec{v} ) ,\; 
s (x , y, \vec{v} ) ,\; \vec{t} (x, y, \vec{v} ))
\end{equation}
that satisfies \eqref{Jet Transformation}. Note that  $ r $  and 
$ s $  determine the infinitesimal transformation and that they may be arbitrary
functions of a finite number of variables.
As before we can associate a first order linear partial differential operator,
\begin{equation}
L = r(x,y, \vec{v} ) \frac{\partial}{\partial x} + s(x,y, \vec{v} ) 
\frac{\partial}{\partial y} + \sum_{i=1}^\infty t_i (x,y, \vec{v} ) 
\partial_i
\end{equation}
with the vector field and that gives the infinitesimal action on functions 
of the variables $ (x,y, \vec{p} ) $.

Some of our formulas can be written more compactly if, as in 
[0], we introduce the operator
\begin{equation} \label{D equals}
D = \frac{\partial}{\partial x} + v_1 
\frac{\partial}{\partial y} + \sum_{k=1}^\infty 
v_{k+1} \frac{\partial}{\partial v_k } \;.
\end{equation}
Then the infinitesimal contact condition \eqref{Jet Transformation} can
be written
\begin{equation}
t_{i+1} = D t_i - v_{i+1} Dr ;'.
\end{equation}

Because all of our definitions of symmetries are based on the notion
of infinitesimal transformations, let us pause to make a complete
definition. \\[.2in]
{\bf Definition.}  Let
\begin{equation}
r(x,y, \vec{v} ) \qquad {\rm and} \qquad s(x,y, \vec{v} )
\end{equation}
be analytic functions of a finite subset the countable infinity of variables
\begin{equation}
(x,y, \vec{v} ) \;,\; \vec{v} = (v_1, v_2, \ldots )
\end{equation}
and let $ D $ be defined as in \eqref{D equals}.
If we set $ t_0 = s $ and then set
\begin{equation}
t_{i+1} (x,y, \vec{v} ) = Dt_i (x,y, \vec{v} ) - v_{i+1} Dr(x,y, \vec{v} )\;,
0 \leq i < \infty \;,
\end{equation}
then
\begin{equation} \label{Infinitesimal 1}
\vec{T} (x, y, \vec{v} ) = (r (x, y, \vec{v} ) \;,\; 
s (x, y, \vec{v} ) \;,\; \vec{t} (x, y, \vec{v} ) )`
\end{equation}
is called a jet vector field while
\begin{equation} \label{Infinitesimal 2}
L = r(x,y, \vec{v} ) \frac{\partial}{\partial x} + s(x,y, \vec{v} ) 
\frac{\partial}{\partial y} + \sum_{i=1}^\infty 
t_i (x,y, \vec{v} ) \frac{\partial}{\partial v_i }
\end{equation}
is called an infinitesimal jet operator.  Both \eqref{Infinitesimal 2}
and \eqref{Infinitesimal 1}
are referred to as infinitesimal jet transformations.

Once we have an infinitesimal jet transformation, then we would like to 
find the group of jet transformations associated with the infinitesimal
via the Lie series mechanism. Unfortunately, in this context, it appears
that the Lie series are rarely well defined. However, this formalism is
so intuitive we will continue to use it in a formal sense, that is,
apply the same rules of manipulation as were valid in our previous
discussions.  Notice that if  $ r $  and  $ s $
depend on only finitely many variables, then each
$ t_i $  depends on only finitely many variables.  However, all of the
$ t_i $'s  together depend, in general, on all of the variables.  It
would appear that the group action usually depends on infinitely many variables.

Let $ L $ be an infinitesimal jet transformation and then define
\begin{align}
\nonumber \xi (\epsilon ,x,y, \vec{v} ) & = e^{\epsilon L } x \;, \\
\eta (\epsilon ,x,y, \vec{v} ) & = e^{\epsilon L } y \;, \\
\nonumber \nu_i (\epsilon ,x,y, \vec{v} ) & = e^{\epsilon L} v_i \quad
0 \leq i < \infty \;.
\end{align}
Then the action of the Lie transformation on function
$ f (x, y, \vec{v} ) $ is given by the Composition Property,
$$ e^{\epsilon L } f(x,y, \vec{p} ) = f(\xi (x,y, \vec{v} ) \;,\;  
\eta (x,y, \vec{v} ) \;,\; \vec{\nu} (x,y, \vec{v} )) \;. $$

The idea of action on a surface that we discussed in Chapter 1
apply to jet transformations.  Thus let 
\begin{equation}
y = g(x)
\end{equation}
be a curve in the plane. This curve corresponds to a curve in the jet
given by the infinity of equations
\begin{align} \label{Values v}
\nonumber
y & = g(x) \;, \\
v_1 & = \frac{dg}{dx} (x) \;, \\
\nonumber
v_2 & = \frac{d^2 g}{dx^2 } \;, \ldots \;.
\end{align}
As we saw in Chapter 1, the infinitesimal action of the group on the curve 
was obtained by computing the action of $ L $ on the function $ y - g (x ) $.
If we call the infinitesimal action on the curve $ S $, then
\begin{equation}
S(f) = L (y - g(x) ) = s(x, y, \vec{v} ) - r(x, y, \vec{v} ) 
\frac{dg}{dx} (x) \;.
\end{equation}
However, the values of $ \vec{v} $ are given by \eqref{Values v}, so $ S $
should, in fact, be given by
\begin{equation}
S (f) = -r(x, g(x), \frac{dg}{dx} (x), \ldots ) 
\frac{dg}{dx} (x) + s(x, g(x), \frac{dg}{dx} (x), \ldots ) \;.
\end{equation}
This is the natural analog of the correspondence between infinitesimal
transformations and infinitesimal actions on curves given in Chapter 1.
If we write the correspondence as 
\begin{equation}
L \rightarrow S
\end{equation}
then it is important to notice that the correspondence is not one to one.
In fact if we define a new infinitesimal by
\begin{equation}
\tilde{r} = 0 \;,\; \tilde{s} = s - r v_1 \;,
\end{equation}
then the new and old infinitesimal transformations have the same action
on curves.
On the other hand, if we consider the operators  $ L $  such that 
$ r \equiv 0 $,
then for these operators the mapping is one to one.
Because we are interested in solutions of differential equations and
such solutions are curves, then any two transformations that have the 
same action on curves will be equivalent. Out of each equivalence class
we prefer the transformation with $ r \equiv 0 $ because the formula for  
$ t_{i+1} $ is particularly simple.

\begin{equation}
t_{i+1} (x,y, \vec{v} ) = Dt_i (x,y, \vec{v} ) \;,\; 1 \leq i < \infty \;.
\end{equation}
\vspace{.2in}
\begin{figure}
\begin{center}
\includegraphics[width=3.0in]{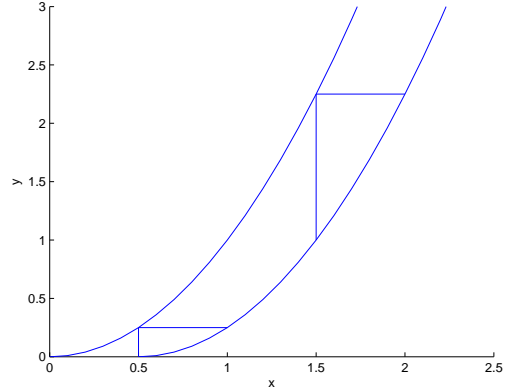}
\caption{Translation Equivlence \label{Figure Equivlence}}
\end{center}
\end{figure}

{\bf Example.}
In the case of translation, it is easy to see
geometrically what the above equivalence means.
Figure \ref{Figure Equivlence} shows that translation of the curve
$y = x^2$ by $1/2$ to the right where it becomes $y = (x-1/2)^2$.
at $x=1/2$ horizontal and vertical lines are drawen beteen the
first and second curves. Next these lines are moved to $x = 3/2$
so we see that the horizontal distance between the two curves
has not changed, but the vertical distance is much larger. The
slope of the first curve at $t = 1/2$ is $s = 1$ while
at $t = 3/2$, $s = 3$.  So translation to the right is equivalent
to a translation downward that increases as the slope of the curve
increases.

This can also be seen analytically. Let
\begin{equation}
L = \frac{\partial}{\partial x} \;,
\end{equation}
that is,
\begin{equation}
r = 1 \;,\; s = 0 \;,\; t_i = 0 \;,\; 1 \leq i < \infty \;.
\end{equation}
Using the formula \eqref{Translation r s}
we see that translation is equivalent to the infinitesimal vector field
\begin{equation}
r = 1 \;,\; s = -v_1 \;,\; t_1 = -v_2 \;,\; \ldots \;,
\end{equation}
that is, the infinitesimal operator
\begin{equation}
\tilde{L} = -v_1 \frac{\partial}{\partial y } - v_2 
\frac{\partial}{\partial v_1} - v_3 \frac{\partial}{\partial v_2}
- \ldots \;.
\end{equation}

It is possible to compress the notation in this section
slightly and this will be an advantage later.  To this end let
\begin{equation}
v_0 = y \;,\; \nu_0 = \eta \;,\; g_0 = g \;,\; t_0 = s \;.
\end{equation}
Also, redefine the vector notation so that 
\begin{align}
\vec{v} & = (y, v_1, v_2, \ldots ) \;,\; &
\vec{\nu} & = (\eta, \nu_1, \nu_2, \ldots ) \;, \\
\nonumber \vec{g} & = (g, g_1, g_2, \ldots ) \;,\;&
\vec{t} & = (s, t_1, t_2, \ldots ) \;.
\end{align}
Let us redo the definition of a jet transformation using this notation.  \\[.2in]
{\bf Definition.}  Let
\begin{equation}
r(x, \vec{v} ) \qquad {\rm and} \qquad s(x, \vec{v} )
\end{equation}
be analytic functions of a finite subset the countable infinity of variables
\begin{equation}
(x, \vec{v} ) \;,\; \vec{v} = (v_0, v_1, v_2, \ldots )
\end{equation}
and let $ D $ be defined as in
\eqref{D equals}.  Set
\begin{equation}
t_{i+1} (x,y, \vec{v} ) = Dt_i (x,y, \vec{v} ) - v_{i+1} Dr(x,y, \vec{v} ) 
\;,\; 0 \leq i < \infty \;,
\end{equation}
then
\begin{equation}
\vec{T} (x, y, \vec{v} ) = (r (x, y, \vec{v} ) , \vec{t} ( x, y, \vec{v} ))
\end{equation}
is called a jet vector field while
\begin{equation}
L = r(x,y, \vec{v} ) \frac{\partial}{\partial x} +
\sum_{i=1}^\infty t_i (x,y, \vec{v} ) \frac{\partial}{\partial v_i}
\end{equation}
is called an infinitesimal jet operator.

\section{General Jet Transformations}

A motivational discussion of jet transformations in several variables
would exactly parallel the discussion in the previous section, the
only difference is that there are many more indicies. For this
reason will not reproduce the motivation and instead we will
concentrate on setting up a notation that is easily understood and
helpful with our computer programs. The formulas we need are ``obvious''
generalizations of the formulas in the previous section.

We will consider the situation in which there are $ n $ independent 
variables and $ m $ dependent variables,
$$ \vec{x} = (x_1, x_2, \ldots , x_n) \;,\;
\vec{u} = (u_1, u_3, \ldots , u_m) \;. $$
A notation will be needed for derivatives of all orders,
\begin{align}
u_i^{ \{ 0 \} } & \Longleftrightarrow u_i \;, \\
\nonumber
u_i^{ \{ j \} } & \Longleftrightarrow \frac{\partial u_i}{\partial x_j} \;, \\
\nonumber u_i^{ \{ j,k \} } & \Longleftrightarrow
\frac{\partial^2 u_i}{\partial x_j \partial x_k} \;, \\
\nonumber u_i^{ \{ j,k, \ell \} } & \Longleftrightarrow
\frac{\partial^3  u_i}{\partial x_j \partial x_k \partial x_\ell } \;, \\
\nonumber \cdots \;.
\end{align}
The superscripts will soon become outrageous so we introduce a notation 
for them,
\begin{equation}
\vec{\sigma} = \{ \sigma_1, \sigma_2, \ldots , \sigma_k \} \;,
\end{equation}
where $ \sigma_i $ is an integer, $ 1 \leq \sigma_i \leq n $
and the sequence $ \vec{\sigma} $ may be of arbitrary length. Then
\begin{equation}
u_i^{ \vec{\sigma} } \;,\; \Longleftrightarrow 
\frac{\partial^k u_i}{\partial x_{\sigma_1} \ldots `
\partial x_{\sigma_k} } \;.
\end{equation}
Because it is possible to interchange the order of partial differentiation,
the sequences will be required to be nondecreasing,
\begin{equation}
0 < \sigma_1 \leq \sigma_2 \leq \ldots \leq \sigma_k \;. 
\end{equation}

The jet space will consist of the infinite set of variables
\begin{equation}
\vec{v} = ( \vec{x} , \vec{u} , \ldots , u_i^{ \vec{\sigma} } , \ldots ) \;.
\end{equation}

It will also be helpful to introduce an operation on the sequences, which
is written as  $ \{ \vec{\sigma}  ,i \} $ where  $ 1 \leq i \leq n $.
The value of $ \{ \vec{\sigma} ,i \} $ is a sequence with one more term than
$ \vec{\sigma} $ obtained by placing
$ i $  in  $ \vec{\sigma} $  in such a way that the result,
$ \{ \vec{\sigma} ,i \} $, remains nondecreasing.

One of the things that we use to help clarify dependency problems are the
{\it chain rule operators},
\begin{equation}
D_i = \frac{\partial}{\partial x_i} + \sum_{k=1}^m u_k^{ \{ i \} } 
\frac{\partial}{\partial u_k} + \sum_{k=1_{s \neq 0}}^m u_k^{ \{ s,i \} } 
\frac{\partial}{\partial u_k^s } \;,\;  1\leq i \leq n \;.
\end{equation}
In certain formulas, it is not clear whether
differentiation is being preformed before or after substitution.
The chain rule operators will clarify this situation. They will
also be useful in the computer programs.
Let $ F ( \vec{v} ) $ be a function on the jet and
$ \vec{u} = \vec{f} ( \vec{x} ) $ be a hyper surface, that is,
a mapping of the independent to the dependent variables.
We will use the same notation as before for the derivatives of $ \vec{f} $;
\begin{equation}
f_i^{ \vec{\sigma} } , \Longleftrightarrow
\frac{\partial^k f_i}{\partial x_{\sigma_{1}} \ldots \partial x_{\sigma_{k}}    } \;.
\end{equation}
Now it is possible to replace the jet variables with the derivatives
for which they stand;
\begin{equation}
F( \vec{f} ( \vec{x} ) ) \equiv
F( \vec{x} , \vec{f} ( \vec{x} ) , \ldots
f_i^{\vec{\sigma}} ( \vec{x} ) , \ldots ) \;.
\end{equation}
With all of the notation defined, the chain rule can be expressed
succinctly as follows. \\[.2in]
{\bf Proposition.}  Let  $ F ( \vec{v} ) $ and $ \vec{f} ( \vec{x} ) $ be given.
Then
\begin{equation}
\frac{\partial}{\partial x_i } (F ( \vec{f} ( \vec{x} ) ) ) =
D_i F ( \vec{v} ) \mid_{ \vec{v} = \vec{f} } \;.
\end{equation}

An infinitesimal jet vector field is given by
\begin{equation}
\vec{T} = ( \vec{r} ( \vec{v} ), \vec{s} ( \vec{v} ), \ldots ,
t_i^{ \vec{\sigma} } ( \vec{v} ), \ldots )
\end{equation}
where
\begin{equation} \label{vec s}
\vec{r} ( \vec{v} ) = (r_1 ( \vec{v} ), \ldots , r_n ( \vec{v} ) ) \;,\;
\vec{s} ( \vec{v} ) = (s_1 ( \vec{v} ), \ldots , s_m ( \vec{v} ) ) \;, 
\end{equation}
and the conditions on $ \vec{r} $, $ \vec{\sigma} $ and
$ t_i^\sigma ( \vec{v} ) $ are given below.  An infinitesimal jet operator is 
given by
\begin{equation} \label{Jet Operator}
L = \sum_{i=1}^n r_i ( \vec{v} ) 
\frac{\partial}{\partial x_i } + \sum_{i=1}^m s_i ( \vec{v} ) 
\frac{\partial}{\partial u_i } + \sum_{i=1_{ \vec{\sigma} \neq 0}}^m
t_i^{ \vec{\sigma} } ( \vec{v} ) 
\frac{\partial}{\partial u_i^{ \vec{\sigma}} }
\end{equation}
where
\begin{align}
\nonumber
& r_i ( \vec{v} ) \;,\; 1 \leq i \leq n \;,\; \\
& s_i ( \vec{v} ) \;,\; 1 \leq i \leq m \;, \\
\nonumber
& t_i^{ \vec{\sigma} } ( \vec{v} ) \;,\; 
1 \leq i \leq m \;,\; \vec{\sigma} \neq 0 \;,
\end{align}
are real analytic functions of a finite subset of the variables $ \vec{v} $.
We introduce the special superscript $ \{ 0 \} $ and define
$ \{ \{ 0 \} , i \} = \{ i \} $ and then if we set 
\begin{equation}
t_i^{ \{ 0 \} } = u_i \;,\; 1 \leq i \leq m \;.
\end{equation}
Now the $ t_i^{ \vec{\sigma} } $ are defined recursively by
\begin{equation}
t_i^{ \{ \vec{\sigma} ,k \} } =
D_k (t_i^{ \vec{\sigma} } ) - \sum_{j=1_{ \vec{\sigma} \neq 0 }}^n
u_i^{ \{ \vec{\sigma} , \ell \} }
D_k (s_\ell ) \;,\; 1 \leq i \leq n \;.
\end{equation}

We can write the group action associated with an infinitesimal jet
operator as a formal Lie series:
\begin{align}
\vec{\xi} (\epsilon) & = e^{\epsilon L} \vec{x} \;,\; \\
\vec{\eta} (\epsilon) & = e^{\epsilon L} \vec{u} \;, \\
\nonumber
\nu_i^{ \vec{\sigma} } (\epsilon) & = e^{\epsilon L} = 
u_i^{ \vec{\sigma} } \;.
\end{align}
The Composition Property of Lie series tells us that
\begin{equation}
e^{\epsilon L} F ( \vec{v} ) = g(e^{\epsilon L} \vec{x} \;,\;
e^{\epsilon L} \vec{u} , \ldots ,
e^{\epsilon L} u_i^{ \vec{\sigma} } , \ldots ) \;.
\end{equation}
This identity will be useful when we study the invariance of differential
equations under jet transformations.

Before we can formulate the invariance of differential equations
we need to know how infinitesimal jet transformations operate on
surfaces (hyper surfaces).  If $ \vec{u} = \vec{f} ( \vec{x} ) $
is a surface then this surface corresponds to a surface in the jet
variable $ \vec{v} $ given by
\begin{equation}
u_i^{ \vec{\sigma} } = 
\frac{\partial^k f_i}{\partial x_{\sigma_{1} } \ldots \partial x_{\sigma_{k} }}
( \vec{x} ) \;.
\end{equation}
It was seen Chapter 1 that if $ L $ is an infinitesimal jet operator, then
the action on surfaces is given by
\begin{align} \label{S Constraint}
\vec{S} ( \vec{f} ) & = L ( \vec{u} - \vec{f} ( \vec{x} ) )  \\
\nonumber
& = \vec{s} ( \vec{v} ) - \sum_{i=1}^n r_i ( \vec{v} ) 
\frac{\partial \vec{f} }{\partial x_i } 
( \vec{x} ) = -( \vec{r} ( \vec{v} ) \cdot \vec{\nabla}_x ) 
\vec{f} ( \vec{x} ) + \vec{s} ( \vec{v} ) \;.
\end{align}
The $ \vec{v} $ arguments are now constrained by
\eqref{S Constraint}, so the final 
form for the action on surfaces is
\begin{equation}
\vec{S} ( \vec{f} ) = - \vec{r} ( \vec{f} ( \vec{x} ) ) \cdot 
\vec{\nabla}_x \vec{f} ( \vec{x} ) + \vec{s} ( \vec{f} ( \vec{x} ) ) \;. 
\end{equation}
This agrees , of course, with the formulas derived in Chapter 1.

It is not possible to give an explicit formula of the exponential
\begin{equation}
e^{\epsilon \vec{S} } \vec{g} ( \vec{x} ) \;.
\end{equation}
However, this action can be described implicitly. One way of doing 
this is to write
\begin{equation}
\vec{g} ( \epsilon , \vec{x} , \vec{u} ) =
e^{\epsilon L}  ( \vec{u} - \vec{f} ( \vec{x} ) )
\end{equation}
and the solve
\begin{equation}
\vec{g} ( \epsilon , \vec{x} , \vec{u} ) = 0 
\end{equation}
for the transformed surface,
\begin{equation}
\vec{u} = \vec{f} ( \epsilon , \vec{x} ) \;.
\end{equation}
It is also possible to specify the evolution of the surface by an
initial value problem for an infinite order partial differential
equation,
\begin{equation}
\frac{\partial \vec{f} (\epsilon \vec{x} ) }{\partial \epsilon } =
\vec{S} ( \vec{x} , \vec{f} ( \epsilon , \vec{x} ) ) \;, 
\vec{f} ( 0, \vec{x} ) = \vec{f} ( \vec{x} ) \;.
\end{equation}
\vspace{.2in}
{\bf Remark.}  
It will be an unusual situation when the initial value problem for
$ f( \epsilon ,x) $  given in the previous proposition will be well posed,
so Lie Backlund transformation will not generate a well defined group
motion on surfaces.

As before, there is a notion of equivalence of infinitesimal
jet transformations. \\[.2in]
{\bf Definition.}  
Two infinitesimal jet transformations are said to be equivalent,
$ L \simeq \tilde{L} $, when the operators $ L $ and $ \tilde{L} $ 
produce the same action on surfaces. \\[.2in]
It is also important to know when jet transformations are, in fact, 
equivalent to a simpler transformation. \\[.2in]
{\bf Proposition.}  A jet transformation  $ L $ of the form \eqref{Jet Operator}
is equivalent to a point transformation if and only if \eqref{vec s}
\begin{equation}
\vec{r} ( \vec{v} ) - \vec{s} 
\end{equation}
is linear in $ v_i^{ \{ k \} } $. \\[.2in]
{\bf Proof.}  An infinitesimal point transformation is equivalent to 
\begin{equation}
L = \sum_{i=1}^n r_i ( \vec{x} , \vec{u} ) 
\frac{\partial}{\partial x_i} + \sum_{i=1}^m s_i ( \vec{x} , \vec{ \vec{u}} ) 
\frac{\partial}{\partial u_i} + \sum_{i=1_{ \vec{\sigma} \neq 0 }}^m
t_i^{ \vec{\sigma} } ( \vec{v} )
\frac{\partial}{\partial u_i^{ \vec{\sigma} } }
\end{equation}
where $ t_i^{ \vec{\sigma} } $ are defined in the usual way.
If the conditions of the proposition hold then the infinitesimal jet
symmetry is equivalent to an infinitesimal of the previous form.  \\[.2in]
{\bf Proposition.}  
A jet transformation  $ L $  is equivalent to a first order contact
transformation if and only if \eqref{vec s}
$ \vec{r} ( \vec{v} ) - \vec{s} $ depends only on 
$ \vec{x} $, $ \vec{u} $ and $ u_i^{ \{ k \} } $. \\[.2in]
{\bf Proof.}  
For such a  $ P $  to generate a contact transformation we must have
$$ P = \eta (x,u,u_1, \ldots , u_n ) - \sum u_i \xi_i (x,u,u_1, \ldots
, u_n ) \;. $$

\section{Invariance of Differential Equations}

The derivation of the conditions that describe the invariance of 
a system of differential equation under a group of jet transformations
is easy now that we have done the ground work in the previous section.
Also the introduction of the jet variables makes the  description of a
system of partial differential equations easy. Thus, $ \vec{F} $
is a system of $ \ell $ partial differential equations provided that
\begin{equation}
\vec{F} ( \vec{v} ) = ( F_1 ( \vec{v} ), \ldots , F_\ell ( \vec{v} ) )
\end{equation}
where $ F_i $, $ 1 \leq i \leq \ell $ are analytic functions
of a finite number of the jet variables $ \vec{v} $.
A surface (function) $ \vec{u} = \vec{f} ( \vec{x} ) $ is a solution of the 
system of partial differential equations provided that
\begin{equation}
\vec{F} ( \vec{f} ( \vec{x} )) = 0 \;.
\end{equation}
\vspace{.2in}
{\bf Theorem.}  The solution space of the differential equation
$$ \vec{F} ( \vec{g} ( \vec{x} )) = 0 $$
is invariant under the group of jet transformation  $ \exp(\epsilon L) $ 
if and only if
$$ L ( \vec{F} ( \vec{v} )) = 0 $$
when the infinite set of conditions written below hold:
$$ \vec{F} ( \vec{v} ) = 0 \;,\;
D_i \vec{F} ( \vec{v} ) = 0 \;,\;
D_i D_j ( \vec{F} ( \vec{v} )) = 0 \;,\; \ldots $$

\newpage \clearpage
\centerline{\bf REFERENCES}

The references are divided into three groups: the first group consists of
books, monographs, reviews and papers that should be of general interest;
the second group contains references to the computer symbol manipulation
literature while the the third group contains references to the research
literature.  We have also included some items of historical interest in
the first group.  The references to the research literature are by no
means complete.  We have attempted to provide a large sample of the 
recent research literature.  No coverage has been give to many 
related topics including similarity methods in engineering, Lie group
and symmetry methods in physics and the currently active mathematical
area of differential equations on Lie groups.  We have tried to emphasize
works that consider the differential equations to be the important given
object and then proceed to study the related symmetries, groups and 
algebras. \\[.2in]
\centerline{\bf GENERAL REFERENCES}
\begin{enumerate}
\item Ablowitz M.J. and H. Sequr, Solitons and the Inverse Scattering Transform,
Siam, Philadelphia, 1981.

\item Anderson, R.L. and N.H. Ibragimov, Lie-Backlund Transformations in 
Applications, SIAM, Philadelphia, 1979.

\item Belinfante, J.G.F. and B. Kolman, Lie Groups and Lie Algebras
with applications and computational methods, SIAM, Philadelphia, 1972.

\item Bluman G.W. and J.D. Cole, Similarity Methods for Differential Equations,
Springer-Verlag, New York, 1974.

\item Campbell, J.E., Introductory Treatise on Lie's Theory of Finite
Continuous Transformation Groups, Chelsea Pub. Co., New York, 1966.

\item Chester, C.R., Techniques in Partial Differential Equations,
McGraw-Hill, New York, 1971.

\item Cohen, A., An Introduction to the Lie Theory of One-parameter Groups
with Applica\-tions to the Solution of Differential Equations. 
D.C. Heath and Co. publishers, New York, 1911.

\item Eisenhart, L.P., Continuous Groups of Transformations,
Dover, New York, 1961.

\item Finkbeiner, D.T., Introduction to matrices and linear transformations,
W.H. Freeman and Co., San Francisco, 1960.

\item Giacaglia, G.E.O., Perturbation Methods in Non-linear Systems 
(See Ch. 1 for Lie series), Springer-Verlag, New York, 1972.

\item Gilmore, R., Lie Groups, Lie Algebras, and Some of Their Applications,
John Wiley and Sons, New York, 1974.

\item Hill, J.M., Solutions of differential equations by means of 
one-parameter groups.  Pitman Advanced Pub. Program, Boston, 1982.

\item Ince, E.L., Ordinary Differential Equations, Dover, New York, 1956.

\item Lie, S., Sophus Lie's 1880 Transformation Group Paper, Translated 
by M. Ackerman, Comments by R. Hermann, Math Sci Press, Brookline, 1975.

\item Lie, S., Sophus Lie's 1884 Differential Invariant Paper, Translated 
by M Ackerman, Comments by R. Hermann, Math Sci Press, Brookline, 1976.

\item Miller, Willard, Jr., Symmetry and Separation of Variables,
Addison-Wesley Pub. Co., London, 1977.

\item Olver, Peter J., Applications of Lie groups to Differetial Equations,
Mathematical Institute, Oxford.

\item Ovsiannikov, L.V., Group Analysis of Differential Equations,
Academic Press, New York, 1982.

\item Page, J.M., Ordinary Differential Equations with an Introduction to 
Lie's Theory of the Group of One-parameter.  Macmillian, New York, 1897.

\item Pommaret, Jean-Francois, Systems of Partial Differential Equations 
and Lie Pseudogroups, Gordon and Breach Science Publishers, New 
York-London-Paris, 1979.

\item Rogers, C. and W.F. Shadwick, Backlund Transformations and Their 
Applications, Academic Press, New York, 1982.

\item Sattinger, D.H., Group Theoretic Methods in Bifurcation Theory,
Springer-Verlag, New York, 1979.

\item Sattinger, D.H., Bifurcation and symmetry breaking in applied mathematics,
Bulletin AMS 3, 1980, 779-819.

\item Winternitz, P., Lie groups and solutions of nonlinear differential 
equations, Centre de recherche de Mathematiques appliquees,
Universite de Montreal, 1982.

\item Wybourne, B.G., Classical Groups for Physicists, John Wiley, New York, 
1974. \\[.2in]
\centerline{\bf SYMBOL MANIPULATION REFERENCES}
\item Char, B.W. and B. McNamara, LCPT: A program for finding linear 
canonical transformations, Lawrence Livermore Laboratory UCID-18185, 1979.

\item Edelen, D.G.B., Isovector Methods for Equations of Balance,
With programs for computer assistence in operator calculations
and an exposition of practiacal topics of the exterior calculus,
M. Nijhoff Pub., Dordrecht, 1880.

\item Kersten, P.H.M., The computation of the infinitesimal symmetries for 
(extended) vacuum Maxwell equations, using REDUCE 2, Technische Hogeschool 
Twente, 7500 AE Enschede, The Netherlands.

\item MACSYMA Reference Manual, The Mathlab Group, Lab. for Comp. Sci., 
MIT, 1983.

\item Reiman, A., Computer-Aided closure of the Lie-algebra associated with a
non-linear partial-differential equation, Computers and Mathematics wth 
Applications, 7, no. 5, 1981, 387-393.

\item Rosenau, P. and J.L. Schwarzmeier, Similarity Solutions of Systems of 
Partial Differential Equations Using MACSYMA, Courant Inst. of Math. Sci. 
Report No. COO-3077-160/MF-94, 1979.

\item Schwarz, F., A Reduce Package for Determining Lie Symmetries of Ordinary 
and Partial-Differential Equations, Computer Physics Comm., Vol. 27, No. 2, 
1982, 179-186.

\item Steinberg, S., Symmetry operators, Proceedings of the 1979 MACSYMA 
User's Conference, E. Lewis (editor), Washington, 1979, pages 408-444.

\item Steinberg, S., Change of variables in partial differential equations,
in preparation.

\item Wester M., S. Steinberg, An extension to MACSYMA's concept of functional 
differentiation, in preparation. \\[.2in]
\centerline{\bf RESEARCH REFERENCES}
\item Abellanas, L., A. Galindo, Conserved densities for non-linear 
evolution equations, 1. even order case, J. of Math. Phys., Vol. 20, 
No. 6, 1979, 1239-1243.

\item Alhassid, V., F. Gursey, F. Iachello, Potential scattering, 
transfer-matrix, and group theory, Physical Review Lett., Vol. 50, 
No. 12, 1983, 873-876.

\item Ames, K.A., W.F. Ames, On group analysis of the vonkarman equations,
Nonlinear Anal.-Theory methods and Appl., Vol. 6, N0. 8, 1982, 845-853.

\item Ames, W.F. and N.H. Ibragimov, Utilization of Group Properties in 
Computation, Paper presented at the International Joint IUTAM/IMU Symposium,
Novosibirsk, 1978.

\item Ames, W.F., R.J Lohner and E. Adams, Group properties of 
$ u_{tt} = [f(u)u_x]_x $, Internat. J. Non-Linear Mech. 16, 1981, 439-447.

\item Anderson, J.T., Lie-Algebra Approach to Symmetry Breaking,
Phys. Rev. D, 23, no. 8, 1981, 1856-1861.

\item Anderson, R.L., J. Harnad and P. Winternitz, Systems of ordinary 
differential equations with nonlinear superposition principles,
Physica 4D, 1982, 164-182.

\item Axford, R.A.  Determination of Invariance Properties of Second Order 
ODE's.  Notes from a lecture at LANL taken by M. Cheney.

\item Axford, R.A., Differential equations invariant under two-parameter 
Lie groups with applications to non-linear diffusion.  Los Alamos Report 
LA-4517 UC-34, 1970.

\item Baumgarte, J., Eine Lie-Algebra, die Delaunay-similar-Elemente in der 
exzentrischen Anomalie erzeugt, J. Phys. A, 13, no. 4, 1980, 1145-1158.

\item Beiglbock, Wolf, Arno Bohm, Eiichi Takasugi, Group theoretical 
methods in physics, Springer-Verlag, Berlin-New York, 1979.

\item Benjamin, T.B., P.J. Oliver, Hamiltonian structure, Symmetries and 
conservation laws for water waves, J. of Fluid Mech., Vol. 125, Dec. 1982, 
137-185.

\item Berman, V.S., I.A. Danilov, On group properties of Landau-Ginzburg 
generalized equation, Doklady Akademii Nauk SSSR, Vol. 258, No. 1, 1981, 67-70.

\item Bhutani, O.P., P. Mital, On the Soluion of Hudrodynamical Equations Via 
Lie-Groups, Inter. J. of Eng. Sci., Vol. 21, No. 5, 1983, 555-562.

\item Bluman, G. and S. Kumei, On the remarkable nonlinear diffusion equation
$ \partial/\partial x [a(u + b)^{-2} (\partial u/\partial x)] - 
\partial u/\partial t $, J. Math. Phys. 21, 1980, 1019.

\item Bouquet, S., M. Feix, Solution of Gravitation Polytrope equations by 
the quasi-invariance group, Comptes Rendus des Seances de L'Academie des 
Sciences Serie II-Mecanique Physiqu, Vol. 295, No. 12, 1982, 993-996.

\item Boyer, Charles P., Symmetries of differenital equations in mathematical 
physics, Lecture Notes in Phys., Vol. 50, Springer, Berlin, 1976, 425-434.

\item Boyer, C.P., E.G. Kalnins and W. Miller Jr., Completely integrable 
relativistic Hamiltonian systems and the separation of variables in Hermitian 
hyperbolic spaces, J. Math. Phys. 24, 1982.

\item Branson, T.P., W.H. Steeb, Symmetries of non-linear diffusion equations,
J. of Phys. A, Vol. 16, No. 3, 1983, 469-472.

\item Brockett, R.W., Lie Algebras and Lie Groups in Control Theory,
in Geometric Methods in Control Theory, (D.A. Mayne and R.W. Brockett, editors),
Reidel, Dordrecht, 1973.

\item Chandler, L.J.,
Separation of variables by the symmetry method for second order linear
partial differential equations,
Dept. of Math. and Stat., Univ. of New Mexico, Albuquerque, 1980.

\item Chattopadhyay, P, Noether's theorem and invariants of certain nonlinear 
systems, Phys. Lett. A, 75, No. 6, 1979, 457-459.

\item Chau, L.L., M.L. Ge, Y.S. Wu, Noether Currents and Albegraic Structure 
of the Hidden Symmetry for Super Chiral-Fields, Physical Rev. D, 25, no. 4, 
1982, 1080-1085.

\item Chern, Shiling Shen, Chia Kuei Peng, Lie groups and KdV equations,
Manuscripta Math. 28, no. 1-3, 1979, 207-217.

\item Chowdhury, A.R., Lie symmetries for  $ SD(2,1) $  Invariant non-linear 
sigma-model, Phys. Lett A., Vol. 93, No. 7, 1983, 317-318.

\item Constantopoulos, J.P., Lie-admissible deformation of selfadjoint systems,
Hadronic J., 3, no. 4, 1979, 1281-1312.

\item Corones, J., A lie group framework for soliton equations. I. Path 
independent case, J. Math. Phys. 18, 1977, 2207-2213.

\item Curtis, W.D., J.D. Logan, W.A. Parker, Dimensional analysis and the 
PI-theorem, Linear Alg. and its Appl., Vol. 47, Oct. 1982, 117-126.

\item Danilov, Y.A., G.I. Kuznetosov, Y.A. Smorodinskii,
On the symmetry of classical and wave equations,
Soviet J. of Nuclear Physics-USSR, Vol. 32, No. 6, 1980, 801-804.

\item Dewanwala, P., Group Theoretic Approach to the Solution of Fluid 
Mechanics Equations and Related Physical Systems.  Thesis, Dept. Math., 
Indian Institute of Technology, Delhi.

\item Dickson, L.E., Differential equations from the group standpoint,
Annal of Math. 25, 1942, 287-378.

\item Dongpei, Zhu., Noether Symmetry of the Single-Particle System,
J. of Phys. A, 14, no. 10, 1981, 2807-2816.

\item Dongpei, Zhu, The dynamical symmetry of isotropic systems,
J. Phys. A, 15, no. 1, 1982, 85-94.

\item Dorodnicyn, V.A., Group properties and invarinat solutions of an 
equation of nonlinear heat transport with a source or a sink,
Akad. Nauk SSSR Inst. Prikl. Mat. Preprint, No. 57, 1979, 31 pp.

\item Dryuma, V.S., Group interpretation of nonlinear wave equations 
integrable by the inverse-problem method from scattering theory.
Differential Equations 13, 1977, 1195-1197.

\item Eliezer, C.J., The symmetries and first integrals of some 
differential equations of dynamics, Hadronic J., 2, no. 5, 1979, 1067-1109.

\item Emets, Y.P., Y.P. Kovbasenko Group properties and invariant solutions 
of force-free magnetic field equations, Dopovidi Akademii Nauk Ukrainskoi 
RSR Seriya A, Vol. 1982, No. 6, 1982, 67-70. 

\item Ernst, M.H., Non-linear model Boltzmann equations and exact solutions,
Phys. Rep. C, Vol. 278, No. 1, 1981, 1-171.

\item Fischer, E., New similarity solutions for the Ernst equations with 
electromagnetic fields, J. of Math. Phys., Vol. 23, No. 7, 1982, 1295-1296.

\item Fokas, A.S., A symmetry approach to exactly solvable evolution equations,
J. Math. Phys. 21, 1980, 1318-1325.

\item Fokas, A.S. and R.L. Anderson, Group theoretic nature of Backlund 
transformations, Lett. Math. Phys. 3, 1979, 117-126.

\item Fokas, A.S. and B. Fuchssteiner, On the structure of symplectic 
operators and hereditary symmetries, Lett. Nuovo Cimento (2), 28, no. 8, 
1980, 299-303.

\item Fokas, A.S. and P.A. Lagerstrom, Quadratic and Cubic Invariants in 
Classical Mechanics, J. Math. Ana. Appl. 74, 1980, 325-341.

\item Fuchssteiner, Benno, The Lie algebra structure of nonlinear evolution 
equations admitting infinite-dimensional abelian symmetry groups,
Progr. Theoret. Phys., 65, no. 3, 1981, 861-876.

\item Fushchich, V.I., On a method of investigating the group properties of 
integro-differential equations, Ukrain. Mat. Zh., 33, n0. 6, 1981, 834-838, 862.

\item Fushchich, V.I., M.M. Serova, The maximal invariance group and the 
general solution of the unidimensional gas dynamic equations,
Doklady Akademii Nauk SSSR, Vol 268, No. 5, 1983, 1102-1104.

\item Fushchich, V.I., V.M. Shtelen, The invariant solutions of non-linear 
Dirac equation, Doklady Akademii Nauk SSSR, Vol. 269, No. 1, 1983, 88-92.

\item Fushchich, V.I., V.M. Shtelen, The symmetry and some exact solutions 
of the relativistic Eikonal equation, Lettre Al Nuovo Cimento, Vol. 34, 
No. 16, 1982, 498-502.

\item Fushchich, V.I., V.A. Vladimirov, Additional invariance of equations 
of motion for vector fields, Doklady Akademii Nauk, SSSR, Vol. 257, No. 5, 
1981, 1105-1109.

\item Garaev, K.G., Group-theoretic approach to the solution of the problem 
of the optimal control of a laminar boundary layer, Izv. Vyssh. Uchebn. 
Zaved. Aviacion. Tehn., no. 2, 1977, 40-44, 150.

\item Garaev, K.G., On a corollary of the Noether theorem for the 
two-dimensional problem of the Mayer type, J. Appl. Math. Mech., 44, 
no. 3, 1980, 316-320.

\item Gazeau, J.P., A. Maquet, Bound states in a Yukawa potential: a 
Sturmian group-theoretical approach, Phys. Rev. A (3), 20, no. 3, 1979, 
727-739.

\item Giachetti, R, Hamiltonian system swith symmetry: an introduction,
Nuovo Cimento (3), 4, no. 12, 1981, 63 pp.

\item Glockner, P.G. and M.C. Singh (Editors), Symmetry, Similarity and 
Group Theoretic Methods in Mechanics, University of Calgary, Calgary, 1974.

\item Gonzalez-Gascon, F., E. Aquerre-Daban, Notes on a paper by Lutzky on 
the non-canonical symmetries of a Hamiltonian-System, Phys. Lett. A 91, 
1982, 284-386.

\item Gonzalez-Gascon, F., F. Moreno-Insertis, E. Rodriguez-Camino,
Geometrical foundations and results on a problem suggested in a paper
by Anderson and Davison: ``A generalization of Lie's 'counting' theorem for
second-order ordinary differential equations", (J. Math. Anal. Appl., 48, 
1974, 301-315), Lett. Nuovo Cimento (2), 21, no. 17, 1978, 595-599.

\item Gue-Zhang, Tu, The Lie Algebra of the invariance group of the KdV, MKdV 
or Burgers equation, Lett. Math. Phys. 3, 1979, 387-393.

\item Guy, J. and B Mangeot, Use of group theory in various integral equations,
SIAM J. Appl. Math. 40, 1981, 390-399.

\item Hainzl, J., On a general concept for separation of variables,
SIAM J. on Math. Anal., Vol. 13, No. 2, 1982, 208-225.

\item Hainzl, J., Uber die Gestalt von gruppeninvarianten linearen partiellen
dieerentialoperatoren, Z. Angew. Math. Mech., 58, no. 7, 1978, T369-T370.

\item Harnad, J. and P. Witernitz, Pseudopotentials and Lie symmetries for 
the generalized nonlinear Schrodinger equation, J. Math. Phys. 23, 1982, 
517-525.

\item Harnad, J., P. Winternitz and R.L. Anderson, Superposition Principles 
for matrix Riccati Equations, J. Math. Phys. 24, 1983.

\item Hazewinkel, Michiel, On Lie algebras of vector fields, Lie algebras of 
differential operators and (nonlinear) filtering, Lecture Notes in Math., 
Springer, Berlin, 894, 1981, 91-106.

\item Hermann, R., C. Martin, Lie and Morse-Theory for Periodic-Orbits of 
Vector-Fields and Matrix Riccati-Equations, 1. General Lie-Theoretic Methods,
Mathematical Systems Theory, Vol. 15, No. 3, 1982, 277-284.

\item Hersh, R. and S. Steinberg, Hyperbolic equations with coefficients in 
an enveloping algebra, J. Diff. Eqns. 34, 1979, 405-426.

\item Hlavaty, L., S. Steinberg and K.B. Wolf, Riccati equations and Lie series,
J. Math. Anal Appl., to appear.

\item Hlavaty, L., S. Steinberg and K.B. Wolf, Nonlinear differential 
equations as invariants under group action of coset bundles, I. Burgers 
adn Korteweg-deVries equation families, in preparation.

\item Hlavaty, L., S. Steinberg and K.B. Wolf, Integral and Backlund 
transforms within symmetry groups of certain families of nonlinear 
differential equations, in preparation.

\item Holm, D.D.H., Symmetry Breaking in Fluid Dynamics: Lie Group Reducible 
Motions for Real Fluids, Los Alamos Report LA-6392-T, 1976.

\item Hou, B.Y., M.L. Ge, Y.S. Wu, Noether Analysis for the Hidden Symmetry 
Responsible for an Infinite Set of Nonlocal Currents, Phys. Rev. D., 24, 
no. 8, 1981, 2238-2244.

\item Ibragimov, N.H., The Equivalence of Evolution-Equations Which Allow an 
Infinite Lie-Backlund Algebra (French), I-Mathematique, Vo. 293, No. 14, 1981, 
657-660.

\item Ibragimov, N.H., On the theory of groups of Lie-Backlund transformations,
Mat. Sb. (N.S.), 109 (151), 1979, 229-253, 327.

\item Infeld, E., Invariants of the 2 dimensional Korteweg-Devries and 
Kadomisev-Petviashvili equations, Phys. Lett. A, Vol. 86, No. 4, 1981, 205-207.

\item Kalinis, E.G., W. Miller Jr., Intrinsic characterization of orthogonal 
R separation for Laplace equations, J. of Phys. A, Vol. 15, No. 9, 1982, 
2699-2709.

\item Kalnins, E.G. and W. Miller Jr., Killing tensors and nonorthogonal 
variable separation for Hamilton Jacobi equations, SIAM J. Math. Anal. 
12, 1981, 617-629.

\item Kaliappan, P., M. Lakshmanan, Connection between the infinite 
sequence of Lie-Backlund symmetries of the Korteweg-Devries ad Sine-Gordon 
equations, J. of Math. Phys., 23, no. 3, 1982, 456-459.

\item Kalotas, T.M., B.G. Wybourne, Dynamical Noether Symmetries,
J. of Phys. A, Vol. 15, No. 7, 1982, 2077-2083.

\item Kamin, S.  Continuous groups of transformations of differential 
equations: applications to free boundary problems, 1st Nax. Alta. Mat. 
Francesco Severi, Rome, 1980, 347-367.

\item Kapitanskii, L.V., Group analysis of the Navier-Stokes and Euler 
equations in the presence of rotation symmetry and new exact solutions 
to these equations, Dokl. Akad. Nauk SSSR, 243, no. 4, 1978, 901-904.

\item Katzin, Gerald, H., Jack Levine, Symmetries of the Hamilton-Jacobi 
equation and concomitant constants of motion, Tensor (N.S.), 34, n0. 2, 
1980, 179-198.

\item Katzin, Gerald, H., Jack Levine, Time-dependent quadratic constants 
of motion, symmetries, and orbit equations for classical particle dynamical 
systems with time-dependent Kepler potentials, Math. Phys., 23, no. 4, 1982, 
552-563.

\item Kaushal, R.S., H.J. Korsch, Dynamical Noether invariants for 
time-dependent nonlinear systems, J. Math. Phys., 22, no. 9, 1981, 1904-1908.

\item Kleinert, Hagen, New symmetries and constants of the motion from 
dynamical groups, Phys. Lett. B, 94, no. 3, 1980, 373-376.

\item Kobayashi, K., Derivation of the Pauli-Lenz vector and its variants,
J. of Phys. A, Vol. 13, No. 2, 1980, 425-430.

\item Kobussen, J.A., On symmetries and 1st integrals, Hadronic J., 5, 1982, 
1451-1478.

\item Koornwinder, T.H., A precise definition of separation of variables,
Stichting Mathematisch Centrum, Amsterdam, 1979.

\item Kondepudi, D.K., Bifurcation, Symmetry and the influence of an electric 
field on chemical dissipative structures, Phys. Lett. A., Vol. 77, No. 2-3, 
1980, 203-206.

\item Konopelchenko, B.G., V.G. Mokhnachev, On the group theoretical analysis 
of differential equations, J. of Phys. A, Vol. 13, No. 10, 1980, 3113-3124.

\item Kosmann-Schwarzback, Y., Generalized symmetries of nonlinear partial 
differential equations, Letters Math. Phys. 3, 1979, 395.

\item Kumei, S., Relationship between conservation laws and invariance 
groups of nonlinear field equations in Hamilton's canonical form,
J. of Math. Phys. Vol. 19, No. 1, 1978, 195-199.

\item Kumei, S. and G.W. Bluman, When nonlinear differential equations 
are equivalent to linear differential equations, SIAM J. Appl. Math. 
42, 1982, 1157-1174.

\item Lakshmanan, M., P. Kaliappan, Lie Transformations, Non-linear 
evolution equations, and Painleve Forms, J. Math. Phys. Vol. 24, 
No. 4, 1983, 795-806.

\item Leach, P.G.L., Application of the Lie theory of extended groups in 
Hamiltonian mechanics: the oscillator and the Kepler problem,
J. Austral. Math. Soc. Ser. B, 23, no. 2, 1981, 173-186.

\item Leach, P.G.L., Comment on a letter: "Noether's theorem and invariants 
of certain nonlinear systems" by Chatopadhyay, Phys. Lett. A, 84, no. 4, 
1981, 161-162.

\item Leo, M., R.A. Leo, G. Soliani, L. Solombrino, L. Martina,
Lie-Backlund Symmetries for the Harry-Dym Equation, Phys. Review 
D-Particles and Fields, Vol. 27, no. 6, 1983, 1406-1408.

\item Leroy, B., Group of invariance of (one-dimensional) sine-Gordon equation,
Lettrere Al Nuovo Cimento, Vol. 22, No. 1, 1978, 17-20.

\item Lloyd, S.P., The infinitesimal group of the Navier-Stokes equations,
Acta Mech. 38, 1981, 85-98.

\item Lonngren, K.E., Self similar solution of plasma equations,
Proc. Indian Acad. Sci. 86, 1977, 125-139.

\item Lounestor, Pertti, Lie groups of motor integrals of generalized Kepler 
motion, Celestial Mech., 17, no. 3, 1978, 207-213.

\item Lutzky, M., New classes of conserved quantities associated with 
non-Noether symmetries, J. Math. Phys. A, 15, no. 3, 1982, L87-L91.

\item Lutzky, M., Symmetry groups and conserved quantities for the harmonic 
oscillator, J. Phys. A, 11, no. 2, 1978, 249-258.

\item Maksimov, B.I., The Symmetry Transformations and Noether Transformations,
Izvestiya Vysshikh Uchebnykh Zavedenii Fizika, Vol. 25, No. 4, 1982, 124-125.

\item Marsden, Jerrold E., Lectures on geometric methods in mathemtical physics,
SIAM, Philadelphia, Pa., 1981.

\item McGuinnes, Mark J., Noether's theorem and infinities of polynomial 
conserved densities, Lecture Notes in Physics, 120, Springer, Berlin, 1980, 
363-368.

\item Meinhardt, J.R.  Symmetries and differential equations,
J. of Physics A, Vol. 14, No. 8, 1981, 1893-1914.

\item Miller, W., Jr., J. Patera, P. Winternitz, Subgroups of Lie groups 
and separation of variables, J. Math. Phys., 22, no. 2, 1981, 251-260.

\item Mitropol's'kii, J.A., V.I. Fushchich, Group theoretic methods in 
mathematical physics, Izdanie Inst. Mat., Akad. Nauk Ukrain. SSR, Kiev, 1978.

\item Munier, A., J.R. Burgan, J. Gutierrez, E. Fijalkow and M.R. Feix,
Group transformations and the non-linear heat diffusion equation.
SIAM J. Appl. Math. 40, 1981, 191-207.

\item Nariboli, G.A., Group invariant solutions of the Fokker-Planck equation,
Stochastic Processes and their Applications 5, 1977, 157-171.

\item Nikitin, A.G., V.V. Nakonechnyi, The invariance algebras of Dirac and 
Schrodinger equations, Ukrain. Fiz. Zh., 25, no. 4, 1980, 618-621, 696.

\item Nono, Takayuki, Fumitake Mimura, Material symmetries, I.
Bull. Fukuoka Univ. Ed. III, 28, 1979, 21-31.

\item Olshanetsky, M.A., A.M. Perelomov, Classical integrable 
finite-dimensional systems related to Lie algebras, Phys. Rep., 71, 
no. 5, 1981, 313-400.

\item Olver, P.J., Symmetry groups and group invariant solutions of 
partial differential equations, J. Diff. Geom. 14, 1979.

\item Orfanidis, S.J., A group-theoretical approach to optimal estimation 
and control, Dept. of Elec. Eng., Rutgers Uni., Piscataway, 1981.

\item Osborne, Anthony, D., Allan E.G. Stuart, On the separability of the 
sine-Gordon equation and similar quasilinear partial differential equations, 
II. Dependent and independent-variable transformations, J. Math. Phys., 21, 
no. 4, 1980, 726-734.

\item Otterson, P., G. Svetlichny, On derivative dependent infinitesimal 
deformations of differentiable maps, J. of Diff. Eq., Vol. 36, No. 2, 1980, 
270-294.

\item Phanthien, N., A method to obtain some similarity soltuions to the 
generalized Newtonian fluid, Zeitscrhrift fur Angewandte Mathematik und 
Physik, Vol. 32, No. 5, 1981, 609-615.

\item Phanthien, N., Invariance group of the plane squeezing flow of a 
viscous fluid, J. of Appl. Mechan., Vol. 47, No. 1, 1980, 213-214.

\item Pommaret, J.-F., A survey of Galois theory for systems of partial 
differential equations and its applications in Physics, Physica A, Vol. 
114, No. 1-3, 1982, 114-123.

\item Prince, G.E., C.J. Eliezer, On the Lie Symmetries of the 
classical Kepler problem, J. of Phys. A, 14, no. 3, 1981, 587-596.

\item Prince, G.E., C.J. Eliezer, Symmetries of the time-dependent 
N-dimensional oscillator, J. Phys. A, 13, no. 3, 1980, 815-823.

\item Prince, G.E., P.G.L. Leach, The Lie theory of extended groups in 
Hamiltonian mechanics, Hadronic J. Vol. 3, No. 3, 1980, 941-961.

\item Prince, G.E., P.G.L. Leach, T.M. Kalotas, C.J. Eliezer, R.M. Santilli,
The Lie and Lie-admissible symmetries of dynamical systems, Hadronic J., 3, 
no. 1, 1979, 390-439.

\item Procopius, Gh., Classification of invariant solutions of RMHD with 
planar symmetry, Bul. Inst. Politehn. Iasi Sect. I, 25(29) no. 1-2, 1979, 
81-85.

\item Ratiu, Tudor, Euler-Poisson equations on Lie algebras and the 
N-dimensional heavy rigid body, Proc. Nat. Acad. Sci. U.S.A., 78, no. 3, 
part 1, 1981, 1327-1328.

\item Ratiu, T. and P. vanMoerbeke, The Lagrange rigid body motion,
Ann. Inst. Fourier 32, 1982, 211-234.

\item Ray, J. R., J.L. Reid, J.J. Cullen, Lie and Noether Symmetry Groups 
of Non-linear Equations, J. of Phys. A. Vol. 15, no. 11, 1982, L575.

\item Reid, J.L., J.R. Ray, Lie Symmetries, Non-linear Equations of Motion 
and New Ermakov Systems, J. of Phys. A, Vol. 15, No. 9, 1982, 2751-2760.

\item Reiman, A.G., Integrable Hamiltonian systems connected with graded 
Lie algebras, Zap. Nauchn. Sem. Leningrad, Otdel. Mat. Inst. Steklov. (LOMI), 
95, 1980, 3-54, 161.

\item Reyman, A.G., M.A. Semenov-Taih-Shansky, Reduction of Hamiltonian 
systems, affine Lie algebra and Lax equations, II., Invent. Math., 63, 
no. 3, 1981, 423-432.

\item Rosen, G., Restricted invariance of the Navier-Stokes equation,
Phys. Rev. A, Vol. 22, No. 1, 1980, 313-314.

\item Rosencrans, S.I., Conservation laws generated by pairs of 
non-Hamiltonian symmetries, Dept. of Math., Tulane Uni., New Orleans, 1980.

\item Rubel, L.A. and B.A. Taylor, An example of a rigid partial 
differential equation, J. Diff. Eqns. 38, 1980, 126-133.

\item Sarlet, W.  Symmetries, 1st integrals and the inverse problem of 
Lagrangian mechanics, J. of Physics A, Vol. 14, No. 9, 1981, 2227-2238.

\item Sarlet, W., F. Cantrijn, Generalization of Noether's theorem in 
classical mechanics, SIAM Rev., 23, no. 4, 1981, 467-494.

\item Sarlet, W., F. Cantrijn, Higher order Noether symmetries and 
constants of the motion, J. of Phys. A, 14, no. 2, 1981, 479-492.

\item Segeda, Y.N., Certain Invariant solutions for a non-linear wave equation,
Ukrainskii Fizicheskii Zhurnal, Vol. 27, No. 5, 1982, 787-788.

\item Shadwick, W.F., Case's conjecture on Backlund transformation and 
conservation laws, Pre\-print, Dept. Appl. Math., University of Waterloo, 1982.

\item Shadwick, W.F., The Hamilton Cartan formalism for rth-order Lagrangians 
and the integrability of the KdV and modified KdV equations, Lett. Math. Phys.,
5, no. 2, 1981, 137-141.

\item Sheftel, M.B., Lie-Backlund Invariance-Groups of the One-Dimensional 
Gas-Dynamics Equations, Vestnik Leningradskogo Universiteta Seriya Matematiki 
Mekhaniki Astronomi, Vol. 1982, No. 2, 1982, 37-41.

\item Shmelev, G.S., Differential-Operators Invariant with Respect to Lie 
Superalgebra $ H(2,2, \lambda ) $ and its Irreducible Representations,
Doklady Bolgarskoi Akademii Nauk., Vol. 35, no. 3, 1982, 287-290.

\item Shtelen, V.M., Group analysis of a system of non-linear differential 
equation equivalent to the Schrodinger equation, Ukrainskii Fizicheskii 
Zhurnal, Vol. 26, No. 2, 1981, 323-326.

\item Sibirskii, K.S., Algebraic invariants of differential equations,
Akademie- Verlag, Berlin, 1977, 269-277.

\item Steeb, W.H., Symmetries and Vacuum Maxwells Equations,
J. of Math. Phys., Vol. 21, No. 7, 1980, 1656-1658.

\item Steeb, W.H., Lie algebras and dynamic nonlinear systems containing 
limit cycles, Internat. J. Theoret. Phys., 16, no. 9, 1977, 671-679.

\item Steeb, W.H., W. Oevel, Backlund Transformation groups of non-linear 
evolution equations and the Painleve property, Zietschrift fur Naturforschung 
Part A, Vol. 38, No. 1, 1983, 86-87.

\item Steeb, W.H., W. Strampp, Diffusion-Equations and Lie and Lie-Backlund 
Transformation Groups, Physica A, Vo. 114, No. 1-3, 1982, 95-99.

\item Steeb, W.H., W. Strampp, Symmetries of Non-linear reaction diffusion 
equations and their solutions, Physica D, vol. 3, No. 3, 1981, 637-643.

\item Steeb, W.H., W. Erig, W. Strampp, Symmetries and the Dirac equation,
J. of Math. Phys., Vol. 22, No. 5, 1981, 970-973.

\item Steinberg, S., Applications of the Lie algebraic formulas of Baker, 
Campbell, Hausdorff and Zassenhaus to the explicit solutions of partial 
differential equations, J. Diff. Eqns. 26, 1977, 404-434.

\item Steinberg, S., Lie theory and differential equations, pages 365-371 in
Information Linkage Between Applied Mathematics and Industry,
P.C.C. Wang (editor), Academic Press, New York, 1979.

\item Steinberg, S., Lie algebras in ill-posed problems, Proceedings of the 
International Symposium on Ill-Posed Problems: Theroy and Practice, M.Z. 
Nashed (editor), Reidel, New York, 1982.

\item Steinberg, S., Lie series and nonlinear ordinary equations,
J. Math. Anal. Appl, to appear.

\item Steinberg, S., Factored product expansions of solutions of nonlinear 
differential equations, SIAM J. Math. Anal., to appear.

\item Steinberg, S. and K.B. Wolf, Groups of integral transforms generated 
by Lie algebras of second-and-higher order differential operators,
Il Nuova Cimento 53A2, 1979, 149-177.

\item Steinberg, S., and K.B. Wolf, Symmetry, conserved quantities and 
moments in diffusive equations, J. Math. Ana. Appl. 80, 1981, 36-45.

\item Stepin, A.M., Polynomial integrals of Hamiltonian systems,
Soc. Math. France, Paris, Asterisque, no. 51, 1978, 429-441.

\item Steudel, H., An infinite set of conservation laws derived by 
Noether's theorem for several nonlinear evolution equations,
Akademie-Verlag, Berlin, 1977, 309-312.

\item Strampp, W., Solution of Partial differential equations using 
transformation groups, Zeitschrift fur Angewandte Mathematik und Mecchanik,
Vol 59, No. 3, 1979, 42.

\item Strampp, W., Invariant group solutions of hydrodynamical equations,
J. of Math. Anal. and Appl., Vol. 78, No. 2, 1980, 618-633.

\item Strauss, W.A., Nonlinear invariant wave equations, Invariant Wave 
Equations (editors G. Velo and A.S. Wightman), Springer-Verlag, New York, 
1977.

\item Suhinin, S.V., A group property and conservation laws of an equation of
transonic motion of a gas, Sinamika Sploshn. Sredy, No. 36 Dinamika Zhidkosti 
so Svobodnymi Granicami, 1978, 130-137, 162.

\item Suyarov, U.S., Noether's theorem for an electron, Dokl. Akad. 
Nauk USSr, 1981, no. 7, 1981, 9-11.

\item Tajiri, M., S. Kawamoto, Reduction of KDV and cylindrical KDV 
equations to Painleve equation, J. of the Physical Soc. of Japan, Vol. 51, 
No. 5, 1982, 1678-1681.

\item Tamizhmani, K.M., M. Lakshmanan, Infinitely many Lie-Backlund 
symmetries for a Quasilinear Evolution Equation, Phys. Lett. A., Vol. 90, 
no. 4, 1982, 159-161.

\item Truax, D.R., Time dependent Schrodinger equations - symmetry 
breaking separation of variables and non-linear effects,
Int. J. of Quantum Chem., Vol. 23, No. 2, 1983, 663-678.

\item Truax, D.R., Dynamical symmetries of rotationally invariant, 
three-dimensional, Schrodinger equations, J. Math. Phys., 21, no. 4, 1980, 
807-817.

\item Truax, D.R., Symmetry of time-dependent Schrodinger equations exact 
solutions for the equation ``not printed", J. of Math. Phys., Vol. 23, 
No. 1, 1982, 43-54.

\item Tu, Gue Zhang, The Lie algebra of invariant group of the KdV, MKdV, or 
Burgers equation, Lett. Math. Phys., 3, no. 5, 1979, 387-393.

\item Vanderbauwhede, A.L., Generic bifurcation and symmetry with an 
application to von Karman equations, Proceedings of the Royal Soc. of 
Edinburgh Section A, Vol. 81, 1978, 3-4, 211-235.

\item Vladimirov, S.A., Symmetry groups of Lagrangians of chiral fields 
taking values in  $ S^2 $, Teoret. Mat. Fiz., 44, no. 3, 1980, 410-413.

\item Vujanovic, B., Conservation laws of dynamical systems via 
d'Alembert's principle, Internat. J. Non-Linear Mech., 13, no. 3, 1978, 
185-197.

\item Vulpe, N.I., K.S. Sibirskii, Affine classification concomitants of a 
quadratic system, Akademie-Verlag, Berlin 1977, 379-382.

\item Wadati, M., Infinitesimal transformations and conservation laws; 
field theoretic approach to the theory of soliton, Accad. Lincei, Rome, 
Pitmann, London, 1978, Res. Notes in Math., 26, 33-63.

\item Winternitz, P., Non-linear action of Lie groups and superposition 
principles for non-linear differential equations, Physica A, Vol. 114, 
No. 1-3, 1982, 105-113.

\item Wolf, K.B. (Editor), Group Theoretical methods in Physics (See 
the section on differential equations), Lecture Notes in Physics, 
Springer-Verlag, New York, 1980.

\item Wolf, K.B. (Editor), School and Workshop on Nonlinear Phenomena,
in preparation.

\item Wulfman, C.E., Dynamical groups in atomic and molecular physics, 
Recent advances in group theory and their application to spectroscopy 
(Proc. NATO Adv. Study Inst., St. Francis Xavier Univ., Antigonish, N.S., 
1978), Plenum, New York, 1979, 329-403.

\item Wulfman, C.E., Limit cycles as invariant functions of Lie groups,
J. Phys. A, 12, no. 4, 1979, L73-L75.

\item Wulfman, C.E. and B.G. Wybourne, The Lie group of Newton's and 
Lagrange's equations for the harmonic oscillator, J. Phys A: Math. Gen. 9, 
1976, 507-518.

\item Wulfman, C.E., Systematic methods for determining the continuous 
transformation groups admitted by differential equations, Symposium on 
Symmetries in Science, S. Ill. Univ., Carbondale, 1979.

\item Zakharov, N.S. and V.P. Korobeinikov, Group analysis of the 
generalized Korteweg-DeVries-Burgers equations, J. Appl. Math. Mech. 44, 
1981, 668-671.

\item Xu, Bo Wei, The maximal kinematical group of the general 
Schrodinger equation, J. Phys. A, 14, no. 5, 1981, L123-L124.
\end{enumerate}

\begin{appendix}
\newpage \clearpage
\setcounter{equation}{0}
\chapter{}
\section{One Term PDE's}
We present a method for solving equations of the form
\begin{equation} \label{Problem}
\frac{\partial^{p_1} }{\partial x_1^{p_1}} 
\frac{\partial^{p_2}}{\partial x_2^{p_2}}
\cdots
\frac{\partial^{p_n}}{\partial x_n^{p_n}}
f(x_1 ,..., x_n ) = 0
\end{equation}
will be described. Here $ p_i \geq 0 $.
Recall that repeated integration gives the general solution of
the one variable equation
\begin{equation}
\frac{d^j }{ dx^j } f(x) = 0
\end{equation}
as
\begin{equation}
f(x) = \sum_{i=0}^{j-1} a_i \, x^i \,,
\end{equation}
where $ a_i, 0 \leq i \leq j-1 $ are constants.
The mulitvariable case is a bit more complicated.

{\bf Proposition.}
The general solution of the partial differential equation \eqref{Problem} is
\begin{equation} \label{General}
f(x_1 ,..., x_n ) = \sum_{k=1}^{n} P_k \,,
\end{equation}
where
\begin{equation} \label{Solution}
P_k = \sum_{i=0}^{k -1 } a_i^{k} x_k^i
\end{equation}
and $ a_i^k $ is an arbitrary function of all of the variables
$ (x_1 ,..., x_n ) $ except $ x_k . $

{\bf Proof}.
If all of the $ p_i = 0 $ then the differential equation is trivial
and the formula \eqref{Solution} gives the correct solution,
$ f \equiv  0 $. Now proceed by induction. Assume that \eqref{Solution}
is true and then try to solve
\begin{equation} \label{Induction}
\frac{\partial^{p_1+1}} {\partial x_1^{p_1+1} } 
\frac{\partial^{p_2}} {\partial x_2^{p_2}} ... 
\frac{\partial^{p_n} } {\partial x_n^{p_n} } \tilde{f} = 0 .
\end{equation}
If we set $ f =\partial \tilde{f} /\partial x_1 , $ then $ f $
satisfies \eqref{Problem} and then the induction hypothesis
says that $ f $ is given by \eqref{General}.

If $ P_k $ is one of the terms in \eqref{Solution}
and $ P_k $ is antidifferentiated with respects to $ x_1 $,
then two possible things happen. If $ k \neq 1 , $ then the
antiderivative of $ P_k $ has has exactly the same form as
$ P_k$ with new coefficients $\tilde{a}^k_i$. If $k = 1$ then
\begin{equation}
\int P_k dx_1 = \sum _{i=0}^ { p_k -1 } 
    a_i^k \frac{ x_1^{i+1} }{i+1} + a
\end{equation}
where $ a $ depends on all of the variables except $ x_1 $.
A change of summation index gives, as was desired,
\begin{equation}
\int P_k dx_1 = \sum_{i=0}^{p_k} \tilde{a}_i^k x_1^i ,
\end{equation}
where $\tilde{a}_i^0 = a $
and $\tilde{a}_i^{k+1} = a_i^k $, or $1 \leq i \leq p_k$.
Because it is possible to interchange the order of partial differentiation,
the argument for $ x_1 $ works for any $ x_j $ and consequently
the induction is finished.

\end{appendix}

\end{document}